\documentclass[11pt,letterpaper]{article}
\usepackage[T1]{fontenc}
\usepackage{lmodern,microtype,amsmath,amssymb,amsthm,mathtools,booktabs,array}
\usepackage[margin=1in]{geometry}
\usepackage[shortlabels]{enumitem}
\usepackage{tikz,needspace,float}
\usetikzlibrary{arrows.meta,positioning,calc}
\usepackage[hidelinks]{hyperref}
\numberwithin{equation}{section}
\newtheorem{theorem}{Theorem}[section]
\newtheorem{lemma}[theorem]{Lemma}
\newtheorem{proposition}[theorem]{Proposition}
\newtheorem{corollary}[theorem]{Corollary}
\theoremstyle{definition}
\newtheorem{definition}[theorem]{Definition}
\theoremstyle{remark}
\newtheorem*{remark}{Remark}
\newcommand{\KG}{\mathrm{KG}}
\newcommand{\C}{\mathcal C}
\newcommand{\N}{\mathcal N}
\newcommand{\F}{\mathbb F}
\newcommand{\Z}{\mathbb Z}
\newcommand{\R}{\mathbb R}
\newcommand{\kk}{\Bbbk}
\newcommand{\one}{\mathbf 1}
\newcommand{\Hr}{\widetilde H}
\newcommand{\sm}{\smallsetminus}
\newcommand{\cK}{\mathcal K}
\newcommand{\cP}{\mathcal P}
\newcommand{\cZ}{\mathcal Z}

\newcommand{\cX}{\mathcal X}
\newcommand{\fT}{\mathfrak T}
\DeclareMathOperator{\Ex}{Ex}

\DeclareMathOperator{\conv}{conv}
\DeclareMathOperator{\aug}{aug}
\newcommand{\step}[1]{\par\smallskip\noindent\emph{#1}\enspace}
\hypersetup{pdftitle={Balanced Fair Division and Generalized Kneser Graphs},
 pdfauthor={Sharareh Alipour and Amir Jafari}}
\title{Balanced Fair Division and\\Generalized Kneser Graphs}
\author{\begin{tabular}{@{}c@{\qquad}c@{}}
Sharareh Alipour & Amir Jafari\\
\normalsize \\
\normalsize \end{tabular}}
\date{}
\begin{document}
\hypersetup{pageanchor=false}
\begin{titlepage}
\maketitle
\thispagestyle{empty}
\begin{abstract}
\normalsize
We study the allocation of indivisible goods between two prescribed
nonempty groups whose members have arbitrary monotone valuations.
We prove that at most five agents always admit an allocation that is
envy-free up to one good (EF1), with bundle sizes differing by at most one.
This balanced guarantee is sharp: every split of six agents can fail.
We also guarantee balanced allocations that are envy-free up to two goods
(EF2) for at most nine agents, improving the previous five-agent
consequence of known chromatic bounds. Both existence guarantees hold
for any number of goods.

Building on the connection established by Kyropoulou, Suksompong, and
Voudouris, we derive these guarantees from chromatic bounds for generalized
Kneser graphs. The graph $\KG(2n,n,r)$ has the $n$-subsets of a $2n$-set
as vertices, with edges joining sets whose intersection has size less
than $r$. We prove that its chromatic number is six for $r=2$ and
$n\ge2$, resolving the conjecture of Jafari and Alipour, and at least
ten for $r=3$ and $n\ge4$, improving the previous lower bound of six;
the known upper bound is fourteen.

We determine the relevant box-complex $\Z_2$-indices exactly, showing
that the standard chromatic lower bound gives only five and nine colors
for $n\ge3$ and $n\ge5$, respectively.
To obtain the extra color, we use restrictions on maps induced by actual
colorings. A hypothetical proper five- or nine-coloring must induce a
nonzero map on homology. We exhibit generators supported on pentagon
or rook configurations and show that local coloring constraints force
every generator to map to zero. The ten-color proof compares integral
and fractional common centers and uses a sharp rounding theorem for
five subset constraints.

A linear-time retraction from the full allocation graph onto its
balanced layers adds cardinality balance to known nonlinear discrepancy
guarantees. For each fixed removal allowance, in the guaranteed-existence
range, finding a balanced fair allocation or an explicit monotonicity
violation from valuation circuits is polynomial-time equivalent to
finding a monochromatic edge in a circuit-specified coloring of the
corresponding graph. The reverse reduction produces normalized monotone
binary valuations using ordinary circuits.
\end{abstract}
\end{titlepage}
\clearpage
\thispagestyle{empty}
\tableofcontents
\clearpage
\pagenumbering{arabic}
\hypersetup{pageanchor=true}
\section{Introduction}\label{sec:intro}

Suppose that indivisible goods must be divided between two prescribed
nonempty groups, such as households sharing their allocated goods.
Every member of a group receives the same bundle, but each agent has
her own monotone valuation. Exact envy-freeness
may be impossible even for one good. An allocation is \emph{envy-free
up to $c$ goods}, or EF$c$, if each agent ceases to envy the other
bundle after at most $c$ goods are removed from it; the removed goods
may depend on the agent. We require also that the two bundle sizes
differ by at most one, and call such an allocation \emph{balanced}.
Our first result determines the number of agents for which balanced
EF1 is always possible and improves the corresponding guarantee for EF2.

\begin{theorem}[Sharp balanced EF1; EF2 for nine agents]\label{cor:fair-main}
Every instance with at most five agents admits balanced EF1, and every
instance with at most nine agents admits balanced EF2, for arbitrary
monotone valuations and every prescribed split into two nonempty groups.
For every split of six agents and every even number of goods at least
four, monotone $0/1$-valued valuations can prevent balanced EF1.
\end{theorem}

Kyropoulou, Suksompong, and Voudouris related balanced EF1 to a
generalized Kneser graph and proved matching allocation obstructions
\cite[Theorem~4.4 and its tightness argument]{KSV}. They observed that
a conjecture of Jafari and Alipour \cite{AJ} would guarantee five
agents. We prove that conjecture, making their sharp threshold
unconditional for both splits $(1,4)$ and $(2,3)$.
The result also extends the four-agent additive guarantee of
G\"olz and Yaghoubizade \cite{GolzYaghoubizade} to five agents with
arbitrary monotone valuations. For EF2, the previous chromatic bound
of Daneshpajouh \cite{Daneshpajouh} guaranteed five agents; ours
guarantees nine. Section~\ref{sec:fair} states and proves the exact
correspondence for every allowance $c$.

\paragraph{The chromatic results.}
Let $\KG(N,k,r)$ have the $k$-subsets of $[N]=\{1,\ldots,N\}$ as
vertices; two distinct sets are adjacent when their intersection has
size less than $r$. A proper coloring partitions these sets into
families in which every pair has at least $r$ common elements.
Write $\chi(G)$ for the chromatic number and put
$G_{n,r}=\KG(2n,n,r)$. For $r=1$, the balanced graph consists of
complementary pairs and has chromatic number two.

\begin{theorem}\label{thm:main}
For every $n\ge2$, $\chi(G_{n,2})=6$. For every $n\ge4$,
\[
                         10\le\chi(G_{n,3})\le14.
\]
In the second statement, equality to fourteen holds when $n=4$.
\end{theorem}

The six-color equality is the conjecture of \cite{AJ} from 2017.
The ten-color bound raises the uniform lower bound from six
\cite[Theorem~1]{Daneshpajouh}; fourteen is the Fano-plane upper bound
of Balogh, Cherkashin, and Kiselev \cite{BCK}.
These are also allocation thresholds: for $2n$ goods with $n>c$,
balanced EF$c$ is guaranteed for every profile of monotone valuations
and a prescribed split of $q$ agents exactly when
$q<\chi(G_{n,c+1})$. A rejecting agent gives an independent family
of allocations, and every independent family can be realized by a
binary monotone valuation. Thus the correspondence gives both
existence and counterexamples.

We also construct a linear-time balancing retraction and extend the
correspondence to allocation and coloring search.

\paragraph{The extra color beyond the index bound.}
Lov\'asz's topological proof of Kneser's conjecture turns graph
colorings into maps to spheres \cite{Lovasz}. In its box-complex
form, $B(G)$ records complete bipartite subgraphs, with an involution
exchanging their sides. Its $\Z_2$-index is the least $d$ for which
a continuous map $B(G)\to S^d$ sends this exchange to $x\mapsto-x$.
The standard bound is
$\chi(G)\ge\operatorname{ind}_{\Z_2}B(G)+2$.

\begin{theorem}\label{thm:index-intro}
For $r\ge2$ and $n\ge2r-1$,
\[
             \operatorname{ind}_{\Z_2}B(G_{n,r})\le4r-5.
\]
Equality holds for $r=2$ and $r=3$.
\end{theorem}

The exact indices are therefore three and seven, giving only five
and nine colors when $n\ge3$ and $n\ge5$, respectively. Thus the
standard index bound stops short of both chromatic conclusions.
Our argument uses the additional restrictions on maps induced by
actual colorings. A putative five- or nine-coloring must induce a
nonzero map on homology. We give generators supported on pentagon
or rook configurations and prove that every such configuration repeats
a color, forcing every generator to map to zero. The local coloring
arguments are a count on five points for the pentagon and a sequence
of forced colors for the rook graph.

Identifying the generators is the topological part of the proof.
We first allow the common centers of a family of sets to be fractional.
The resulting complex has a geometric model made of coordinate faces
of a hypersimplex, a slice of a cube with fixed coordinate sum.
Boundaries of the first omitted faces generate its relevant homology;
the pentagon and rook cycles map with degree one onto these boundaries.
For radius one, rounding five labels preserves the chain groups and
boundary maps that compute third homology.

For radius two, the relevant seven-dimensional cycles require control
of eight-dimensional fillings, which five-label rounding does not provide. We construct a chain map that
returns fractional fillings to integral ones while fixing the original
cycle. The local extension steps reduce to integral homology calculations
in two fixed complexes with $84$ and $70$ labels. This comparison,
proved in Section~\ref{sec:comparison}, is what permits the same
coloring argument to reach ten colors.

\paragraph{General allowances and computation.}
The full allocation graph allows bundles of all sizes. We construct
a complement-preserving retraction onto its central layers, in linear
time, and prove that its chromatic number equals that of the balanced
graph (Theorem~\ref{thm:exact-balancing}). Combining this with the
nonlinear discrepancy bounds obtained independently and concurrently
in \cite{HMMS26,DF25} gives balanced EF$c$ for $q$ agents with
$c=O(\sqrt{q\log q})$.
The discrepancy estimate is inherited; the retraction supplies balance.
In chromatic terms it gives $\chi(G_{n,r})=\Omega(r^2/\log r)$,
uniformly for $n\ge r$. For $n\ge2r-1$ this exceeds the linear index
ceiling when $r$ is large (Section~\ref{ss:chromatic-growth}).

For fixed $c$, the correspondence also extends to search with succinct
inputs when $n>c$ and $q<\chi(G_{n,c+1})$
(Theorem~\ref{thm:search-equivalence}). Given rational valuation
circuits, finding a balanced EF$c$ allocation or an explicit
monotonicity violation is polynomial-time equivalent to finding a
monochromatic edge in a circuit-specified $q$-coloring of $G_{n,c+1}$.
The reductions preserve the goods, allowance, and number of agents.
The reverse reduction produces normalized monotone binary valuations
using ordinary circuits; bounded completion searches make this possible.

\paragraph{Rounding five constraints.}
The rounding input also gives a standalone statement
(Theorem~\ref{thm:rounding}). With a fixed integer coordinate sum,
any family of at most five subset inequalities with integer lower bounds
that is feasible in $[0,1]^N$ has a $0/1$ solution. Six constraints can fail.
The statement applies to arbitrary subsets, beyond the common-center
constraints used in the coloring proofs.

\paragraph{Related perspectives.}
Colorful subgraph theorems of Simonyi and Tardos \cite{ST06} also
extract color patterns from topology. Our separation concerns the
standard box-complex index bound, within the hierarchy discussed in
\cite{DM24}. In fair division, consensus halving for sets of items
\cite{GHIMS22} permits fractional items and seeks two bundles that
each agent values equally. Here the goods remain indivisible, the groups are prescribed,
and balance constrains the numbers of goods rather than their values.

\paragraph{Open problems.}
What is $\chi(G_{n,3})$ for $n\ge5$, between ten and fourteen?
The discrepancy consequence already gives $\chi(G_{n,r})\ge4r-2$
for $r\ge56$ and $n\ge r$. Can the explicit-generator and local-coloring
method prove this bound for small fixed $r\ge4$, beginning with $r=4$,
uniformly for $n\ge2r-1$? Can the five-agent balanced EF1 and nine-agent balanced
EF2 search problems be solved in polynomial time, and do they admit
membership or hardness results for PPA, the class based on undirected parity? The search equivalence transfers
these computational questions without resolving them.

\paragraph{Reading the paper.}
Section~\ref{sec:fair} gives the fair-division correspondence.
Sections~\ref{sec:pentagon} and~\ref{sec:rook} explain the chromatic
proofs; Sections~\ref{sec:balancing} and~\ref{sec:search} give the
balancing and search results. The detailed reductions, rook coloring
argument, and topological proofs follow. Appendices~\ref{sec:filling}--\ref{sec:eight}
prove the finite integral homology inputs. No theorem depends on computation.

\section{Balanced fair division and the chromatic threshold}\label{sec:fair}

An agent's rejection family is independent in the allocation graph,
and every independent family arises in this way. This correspondence
turns a chromatic number into an exact threshold for every prescribed
split of the agents.

Table~\ref{tab:fair-comparison} compares the small-agent guarantees.
The assumptions differ: G\"olz and Yaghoubizade obtain an efficient
algorithm and balance under additivity
\cite[Theorem~3.1 and Corollary~3.2]{GolzYaghoubizade}, while the
revised monotone theorem of Bu, Li, Liu, Song, and Tao concerns two
groups of two without asserting cardinality balance
\cite[Theorem~3.1]{BuEtAl}.

\begin{table}[ht]
\centering
\setlength{\tabcolsep}{4pt}
\begin{tabular}{@{}>{\raggedright\arraybackslash}p{1.52in}
>{\raggedright\arraybackslash}p{1.02in}
>{\raggedright\arraybackslash}p{2.45in}
>{\raggedright\arraybackslash}p{.82in}@{}}
\toprule
Work and guarantee & Valuations & Prescribed agents & Balanced\\
\midrule
G\"olz--Yaghoubizade \cite{GolzYaghoubizade}, EF1
 & additive & at most four, every split & yes\\[3pt]
\cite{BuEtAl}, EF1
 & arbitrary monotone & two groups of two & not asserted\\[3pt]
\cite{Daneshpajouh} and the chromatic correspondence, EF2
 & arbitrary monotone & at most five, every split & yes\\[3pt]
\textbf{This paper, EF1}
 & arbitrary monotone & \textbf{at most five, every split} & \textbf{yes; sharp}\\[3pt]
\textbf{This paper, EF2}
 & arbitrary monotone & \textbf{at most nine, every split} & \textbf{yes}\\
\bottomrule
\end{tabular}
\caption{Group fair-division guarantees. All memberships are fixed
in advance. Balance means that the two bundle sizes differ by at most
one. The Daneshpajouh row is the consequence of his chromatic bound
under the correspondence proved below. The EF1 sharpness is witnessed
by six agents on every even number of goods at least four.}
\label{tab:fair-comparison}
\end{table}

\paragraph{The exact chromatic correspondence.}
We give the proof for general EF$c$. For EF1, both directions and the
$0/1$-valued construction are due to \cite{KSV}; they also noted the
extension of the existence argument to EF$c$.

Let $M=[2n]$ be a set of goods, and fix two nonempty groups of sizes
$q_1,q_2$. Write $q=q_1+q_2$. Each agent has a normalized monotone
valuation $u:2^M\to\R_{\ge0}$: $u(\varnothing)=0$, and $u(S)\le u(T)$
when $S\subseteq T$. Subtracting $u(\varnothing)$ from every value
does not change any envy comparison, so normalization costs nothing.
The group sizes and memberships are fixed before
the goods are allocated. A balanced allocation is specified by the
$n$-set $A$ given to the first group; the second receives $A^c=M\sm A$.
An agent receiving $P$ accepts the allocation as EF$c$ if
\[
            u(P)\ge u(Q\sm T)\quad\text{for some }T\subseteq Q,
                              \quad |T|\le c,
\]
where $Q$ is the other bundle. Otherwise she \emph{rejects} it.

\begin{theorem}[The balanced threshold]\label{thm:balanced-threshold}
Let $n>c\ge1$ and $q_1,q_2\ge1$. Every profile of normalized monotone
valuations admits a balanced EF$c$ allocation for this split if and only if
\[
                         q_1+q_2<\chi(G_{n,c+1}).
\]
If the inequality fails, a counterexample exists with $0/1$-valued
valuations.
\end{theorem}

The proof follows the chromatic framework of \cite{KSV}. We include
both directions, since one gives existence and the other gives sharpness.

\begin{lemma}\label{lem:rejection-families}
A family of balanced allocations is the rejection family of a monotone
agent if and only if it is independent in $G_{n,c+1}$. Every independent
family is realizable by a normalized $0/1$-valued agent in either group.
\end{lemma}

\begin{proof}
Suppose a first-group agent rejects adjacent allocations $A,B$.
Deleting $A^c\cap B^c$ from the other bundle at $A$ is allowed, since
$|A^c\cap B^c|=|A\cap B|\le c$. Rejection and monotonicity imply
\[
                     u(A)<u(A^c\cap B)\le u(B).
\]
Reversing $A$ and $B$ gives the opposite inequality, a contradiction.
For a second-group agent, apply the same argument to complements.

Conversely, let $\mathcal I$ be a nonempty independent family. Define
\[
 u_{\mathcal I}(S)=
 \begin{cases}0,&S\subseteq B\text{ for some }B\in\mathcal I,\\
               1,&\text{otherwise}.
 \end{cases}
\]
This is normalized and monotone. An $n$-set outside $\mathcal I$ has
value one, so it is accepted by a first-group agent.
For $A\in\mathcal I$, suppose that $A^c\sm T$ has value zero for some
$|T|\le c$. Then $A^c\sm T\subseteq B$ for some $B\in\mathcal I$, and
\[
                      |A\cap B|=|A^c\cap B^c|\le |T|\le c.
\]
Since $n>c$, the set $A^c\sm T$ is nonempty, so $B\ne A$.
This contradicts independence. Thus exactly the allocations in
$\mathcal I$ are rejected. For the second group, use the same construction
with $\{A^c:A\in\mathcal I\}$. The empty family is realized by the
identically zero valuation.
\end{proof}

\begin{proof}[Proof of Theorem~\ref{thm:balanced-threshold}]
If every allocation were rejected, color it by the name of one rejecting
agent. The lemma makes this a proper $q$-coloring. Hence $q<\chi(G_{n,c+1})$
guarantees an accepted allocation.
Conversely, given a proper coloring with at most $q$ colors, assign one
color class to each agent and realize it as her rejection family in the
prescribed group. Unused colors give zero valuations. Every allocation
is rejected, which proves the converse.
\end{proof}

\paragraph{Consequences and boundary cases.}
The equality $\chi(G_{n,2})=6$ gives balanced EF1 for five agents and,
by the tightness construction of \cite{KSV}, a counterexample for every
split of six agents on every even number of goods at least four.
The bound $\chi(G_{n,3})\ge10$ gives balanced EF2
for nine agents when $n\ge4$. For $n=3$, $G_{3,3}=K_{20}$, so the same
guarantee holds. If $n\le c$, any balanced allocation is EF$c$, since
the entire comparison bundle may be deleted.

For $2n-1$ goods, add a dummy good $d$ with
$u'(S)=u(S\sm\{d\})$. Remove $d$ from a balanced EF$c$ allocation on
the enlarged set. The bundle sizes now differ by one, and replacing
each deletion witness $T$ by $T\sm\{d\}$ preserves EF$c$.
This proves Theorem~\ref{cor:fair-main}. The dummy-good argument
transfers existence, not the even-size counterexamples.

Together with \cite[Propositions~3.3 and~3.4, Corollary~3.6]{KSV},
Theorem~\ref{cor:fair-main} classifies the group-size pairs for
unrestricted monotone EF1, except $(5,1)$, up to exchanging groups.
At most five agents always admit EF1; every larger split other than
$(5,1)$ can fail, by those counterexamples and adding zero-valued agents.
For $(5,1)$, balanced EF1 can fail \cite[Proposition~4.7]{KSV}, whereas
EF1 without the balance requirement is guaranteed for binary additive
valuations \cite[Theorem~3.1]{KSV}. The general monotone case is not
resolved by these results.

For EF2, the known fourteen-color
construction gives counterexamples for every split of fourteen agents
on every even number of goods at least eight. Thus the largest universal
balanced EF2 agent threshold $Q_2$ satisfies $9\le Q_2\le13$.
The upper side is inherited from the known coloring; the new result is
the lower side. On exactly eight goods the threshold is $13$, since
$\chi(G_{4,3})=14$.
By Lemma~\ref{lem:balanced-deletion}, $\chi(G_{n,3})$ is nonincreasing
for $n\ge3$ and therefore eventually constant, with value $\chi_\infty$.
Theorem~\ref{thm:balanced-threshold}, the small-bundle cases and the
dummy-good argument give $Q_2=\min_{n\ge3}\chi(G_{n,3})-1=\chi_\infty-1$.
Thus the universal balanced EF2 threshold is determined by the eventual
chromatic number; the argument gives no bound on when stabilization occurs.

\section{The six-color theorem}\label{sec:pentagon}\label{sec:framework}

We show why five colors cannot suffice for $G_{n,2}$ when $n\ge3$.
There are three steps: a coloring gives a map to a sphere; pentagons
represent the homology that this map must preserve; and a count of
pairs shows that every pentagon loses a color.

\subsection{The coloring map}
The \emph{neighborhood complex} $\N(G)$ has as its nonempty faces the
vertex families with a common neighbor. A simplicial complex records
a collection of finite sets, closed under taking subsets; its geometric
realization fills each nonempty set by a simplex. We write $|K|$ for
the realization of a complex $K$. For a finite vertex set $\sigma$,
$|\sigma|$ is its cardinality, $\Delta^\sigma$ is the full simplex on it,
and $|\Delta^\sigma|$ is that simplex's realization. For example, three
vertices with a common neighbor span a filled triangle in $\N(G)$.

Suppose that $\kappa:V(G)\to[5]$ is a proper coloring.
A face of $\N(G)$ cannot use the color of its common neighbor.
Thus the coloring defines a simplicial map
\begin{equation}\label{eq:color-map}
       \kappa_\N:\N(G)\longrightarrow\partial\Delta^4\cong S^3.
\end{equation}
Here $\Delta^4$ is the simplex on the five colors; its boundary consists
of all proper subsets of those colors.

We use homology with coefficients in $\F_2$ in the coloring argument.
A $d$-cycle is a sum of $d$-dimensional simplices whose boundary is zero;
two cycles represent the same homology class if their difference is a
boundary. We write $H_j(K;\kk)$ and $\Hr_j(K;\kk)$ for ordinary and
reduced homology with coefficients $\kk=\Z$ or $\F_2$. They agree in
positive degrees; for a nonempty complex, $\Hr_0(K;\kk)=0$ means that
$|K|$ is connected. A simplicial map $f$ induces a homomorphism $f_*$
on homology. A space is $t$-connected if every map from $S^j$ into it
extends over a ball for $0\le j\le t$.

The following fact is the topological constraint on a coloring.
We state it for arbitrary $d$, but use only $d=3$ in this section.
Its proof is given in Section~\ref{ss:nonvanishing-proof}.

\begin{lemma}[Nonvanishing]\label{lem:nonvanishing}
Let $d\ge1$, and suppose that $\N(G)$ is $(d-1)$-connected. Every proper
coloring with colors in $[d+2]$ induces a nonzero homomorphism
\[
 (\kappa_\N)_*:H_d(\N(G);\F_2)
       \longrightarrow H_d(\partial\Delta^{d+1};\F_2)\cong\F_2.
\]
\end{lemma}

For five colors, the induced map must therefore preserve some class
in $H_3$. Our task is to exhibit generators and show that it kills
every one. A useful observation is that a cycle supported on five
vertices maps to zero whenever two of its vertices have the same
color: its image lies in one face of $\Delta^4$, which is contractible
(it deforms continuously to a point).

\subsection{Pentagons give the relevant cycles}
For $k$-subsets $A,Z$ of $[N]$, the \emph{Johnson distance}
$d_J(A,Z)=k-|A\cap Z|$ is the number of replacements needed to turn
$A$ into $Z$. Let $\C_1(N,k)$ be the complex of families with a common
center at Johnson distance at most one. In $G_{n,2}$, complementation
turns a common neighbor into a center:
\begin{equation}\label{eq:nbhd-one}
                       \N(G_{n,2})=\C_1(2n,n).
\end{equation}
Indeed, $|A\cap B|\le1$ is equivalent to
$d_J(A,[2n]\sm B)\le1$. We call the vertices of this complex
\emph{labels} when distinguishing them from centers.

On a five-point set $D$, arrange a pentagon. Let $A_i$ be the two
neighbors of point $i$, and let $Z_i$ be its two nonneighbors.
Then
\[
             |A_i\cap Z_i|=0,\qquad |A_i\cap Z_j|=1\quad(i\ne j).
\]
Thus $Z_j$ is a radius-one center of all four labels other than $A_j$.
Every proper subfamily of $A_1,\ldots,A_5$ is a face of $\C_1(5,2)$.
The five tetrahedra obtained by omitting one label at a time form the
boundary of a four-dimensional simplex, hence a three-cycle.

\begin{figure}[ht]
\centering
\begin{tikzpicture}[scale=.78,every node/.style={font=\normalsize}]
\foreach \j/\angle/\col in {1/90/gray!15,2/18/blue!20,3/-54/orange!25,4/-126/orange!25,5/162/blue!20}
 \node[draw,circle,fill=\col,minimum size=7mm,inner sep=0pt] (p\j) at (\angle:1.45) {$\j$};
\foreach \u/\v in {1/2,2/3,3/4,4/5,5/1} \draw (p\u)--(p\v);
\node[align=left,anchor=west] at (2.1,.3) {$A_1=\{2,5\}$: blue points\\$Z_1=\{3,4\}$: orange points\\$Z_j$ centers the four labels other than $A_j$};
\end{tikzpicture}
\caption{The pentagon provides five labels and five centers.
The cycle is the boundary on the five set-labels $A_i$, not the
drawn polygon on the five ground-set points.}
\label{fig:pentagon}
\end{figure}
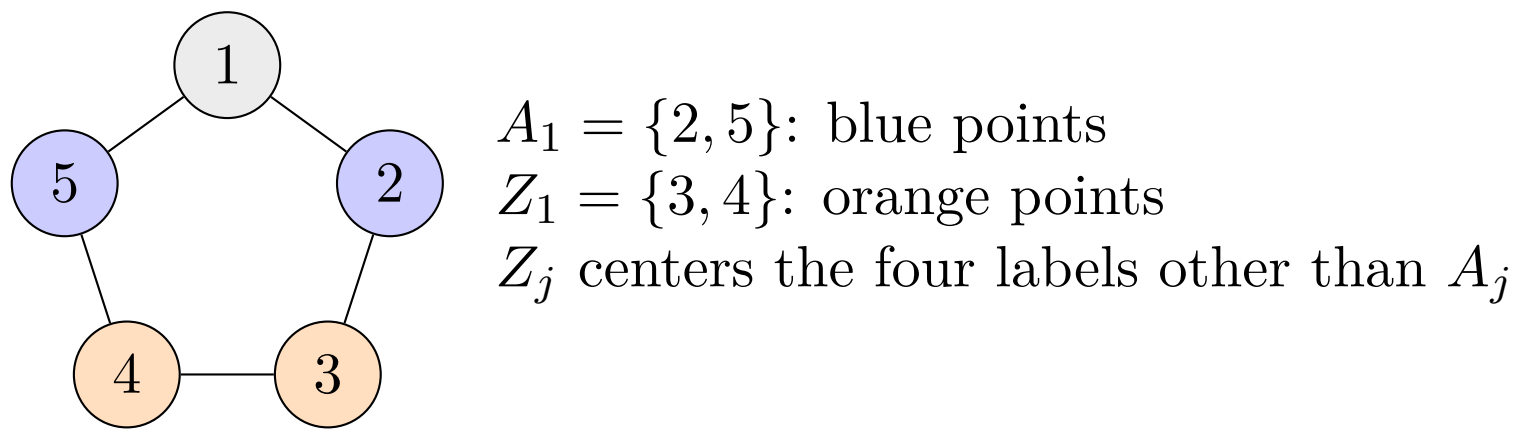

To obtain $n$-sets, choose disjoint $I,D\subseteq[2n]$ with
$|I|=n-2$ and $|D|=5$, and use the labels $I\cup A_i$.
Their centers are $I\cup Z_j$. We also use the cycles obtained by
complementing all five $n$-sets.

\begin{proposition}[Pentagon generators]\label{prop:pentagon-topology}
For $n\ge3$, the complex $\N(G_{n,2})$ is $2$-connected, and its
$H_3(-;\F_2)$ is spanned by the pentagon cycles just described.
\end{proposition}

\begin{figure}[ht]
\centering
\begin{tikzpicture}[scale=.9,every node/.style={font=\normalsize}]
\draw[thick] (-2,-1)--(0,-1.65)--(2,-1)--(2,1)--(0,1.65)--(-2,1)--cycle;
\draw[thick,blue!65!black,fill=blue!5] (0,1.05)--(-1.15,-.65)--(1.15,-.65)--cycle;
\fill (0,0) circle (1.4pt) node[left] {$b_F$};
\draw[->,thick,orange!85!black] (0,1.05)--(0,1.65);
\draw[->,thick,orange!85!black] (-1.15,-.65)--(-1.853,-1.0476);
\draw[->,thick,orange!85!black] (1.15,-.65)--(1.853,-1.0476);
\node[align=left,anchor=west] at (2.55,.65) {inner boundary: configuration cycle\\outer boundary: coordinate face};
\node[align=left,anchor=west] at (2.55,-.5) {each ray from $b_F$ meets\\each boundary exactly once};
\end{tikzpicture}
\caption{A planar illustration of the degree-one argument. For the pentagon, the inner simplex and coordinate face
have dimension four; for the rook graph they have dimension eight.
Here $b_F$ is their common barycenter, contained in both interiors.}
\label{fig:radial}
\end{figure}
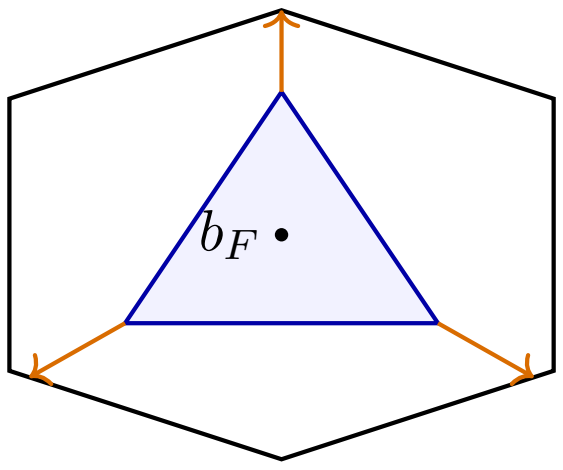

There are two tasks: compute the homology after allowing fractional
centers, then return to integral centers. Replace a set by its $0/1$
indicator vector and allow centers to be fractional: $x\in[0,1]^{2n}$ with
$\sum_i x_i=n$. The fractional complex has a geometric model built
from coordinate faces of this hypersimplex, obtained by fixing some
coordinates to zero or one. The model contains every face of dimension
at most three. The first missing faces have five free coordinates,
of total weight two or three; their boundaries generate the third
homology. Each pentagon cycle projects once onto one of these boundaries,
as in Figure~\ref{fig:radial}.

To return to integral centers, an elementary rounding lemma for five
labels shows that the relaxation preserves all simplices of dimension at most
four, hence the chain groups and boundary maps that compute $H_3$.
Section~\ref{sec:geometry} proves these statements; its short
radius-one rounding proof is in Section~\ref{ss:rounding-one}.
The six-color argument uses no radius-two comparison theorem.

\subsection{A count of ten pairs}
Call a family \emph{rainbow} if its labels have distinct colors.
For a pair $Z\subseteq D$, let
\[
       \mathcal B_1(Z)=\{A\in\tbinom D2:A\cap Z\ne\varnothing\}.
\]
We first prove a local statement on just five ground-set points.

\begin{lemma}[The pentagon loses a color]\label{lem:pentagon-local}
Let $\varphi:\binom D2\to[5]$. If each ball $\mathcal B_1(Z)$ misses
a color, then the five pentagon labels $A_i$ are not rainbow.
\end{lemma}
\begin{proof}
Suppose they are rainbow, and name their colors by $i$.
For every pair $Z$, choose a missing color $\psi(Z)$.
The incidence rule forces $\psi(Z_i)=i$.
Thus both $\varphi$ and $\psi$ use all five colors, and
\[
             \varphi(A)\ne\psi(Z)\quad\hbox{if }A\cap Z\ne\varnothing.
\]
For each color $i$, let $U_i$ be the union of the pairs given that color
by $\varphi$. At least one pair has that color, so $|U_i|\ge2$.
A pair colored $i$ by $\psi$ must avoid $U_i$, so $|U_i|\le3$.

Every pair is contained in some $U_i$, by its $\varphi$-color, and in
some $D\sm U_i$, by its $\psi$-color. If $a$ of the five sets $U_i$
have size three, counting the ten pairs gives
\[
                         10\le5+2a,\qquad 10\le15-2a.
\]
The first inequality requires $a\ge3$ and the second $a\le2$.
\end{proof}

\subsection{Finishing the six-color theorem}
Suppose that $\kappa$ is a proper five-coloring of $G_{n,2}$, $n\ge3$.
For a pentagon configuration with fixed core $I$, put
$\varphi(A)=\kappa(I\cup A)$. If $A\in\mathcal B_1(Z)$, then
\[
       |(I\cup A)\cap([2n]\sm(I\cup Z))|=2-|A\cap Z|\le1.
\]
These two $n$-sets are distinct because $I\ne\varnothing$.
Thus the entire ball misses the color of $[2n]\sm(I\cup Z)$,
and Lemma~\ref{lem:pentagon-local} makes the configuration nonrainbow.
The same is true of a complemented configuration: apply the argument
to the proper coloring $A\mapsto\kappa(A^c)$, using
$|A^c\cap B^c|=|A\cap B|$ for $n$-sets in $[2n]$.

Every generating three-cycle therefore maps into a proper face of the
simplex on the five colors, so $(\kappa_\N)_*$ is zero on every
generator. Proposition~\ref{prop:pentagon-topology} makes it zero on
all of $H_3$, contradicting Lemma~\ref{lem:nonvanishing} with $d=3$.
Hence $\chi(G_{n,2})\ge6$.

For the upper bound, fix three points of $[2n]$. Every $n$-set $A$ or
its complement contains at least two of them. Choose one such pair
and record the pair and its side. This gives six colors. Sets with
the same positive color contain a common pair; for a negative color,
their complements do, which gives the same intersection bound.
Thus the coloring is proper. When $n=2$, every two distinct $2$-sets
of $[4]$ are adjacent, so $G_{2,2}=K_6$. This proves
$\chi(G_{n,2})=6$ for every $n\ge2$.

\paragraph{The general coloring criterion.}
The preceding argument has the following form for an arbitrary graph.
It packages the two steps that will also finish the rook proof.

\begin{proposition}\label{prop:coloring-criterion}
Let $d\ge1$. Suppose that $\N(G)$ is $(d-1)$-connected and that cycles
$z_\alpha$ generate $H_d(\N(G);\F_2)$. If every proper $(d+2)$-coloring
uses at most $d+1$ colors on the support of each $z_\alpha$, then
$\chi(G)\ge d+3$.
\end{proposition}
\begin{proof}
Each generator maps into a contractible proper face of the simplex on
the colors. The induced homomorphism is therefore zero, contradicting
Lemma~\ref{lem:nonvanishing}.
\end{proof}

\section{The ten-color lower bound}\label{sec:rook}

The coloring map and the final contradiction are the same. There are
two new ingredients: nine rook labels replace the five pentagon labels,
and identifying the seventh homology requires more than five-constraint
rounding. The rook labels define the candidate generators.
A comparison between integral and fractional centers establishes their
topological role; the local coloring argument then gives the contradiction.

\subsection{The rook cycle and the local statement}
Number a $3\times3$ array by $1,\ldots,9$, row by row, and let $D$ be
its set of cells. Let $A_i$ consist of the four other cells in the row
or column of $i$, and put $Z_i=D\sm(A_i\cup\{i\})$. Then
\begin{equation}\label{eq:incidence}
          |A_i\cap Z_i|=0,\qquad |A_i\cap Z_j|=2\quad(i\ne j).
\end{equation}
Define $\C_2(N,k)$ using common Johnson centers of radius two.
Equation~\eqref{eq:incidence} says that every proper subfamily of the
nine labels is a face of $\C_2(9,4)$, so its formal boundary is a
seven-cycle. For $n\ge5$, adjoining a fixed $(n-4)$-set $I$ disjoint
from $D$ gives a cycle in $\N(G_{n,3})=\C_2(2n,n)$.
Again include the complemented configurations.

\begin{figure}[ht]
\centering
\begin{tikzpicture}[scale=.78,every node/.style={font=\normalsize}]
\foreach \j/\x/\y/\col in {1/0/2/gray!15,2/1/2/blue!20,3/2/2/blue!20,4/0/1/blue!20,5/1/1/orange!25,6/2/1/orange!25,7/0/0/blue!20,8/1/0/orange!25,9/2/0/orange!25}
 \node[draw,fill=\col,minimum size=8mm] at (\x-1,\y-1) {$\j$};
\node[align=left,anchor=west] at (2.1,.2) {$A_1=\{2,3,4,7\}$: blue cells\\$Z_1=\{5,6,8,9\}$: orange cells\\$Z_j$ centers the eight labels other than $A_j$};
\end{tikzpicture}
\caption{The rook configuration plays the role of the pentagon.
Its cycle is the boundary on the nine set-labels $A_i$.}
\label{fig:rook}
\end{figure}

\begin{proposition}[Rook generators]\label{prop:rook-topology}
For $n\ge5$, the complex $\N(G_{n,3})$ is $6$-connected, and its
$H_7(-;\F_2)$ is spanned by the rook cycles just described.
\end{proposition}

This is the radius-two case of Proposition~\ref{prop:topology}.
Its extra comparison step is explained in Section~\ref{ss:why-seven}.
The local coloring statement is as follows.

\begin{lemma}[The rook configuration loses a color]\label{lem:local-coloring}
Let $\varphi:\binom D4\to[9]$. If each ball
$\mathcal B_2(Z)=\{A\in\tbinom D4:|A\cap Z|\ge2\}$ misses a color,
then the nine labels $A_i$ are not rainbow.
\end{lemma}

The local proof, in Section~\ref{sec:rook-local}, uses two lists of
available colors and a short sequence of forced exclusions.
Given the lemma, the lower bound follows exactly as before.
In a proper nine-coloring, the entire ball around $Z$ misses the color
of $[2n]\sm(I\cup Z)$, because
$|(I\cup A)\cap([2n]\sm(I\cup Z))|=4-|A\cap Z|\le2$.
The fixed core is nonempty, so this is an edge; complemented
configurations are handled by complementing the coloring.
Each generating seven-cycle then uses at most eight colors and maps
to zero, contradicting Lemma~\ref{lem:nonvanishing} with $d=7$.
Therefore $\chi(G_{n,3})\ge10$ for $n\ge5$.

\paragraph{The small case and the upper bound.}
For $n=4$, identify the ground set with $\F_2^3$. Its fourteen affine
planes have four points each, and two distinct planes meet in zero or
two points. They form a clique in $G_{4,3}$.
For every $n\ge4$, the Fano-plane construction gives fourteen colors
\cite{BCK}. Label seven points by the nonzero vectors of $\F_2^3$.
The seven triples $\{u,v,u+v\}$ are the Fano lines. Every two-coloring
of these points has a monochromatic line. Thus some line lies in $A$
or in $A^c$; record a chosen line and its side. Two sets with the same
color meet in at least three points, using complementation for the
negative side. This proves the upper bound and $\chi(G_{4,3})=14$.

\subsection{The radius-two comparison}\label{ss:why-seven}
The topological input is Proposition~\ref{prop:rook-topology}.
Its proof requires a comparison between integral and fractional
centers: allowing fractional centers must not kill the seventh
homology used by the coloring argument.

Replace a $k$-set by its indicator vector in the hypersimplex
\[
             \Delta(N,k)=\{x\in[0,1]^N:\textstyle\sum_i x_i=k\}.
\]
A common radius-two center satisfies $x(A)\ge k-2$ for every label $A$,
where $x(A)=\sum_{i\in A}x_i$. Requiring $x\in\{0,1\}^N$ gives the
integral complex $\C_2(N,k)$. Allowing arbitrary $x\in\Delta(N,k)$ gives
a larger complex $Y_2$, the complex of families with a fractional center.
This is a linear relaxation of the common-center condition.

The fractional complex has the coordinate-face model already used
for radius one; the rook configurations map onto its seventh-homology
generators (Section~\ref{sec:geometry}). We must show that the
relaxation has not killed any integral homology class in that degree.

The five-constraint rounding theorem (Section~\ref{sec:rounding})
implies that $\C_2(N,k)$ and $Y_2$ have the same four-skeleton: their
faces with at most five labels agree. At radius one, the shorter lemma
in Section~\ref{ss:rounding-one} already identifies the chain groups and
boundary maps computing $H_3$.

For radius two, a seven-cycle has eight-dimensional fillings. Five-label
rounding does not control such fillings. We instead prove
\begin{equation}\label{eq:overview-injection}
 H_j(\C_2(N,k);\kk)\longrightarrow H_j(Y_2;\kk)
 \text{ is injective for }0\le j\le7,
 \qquad\kk=\Z,\F_2.
\end{equation}
Here is the construction. Given a cycle $z$ in the integral complex,
we attach compatible center data to obtain a cycle $\widetilde z$.
Contractibility of the fractional center spaces supplies a filling $w$,
meaning $\partial w=\widetilde z$, whenever $z$ bounds in $Y_2$.
A chain map is a degree-preserving linear map on chains that commutes
with the boundary operator: $\phi\partial=\partial\phi$.
We construct such a map $\phi$ returning the filling to the integral
complex and satisfying $\phi(\widetilde z)=z$. Therefore
\[
                         z=\phi(\partial w)=\partial\phi(w).
\]
The equality on the original cycle is essential. Independently rounding
the simplices of a filling would not make their boundaries agree.

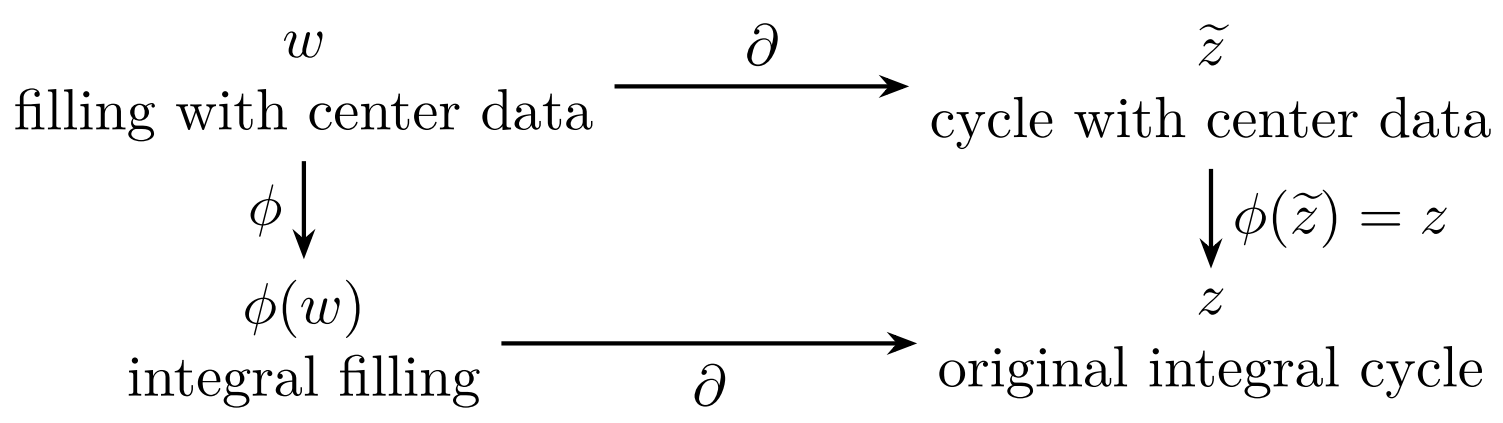
\begin{figure}[ht]
\centering
\begin{tikzpicture}[every node/.style={font=\normalsize},
 arr/.style={-{Stealth[length=2mm]},thick}]
\node[align=center] (w) at (0,1.0) {$w$\\filling with center data};
\node[align=center] (z) at (6,1.0) {$\widetilde z$\\cycle with center data};
\node[align=center] (pw) at (0,-.7) {$\phi(w)$\\integral filling};
\node[align=center] (pz) at (6,-.7) {$z$\\original integral cycle};
\draw[arr] (w)--node[above] {$\partial$}(z);
\draw[arr] (pw)--node[below] {$\partial$}(pz);
\draw[arr] (w)--node[left] {$\phi$}(pw);
\draw[arr] (z)--node[right] {$\phi(\widetilde z)=z$}(pz);
\end{tikzpicture}
\caption{Returning a fractional filling to the integral complex.
The chain map is constructed through degree eight and agrees exactly
with the prescribed map on the lifted cycle.}
\label{fig:return}
\end{figure}

There are two inputs to this construction. First, a mediator set fills
triangles among integral centers. It makes the spaces used for large
label families simply connected, meaning path connected with every
loop contractible; small families use the contractible fractional
center spaces and five-label rounding. Second, the return map
is extended one simplex at a time inside local complexes of integral
centers. The required vanishings reduce to
\begin{equation}\label{eq:local-overview}
 \Hr_j(\C_2(9,3);\Z)=\Hr_j(\C_2(8,4);\Z)=0
                      \qquad(0\le j\le7).
\end{equation}
Here is why these two complexes suffice. A coordinate face with $m$
free coordinates of total weight $\ell$ is retained in the geometric
model exactly when $\ell(m-\ell)\le2m$. Its local complex is homotopy
equivalent to $\C_2(m,\ell)$, which is a full simplex if
$\min(\ell,m-\ell)\le2$. Otherwise, after complementation, the inequality
leaves only $\ell=3$ with $6\le m\le9$, or $(m,\ell)=(8,4)$.
Moreover, $\C_2(m,3)$ is a retract of $\C_2(9,3)$ by
Lemma~\ref{lem:face-retraction}. Thus the local homology reduces to
$\C_2(9,3)$ and $\C_2(8,4)$, with $\binom93=84$ and $\binom84=70$
labels. Lemma~\ref{lem:carriers} proves this reduction.
Appendices~\ref{sec:nine} and~\ref{sec:eight} prove
\eqref{eq:local-overview} over the integers, as needed to deduce
connectivity. The final coloring argument uses only $\F_2$.

\section{Balancing the full allocation graph}\label{sec:balancing}

The graph on all allocations has the same chromatic number as its
balanced layers. This permits discrepancy bounds proved without a
balance requirement to yield balanced fair allocations.

For $m$ goods, let $\Gamma_{c,c}(m)$ have all subsets of $[m]$ as
vertices, with distinct $A,B$ adjacent when
\[
                    |A\cap B|\le c,\qquad |A^c\cap B^c|\le c.
\]
Complements are taken in the current ground set. For $c=1$, this is
the graph used in the revised proof of
\cite[Definition~3.3]{BuEtAl}. The \emph{central layers} consist of
sets of sizes $\lfloor m/2\rfloor$ and $\lceil m/2\rceil$.
A graph retraction onto these layers preserves edges and fixes every
central vertex.

\begin{theorem}[Exact balancing]\label{thm:exact-balancing}
For $c\ge1$ and $m\ge2c+1$, there is a complement-preserving graph
retraction from $\Gamma_{c,c}(m)$ onto its central layers, computable
in $O(m)$ time. These layers induce a graph isomorphic to
$G_{\lceil m/2\rceil,c+1}$. Consequently,
\[
             \chi\bigl(\Gamma_{c,c}(m)\bigr)
             =\chi\bigl(G_{\lceil m/2\rceil,c+1}\bigr).
\]
\end{theorem}

On $[2n]$, define $\beta(X)$ by adding the smallest missing elements
if $|X|<n$, or deleting the smallest present elements if $|X|>n$,
until its size is $n$. This fixes every $n$-set and preserves
complements: the additions for $X$ are precisely the deletions for $X^c$.

\begin{lemma}[The balancing scan is nonexpanding]\label{lem:balancing-scan}
For $\delta(X,Y)=\max\{|X\sm Y|,|Y\sm X|\}$, one has
$\delta(\beta(X),\beta(Y))\le\delta(X,Y)$.
\end{lemma}
\begin{proof}
Implement $\beta$ by one scan through the coordinates, changing a
coordinate only if this moves the set toward size $n$. Apply each
scan step to both sets. If $|X\sm Y|$ increases, it increases by one.
This can happen only when $X$ adds that coordinate, $Y$ deletes it,
or both do so. The three cases imply, respectively,
$|X|<n\le|Y|$, $|X|\le n<|Y|$, or $|X|<n<|Y|$ before the step.
Thus
\[
              |X\sm Y|+1\le|Y\sm X|,
\]
because $|Y\sm X|-|X\sm Y|=|Y|-|X|$. The increased count does not
exceed the old maximum. Interchanging $X,Y$ treats the other count,
so no scan step increases $\delta$.
\end{proof}

\begin{proof}[Proof of Theorem~\ref{thm:exact-balancing}]
First let $m=2n$. For $n$-sets,
$|A^c\cap B^c|=|A\cap B|$, so the central layer induces $G_{n,c+1}$.
Adjacency in the full graph means $\delta(A,B^c)\le c$.
Lemma~\ref{lem:balancing-scan} and complement preservation give
\[
 |\beta(A)\cap\beta(B)|
 =\delta(\beta(A),\beta(B)^c)
 =\delta(\beta(A),\beta(B^c))\le c.
\]
The images are distinct because $n>c$. Hence $\beta$ is the required
retraction.

For $m=2n-1$, adjoin a coordinate $\infty$ and put
\[
 \theta(A)=
 \begin{cases}
 A\cup\{\infty\},&|A|<n,\\
 A,&|A|\ge n.
 \end{cases}
\]
This preserves complements. To check edges, write
$a=|A\cap B|$ and $b=|A^c\cap B^c|$. Then
$a-b=|A|+|B|-(2n-1)$. If both sets have size less than $n$, then
$a<b$ and their new intersection counts are $a+1,b$; if both have
size at least $n$, then $b<a$ and the counts are $a,b+1$.
Otherwise the counts are unchanged. Thus $\theta$ preserves edges.

On the central layers, $\theta$ is a bijection onto the $n$-sets of
$[2n]$. The same identity shows that the image intersection has size
$\max\{a,b\}$, so this bijection is a graph isomorphism onto
$G_{n,c+1}$. Compose $\theta$ with the even balancing map and then
delete $\infty$. This is a complement-preserving graph map into the
central layers and fixes every central vertex. Size counts and one
scan take $O(m)$ time. Finally, the retraction and the inclusion of
the central layers give the two chromatic inequalities.
\end{proof}

The retraction transfers chromatic existence guarantees. It need not
preserve the fairness of an allocation for fixed valuations, so it
does not give a procedure for rebalancing an arbitrary fair allocation.

\subsection{General removal allowances}
Hollender, Manurangsi, Meka, and Suksompong guarantee EF$c$ with
$c=O(\sqrt{q\log q})$ for $q$ agents with monotone valuations,
without requiring balance \cite[Corollary~11]{HMMS26}.
Their nonlinear discrepancy estimate was proved independently and
concurrently by Dupr\'e la Tour and Fujii \cite[Theorem~1]{DF25}.
Theorem~\ref{thm:exact-balancing} adds balance to this guarantee.

\begin{corollary}\label{cor:general-allowance}
Put $D(q)=\sqrt{2q(1+\ln(2q))}$. If $q\ge2$, $c\ge1$, and $n>c$
are integers with $D(q)\le c+1$, then
$\chi(G_{n,c+1})>q$. Consequently, for every prescribed split of
$q\ge2$ agents with monotone valuations and any number of goods,
balanced EF$c$ is guaranteed with
$c=\lceil D(q)\rceil-1=O(\sqrt{q\log q})$.
\end{corollary}
For the chromatic claim, compose a putative coloring with the retraction
and apply nonlinear discrepancy to the distances from its color classes.
This forces a monochromatic edge; Section~\ref{ss:chromatic-growth}
gives the proof. Reading the inequality in the other direction yields
$\chi(G_{n,r})=\Omega(r^2/\log r)$ uniformly for $n\ge r$.
For sufficiently large $n$ in terms of $r$, the Hadamard construction
of \cite{BCK} gives the upper bound $O(r^2)$.

\section{Equivalent search problems}\label{sec:search}

The allowance $c$ is fixed throughout this section. Let $n>c$,
fix positive group sizes $q_1,q_2$, and put $q=q_1+q_2$ and $M=[2n]$.
Agents $1,\ldots,q_1$ are in the first group. Write $P_i(X)=X$ for
these agents and $P_i(X)=X^c$ for the others; thus $P_i(X)$ is
agent $i$'s bundle at allocation $(X,X^c)$.

\begin{definition}[The circuit search problems]\label{def:search}
In \emph{balanced Kneser search}, the input is a Boolean circuit
computing a color $\varphi(X)\in[q]$ for each $n$-set $X$.
Outputs outside $[q]$ are interpreted as color one. A solution is
two $n$-sets $X,Y$ with $\varphi(X)=\varphi(Y)$ and
$|X\cap Y|\le c$.

In \emph{balanced EF$c$ search}, the input is one rational-valued
circuit $u_i:2^M\to\mathbb Q$ per agent, with no monotonicity promise.
A solution is either
\begin{enumerate}[(i),leftmargin=2em]
\item an $n$-set $X$ and, for each agent $i$, a set
$D_i\subseteq P_i(X)^c$ with $|D_i|\le c$ and
$u_i(P_i(X))\ge u_i(P_i(X)^c\sm D_i)$; or
\item an agent $i$ and sets $S\subsetneq T$ with $u_i(S)>u_i(T)$.
\end{enumerate}
\end{definition}

Let $L$ be the total binary encoding length of the circuit descriptions,
including their input and output widths;
Section~\ref{sec:search-proofs} specifies the encoding.
Solutions have polynomial length and can be checked in polynomial time.
The violation alternative, as in \cite[Remark~6.6]{Haviv24}, makes
totality independent of a monotonicity promise.

\begin{theorem}[Search equivalence]\label{thm:search-equivalence}
For fixed $c\ge1$ and positive $q_1,q_2$, suppose that $n>c$ and
$q=q_1+q_2<\chi(G_{n,c+1})$. The two problems in
Definition~\ref{def:search} are equivalent under deterministic
polynomial-time many-one search reductions. Both reductions preserve
$n,c,q$, take time $L^{O(1)}n^{O(c)}$, and produce circuits of that size.
The reduction from Kneser search produces normalized monotone
$\{0,1\}$-valued valuations.
\end{theorem}

The forward reduction colors an allocation by a rejecting agent;
a monochromatic edge yields a fair allocation or a monotonicity violation.
The reverse reduction uses binary valuations. Assigning zero below
size $n-c$ makes every other relevant bundle completable to an $n$-set
by at most $c$ additions, so ordinary circuits can perform the required
color-class searches. Section~\ref{sec:search-proofs} gives both
reductions and their decoders.

Thus five-agent balanced EF1 ($n\ge2$) and nine-agent balanced EF2
($n\ge3$) lie in TFNP, the class of total search problems with
polynomially bounded, polynomially verifiable certificates.
Polynomial-time algorithms and PPA membership or hardness remain open.
\label{end:overview}

\clearpage
\section{The search reductions}\label{sec:search-proofs}

We now prove Theorem~\ref{thm:search-equivalence}.
The forward construction tests at most $c$ deletions from a bundle.
The reverse construction tests at most $c$ additions to one.
These bounded searches make both reductions polynomial for fixed $c$.
We use the notation of Section~\ref{sec:search}.

Subsets are encoded by indicator vectors. A valuation circuit outputs
a sign bit $s$ and unsigned integers $a,b$, decoded as
$(-1)^s a/(b+1)$; hence every output string has a value.
Comparisons use exact integer arithmetic. A many-one search reduction
constructs one target instance and converts every valid target solution
into a source solution. On monotone inputs only allocations can be
returned; subtracting $u_i(\varnothing)$ normalizes such inputs without
changing fairness. Both problems are total under the stated chromatic
inequality: Theorem~\ref{thm:balanced-threshold} applies to monotone
inputs, and a nonmonotone input has a violation certificate.
The polynomial bound requires fixed $c$; it is not a bound of the form
$f(c)\operatorname{poly}(L,n)$.

Haviv's reverse reduction from classical Kneser search to
\textsc{Agreeable-Set} uses subset queries asking whether a set
contains a vertex of a specified color \cite[Theorem~1.5]{Haviv24}.
Here the bounded completion search implements the reverse reduction
with ordinary circuits.

\subsection{From valuations to a monochromatic edge}
Given the valuation circuits, color a balanced allocation $X$ by its
first rejecting agent, or by color one if all agents accept it.
For each agent, acceptance can be tested by enumerating all
$\sum_{j=0}^c\binom nj=O(n^c)$ deletion sets from the other bundle.
Composing these tests with the valuation circuits gives a coloring
circuit of size $L^{O(1)}n^{O(c)}$, without testing monotonicity.

Let $X,Y$ be any monochromatic edge returned by the target problem.
Test both allocations. If either is fair, return it with the deletion
witnesses found by the tests. Otherwise their common color identifies
an agent $i$ who rejects both, including when this color is one.
Put $P_X=P_i(X)$, $P_Y=P_i(Y)$ and $Q_X=P_X^c$, $Q_Y=P_Y^c$.
Interchange $X,Y$ if necessary so that $u_i(P_X)\ge u_i(P_Y)$.
Balance and adjacency give
\[
                  |Q_X\cap Q_Y|=|X\cap Y|\le c.
\]
Rejection at $X$ applies in particular to deleting $Q_X\cap Q_Y$
from $Q_X$. The remaining set $S=Q_X\cap P_Y$ satisfies
\[
              S\subseteq P_Y,\qquad
              u_i(S)>u_i(P_X)\ge u_i(P_Y).
\]
The values differ, so $S\subsetneq P_Y$. Thus $(i,S,P_Y)$ is a
valid monotonicity-violation certificate. Every target solution has
been decoded in polynomial time, proving the forward reduction.

\subsection{From a coloring to binary valuations}
Let $\varphi$ be the input coloring. Assign color $i$ to agent $i$
in the prescribed split, and define $u_i:2^M\to\{0,1\}$ by
\begin{equation}\label{eq:search-valuation}
u_i(S)=0\quad\Longleftrightarrow\quad
\begin{cases}
 |S|<n-c,\quad\text{or}\\
 |S|\le n\ \text{and some }Z\in\binom Mn
 \text{ contains }S\text{ and satisfies }\varphi(P_i(Z))=i.
\end{cases}
\end{equation}
Otherwise $u_i(S)=1$. The zero sets are downward closed: a subset
of a small zero set is still small, and a subset of a set admitting
the displayed completion admits that same completion.
Thus $u_i$ is monotone for every coloring, even when its color classes
are not independent. It is normalized because $n>c$.

For $|S|<n-c$ or $|S|>n$, the value is immediate. Otherwise enumerate
all sets of at most $c$ additional goods, retain those whose union
with $S$ has size $n$, and evaluate $\varphi$ on the corresponding
$P_i(Z)$. There are $O(n^c)$ candidates. This enumeration can be
built into an ordinary Boolean circuit, using one copy of the coloring
circuit for each candidate and taking the disjunction of the tests.
The circuit size and construction time are $L^{O(1)}n^{O(c)}$.
The small-bundle clause is what makes the test bounded; it does not
affect any comparison of a balanced bundle with the other bundle
after at most $c$ deletions.

Every valid solution for these monotone valuations is a fair allocation
with deletion witnesses. Let $X$ be its first-group bundle and take
$i=\varphi(X)$. Agent $i$ receives $P=P_i(X)$, and $u_i(P)=0$:
in \eqref{eq:search-valuation}, choose $Z=P$ and use
$P_i(P_i(X))=X$. Her deletion witness $D_i\subseteq P^c$ gives
$u_i(P^c\sm D_i)=0$. Since $|P^c\sm D_i|\ge n-c$, the small-bundle
clause cannot explain this value. Enumerating completions therefore
finds an $n$-set $Z\supseteq P^c\sm D_i$ with $\varphi(P_i(Z))=i$.
At most $|D_i|$ elements of $Z$ lie in $P$, so $|P\cap Z|\le c$.

Output $Y=P_i(Z)$. Complementation preserves intersection sizes of
two $n$-sets in $[2n]$, and hence
\[
              |X\cap Y|=|P\cap Z|\le c,
              \qquad \varphi(Y)=i=\varphi(X).
\]
The vertices are distinct because $n>c$. The decoder uses $O(n^c)$
color evaluations, completing the proof of
Theorem~\ref{thm:search-equivalence}.

For EF1, Theorem~\ref{thm:main} gives totality for $2\le q\le5$
and $n\ge2$. For EF2 it gives totality for $2\le q\le9$ and $n\ge4$;
the remaining case $n=3$ follows from $G_{3,3}=K_{20}$.
These equivalences hold for every positive split and also for
binary-valued input circuits, with the same violation alternative.
The forward reduction accepts such inputs, and the reverse reduction
already produces them.

\section{The local rook coloring obstruction}\label{sec:rook-local}

We prove Lemma~\ref{lem:local-coloring}. If the nine rook labels had
distinct colors, each four-set would have a short list of available
colors on each side of the incidence relation. An attachment
restriction reduces these lists, after which four forced values
give a contradiction.

\subsection{Two lists}
We use strings for small sets:
$1234$ means $\{1,2,3,4\}$. The count of pairs from the pentagon proof
is replaced by restrictions on the available colors.

Let $\varphi$ be as in Lemma~\ref{lem:local-coloring}, and suppose that the
values $\varphi(A_1),\dots,\varphi(A_9)$ are distinct. Renaming the
values, we may assume $\varphi(A_i)=i$. For every $4$-set
$Z\subseteq D$ choose a value $\psi(Z)$ not occurring on $\mathcal B_2(Z)$.
Then
\begin{equation}\label{eq:rule}
       \varphi(S)\ne\psi(T)\qquad\text{whenever }|S\cap T|\ge2 .
\end{equation}
By the incidence rule \eqref{eq:incidence}, $\mathcal B_2(Z_i)$ contains every
$A_j$ with $j\ne i$, hence every value other than $i$; so $\psi(Z_i)=i$.
Applying \eqref{eq:rule} with $T=Z_i$, and with $S=A_i$, gives
\begin{equation}\label{eq:lists}
 \varphi(S)\in L(S)=\{i:\ |S\cap Z_i|\le1\},\qquad
 \psi(T)\in L^*(T)=\{i:\ |T\cap A_i|\le1\}.
\end{equation}
Lists are easily computed. For a cell $i$ in row $a$ and column $b$, let
$r_a$ and $c_b$ be the numbers of elements of $S$ in row $a$ and in column
$b$, and $e_i=1$ if $i\in S$, $e_i=0$ otherwise. Then
$|S\cap A_i|=r_a+c_b-2e_i$ and $|S\cap Z_i|=4-r_a-c_b+e_i$, so that
\begin{equation}\label{eq:list-rule}
 i\in L(S)\iff r_a+c_b-e_i\ge3,\qquad
 i\in L^*(T)\iff r_a+c_b-2e_i\le1 \ \ (\text{counts taken in }T).
\end{equation}
For example, $L(1234)=123$: in $1234$ the row counts are $3,1,0$ and the
column counts $2,1,1$.

We write $F_S=\varphi(S)$ and $G_T=\psi(T)$, and denote sets by strings
of digits; thus $F_{1234}=\varphi(\{1,2,3,4\})$. A \emph{solution} is a
pair of maps $F,G\colon\binom{D}4\to[9]$ satisfying \eqref{eq:rule}
with $F_{A_i}=i$ for every $i$; as shown above, it also satisfies
$G_{Z_i}=i$ and \eqref{eq:lists}. We shall show that there is no
solution. All lists and intersections used below are read off from
\eqref{eq:list-rule}.

\subsection{Symmetries}
Two symmetries avoid repeating the case analysis. Grid symmetries
reduce the attachment restriction to one representative; a duality
then transfers the restriction from $F$ to $G$.

\emph{Automorphisms.} An automorphism $g$ of the rook graph permutes the
cells and satisfies $g(A_i)=A_{g(i)}$ and $g(Z_i)=Z_{g(i)}$. If $(F,G)$
is a solution, so is $(F',G')$ with $F'_S=g(F_{g^{-1}S})$ and
$G'_T=g(G_{g^{-1}T})$. The automorphisms include the permutations of
rows, the permutations of columns, and transposition. They act
transitively on the triples $(\Lambda,e,i)$ in which $\Lambda$ is a line
(a row or a column), $e$ is a cell outside $\Lambda$, and $i$ is the cell
of $\Lambda$ lying in a common line with $e$.

\emph{Duality.} Write the cell in row $a$ and column $b$ as $(a,b)$, and
put $\pi(a,b)=(a+b,\,a-b)$ modulo $3$. Then $\pi$ is a bijection which
interchanges adjacency and nonadjacency of distinct cells: if two cells
differ by $(\alpha,\beta)$, their images differ by
$(\alpha+\beta,\alpha-\beta)$, and modulo $3$ exactly one of $\alpha,\beta$
vanishes if and only if neither of $\alpha+\beta,\alpha-\beta$ does.
Consequently $\pi(A_i)=Z_{\pi(i)}$ and $\pi(Z_i)=A_{\pi(i)}$. In the
numbering of $D$,
\[
 \pi=\begin{pmatrix}1&2&3&4&5&6&7&8&9\\1&6&8&5&7&3&9&2&4\end{pmatrix}.
\]
If $(F,G)$ is a solution, so is $(F^\vee,G^\vee)$, where
\[
   F^\vee_S=\pi^{-1}\bigl(G_{\pi(S)}\bigr),\qquad
   G^\vee_T=\pi^{-1}\bigl(F_{\pi(T)}\bigr).
\]
Indeed \eqref{eq:rule} is preserved because $|\pi S\cap\pi T|=|S\cap T|$,
and $F^\vee_{A_i}=\pi^{-1}(G_{Z_{\pi(i)}})=i$.

\subsection{The attachment restriction}

\begin{lemma}\label{lem:attachment}
For every solution, $F_{\Lambda\cup\{e\}}\ne i$ whenever $\Lambda$ is a
line, $e$ a cell outside $\Lambda$, and $i$ the cell of $\Lambda$ lying in
a common line with $e$. In particular $F_{1234}\ne1$ and $F_{1247}\ne1$.
\end{lemma}

\begin{proof}
By transitivity it suffices to prove $F_{1234}\ne1$ for every solution.
(The second example is the image of the first under transposition.)
Suppose that $F_{1234}=1$. One has $F_{5689}\in L(5689)=5689$. The
column exchange $\tau=(2\,3)(5\,6)(8\,9)$ fixes the cell $1$ and the sets
$1234$ and $5689$; replacing the solution by its image under $\tau$ if
necessary, we may assume $F_{5689}\in\{5,8\}$. In each case the table
below lists, in order, variables whose values are forced. A removed value
$v$ is followed, in parentheses, by a variable already known to take the
value $v$ on a set meeting the current set in at least two cells; by
\eqref{eq:rule} the current variable cannot take the value $v$. The
symbol $\varnothing$ marks a contradiction.
\begin{center}
\renewcommand{\arraystretch}{1.12}
\begin{tabular}{@{}llll@{}}
\toprule
variable & list & removed values & value\\
\midrule
\multicolumn{4}{@{}l}{\emph{Case} $F_{1234}=1$, $F_{5689}=5$}\\
$G_{1259}$ & $159$ & $1\ (F_{1234})$, $5\ (F_{5689})$ & $9$\\
$G_{1369}$ & $158$ & $1\ (F_{1234})$, $5\ (F_{5689})$ & $8$\\
$F_{2679}$ & $389$ & $8\ (G_{1369})$, $9\ (G_{1259})$ & $3$\\
$F_{1789}$ & $789$ & $8\ (G_{1369})$, $9\ (G_{1259})$ & $7$\\
$G_{3579}$ & $357$ & $3\ (F_{2679})$, $5\ (F_{5689})$, $7\ (F_{1789})$ &
   $\varnothing$\\
\midrule
\multicolumn{4}{@{}l}{\emph{Case} $F_{1234}=1$, $F_{5689}=8$}\\
$G_{1456}$ & $189$ & $1\ (F_{1234})$, $8\ (F_{5689})$ & $9$\\
$G_{1369}$ & $158$ & $1\ (F_{1234})$, $8\ (F_{5689})$ & $5$\\
$G_{1368}$ & $168$ & $1\ (F_{1234})$, $8\ (F_{5689})$ & $6$\\
$F_{3468}$ & $569$ & $5\ (G_{1369})$, $6\ (G_{1368})$, $9\ (G_{1456})$ &
   $\varnothing$\\
\bottomrule
\end{tabular}
\end{center}
For instance, the first line uses $1259\cap1234=12$ and
$1259\cap5689=59$. Both cases are impossible. 
\end{proof}

\begin{corollary}\label{cor:dual}
For every solution, $G_{1249}\ne9$ and $G_{1267}\ne6$.
\end{corollary}

\begin{proof}
Apply Lemma~\ref{lem:attachment} to the dual solution $(F^\vee,G^\vee)$:
with $\Lambda=789$, $e=1$, $i=7$, and with $\Lambda=258$, $e=1$, $i=2$, it
gives $F^\vee_{1789}\ne7$ and $F^\vee_{1258}\ne2$. Since
$\pi(1789)=1249$, $\pi(7)=9$, $\pi(1258)=1267$ and $\pi(2)=6$, these read
$G_{1249}\ne9$ and $G_{1267}\ne6$. 
\end{proof}

\subsection{Completion of the local coloring argument}
Let $(F,G)$ be a solution. By Lemma~\ref{lem:attachment},
$F_{1234}\in L(1234)\sm\{1\}=\{2,3\}$. The exchange $\tau$ of the
preceding proof fixes $1234$ and interchanges the values $2$ and $3$, and
Lemma~\ref{lem:attachment} and Corollary~\ref{cor:dual} hold for every
solution; so we may assume $F_{1234}=2$. Then successively
\begin{align*}
 G_{1249}&\in L^*(1249)\sm\{9,2\}=249\sm\{9,2\}=\{4\},\\
 F_{1247}&\in L(1247)\sm\{1,4\}=147\sm\{1,4\}=\{7\},\\
 G_{1267}&\in L^*(1267)\sm\{6,7\}=267\sm\{6,7\}=\{2\}.
\end{align*}
In each line the first removed value comes from Corollary~\ref{cor:dual}
or Lemma~\ref{lem:attachment}, and the second from the preceding line
together with \eqref{eq:rule}, using $1249\cap1234=124$,
$1247\cap1249=124$ and $1267\cap1247=127$. But $1267\cap1234=12$, so
\eqref{eq:rule} forbids $G_{1267}=F_{1234}=2$. Hence there is no solution,
and the values $\varphi(A_1),\dots,\varphi(A_9)$ are not distinct. 

\begin{figure}[ht]
\centering
\begin{tikzpicture}[every node/.style={font=\normalsize},
 box/.style={draw,rounded corners=2pt,align=center,minimum width=27mm,minimum height=9mm},
 arr/.style={-{Stealth[length=2mm]},thick}]
\node[box] (a) at (0,1.05) {$F_{1234}=2$};
\node[box] (b) at (6.5,1.05) {$G_{1249}=4$};
\node[box] (c) at (6.5,-.8) {$F_{1247}=7$};
\node[box] (d) at (0,-.8) {$G_{1267}=2$};
\draw[arr] (a)--node[above] {remove $9$ and $2$}(b);
\draw[arr] (b)--node[right] {remove $1$ and $4$}(c);
\draw[arr] (c)--node[below] {remove $6$ and $7$}(d);
\draw[very thick,red!65!black,dashed] (d)--node[left,align=right] {forbidden:\\$|1234\cap1267|=2$}(a);
\end{tikzpicture}
\caption{The last four forced values in the rook argument. Each
solid arrow uses an already proved restriction and the preceding value.
The dashed edge contradicts the rule that intersecting opposite-side
sets must receive different colors.}
\label{fig:forced-cycle}
\end{figure}
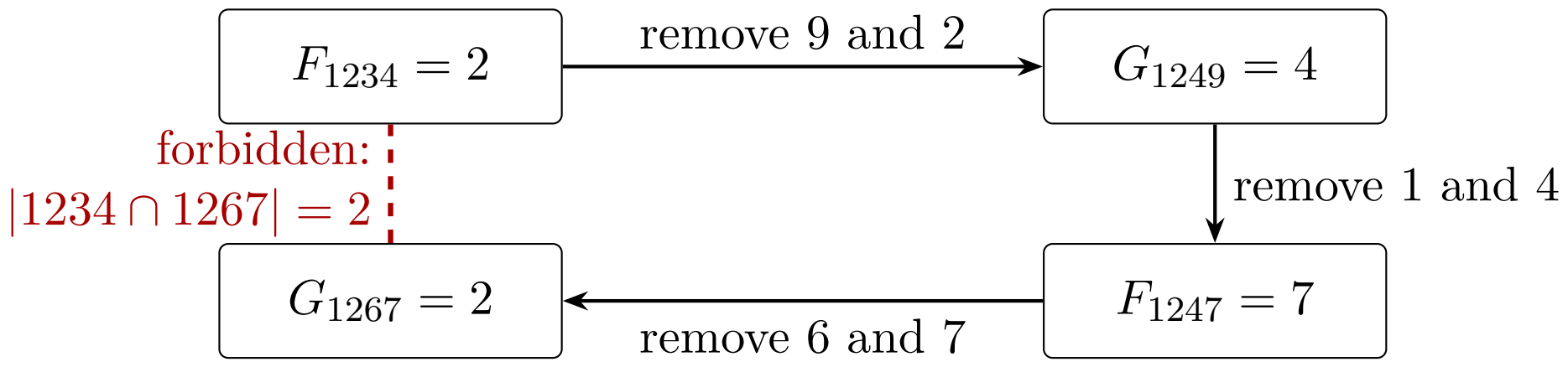

\section{The common topological and geometric tools}\label{sec:geometry}

We prove the statements used in the common coloring argument.
The nonvanishing lemma applies to arbitrary graphs, and the geometric
construction applies to every integral radius. Radius one has a short
rounding argument; radius two uses the general rounding theorem and
an additional comparison. A reader following the six-color proof may
set $c=1$ throughout and skip the indicated radius-two material.

\begin{figure}[H]
\centering
\begin{tikzpicture}[every node/.style={font=\normalsize},
 box/.style={draw,rounded corners=2pt,align=center,text width=35mm,minimum height=9mm},
 arr/.style={-{Stealth[length=1.7mm]},thick}]
\node[box] (g) at (0,0) {geometry\\and rounding\\Section~\ref{sec:geometry}};
\node[box] (l) at (4.6,0) {finite integral inputs\\Appendices~\ref{sec:nine}, \ref{sec:eight}};
\node[box] (c) at (2.3,-2) {radius-two\\comparison\\Section~\ref{sec:comparison}};
\node[box] (h) at (2.3,-4) {connectivity\\and generators\\Proposition~\ref{prop:topology}};
\node[box] (k) at (7,-4) {local coloring\\lemmas\\Sections~\ref{sec:pentagon}, \ref{sec:rook-local}};
\node[box] (x) at (4.65,-6) {chromatic\\lower bounds\\Theorem~\ref{thm:main}};
\node[box] (s) at (10,0) {explicit sphere map\\Section~\ref{sec:general-index}};
\node[box] (i) at (10,-2) {index upper bound\\Theorem~\ref{thm:index-intro}};
\draw[arr] (g)--(c);\draw[arr] (l)--(c);\draw[arr] (c)--(h);
\draw[arr] (g.west)--++(-.45,0)|-node[pos=.8,above] {$c=1$}(h.west);
\draw[arr] (h)--(x);\draw[arr] (k)--(x);\draw[arr] (s)--(i);
\end{tikzpicture}
\caption{Proof dependencies. Radius one uses the common geometry and
rounding. Radius two also uses the comparison and finite integral
inputs. The sphere map proves the index upper bound independently;
connectivity supplies the matching lower bounds for $r=2,3$.}
\label{fig:dependencies}
\end{figure}
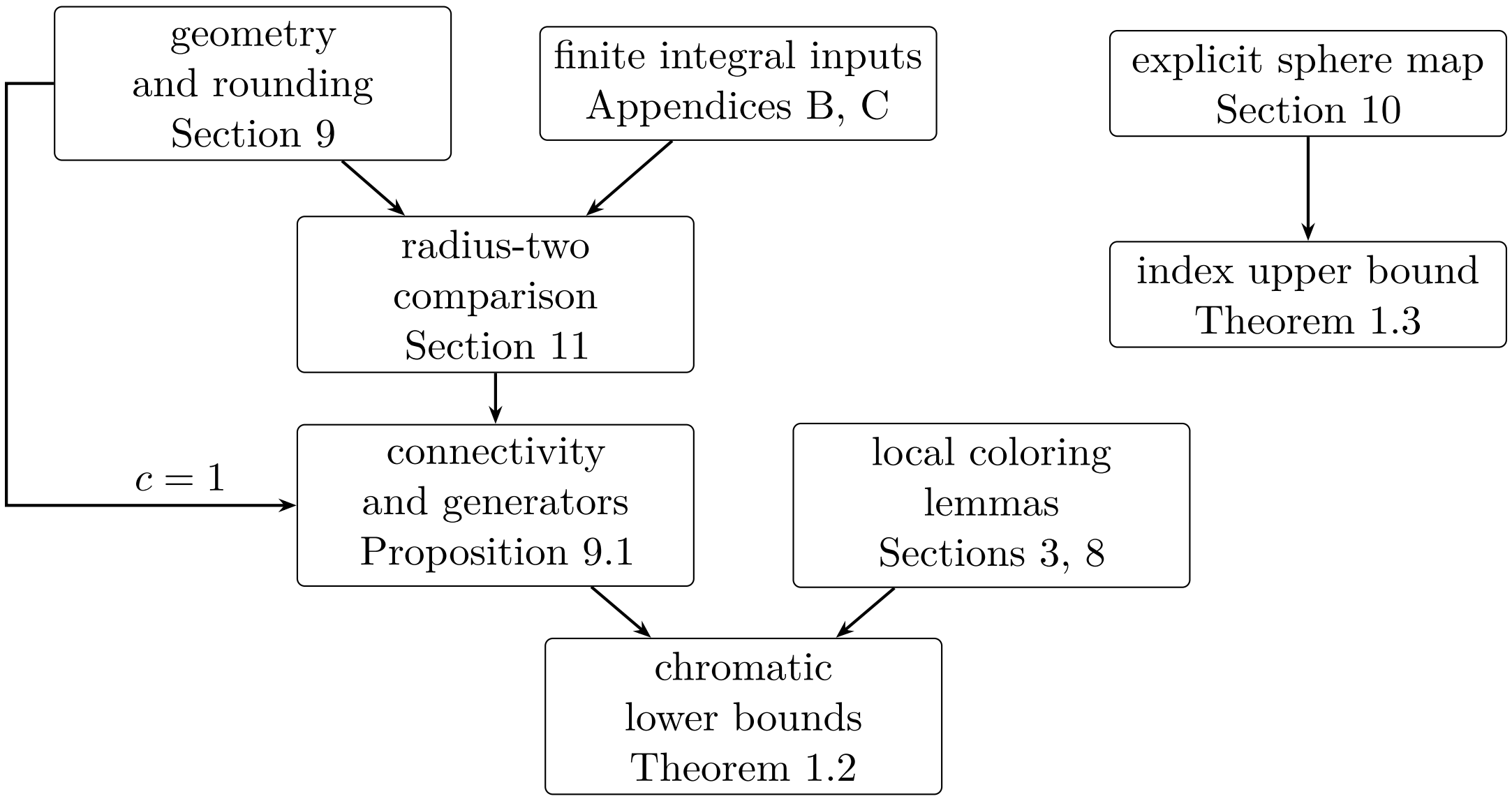

\subsection{Why the coloring map is nonzero}\label{ss:nonvanishing-proof}
\begin{proof}[Proof of Lemma~\ref{lem:nonvanishing}]
Let $B(G)$ be the geometric realization of $\operatorname{Hom}(K_2,G)$.
Concretely, its points are pairs $(\lambda,\mu)$ of probability
distributions on $V(G)$ such that every vertex in the positive support
of $\lambda$ is adjacent to every vertex in the positive support of
$\mu$. Each complete bipartite subgraph gives a cell, the product of
the two simplices of weights on its sides. This is a finite free $\mathbb Z_2$-CW
complex, with the involution swapping the two distributions.
Babson and Kozlov prove that $B(G)\simeq|\N(G)|$
\cite[Proposition~4.2]{BabsonKozlovHom}; it is therefore
$(d-1)$-connected.

A map between spaces with involutions is \emph{equivariant} if it
commutes with the involutions; on $S^d$ the involution is $x\mapsto-x$,
and on $B(G)$ it exchanges the two distributions.
There is consequently an equivariant map $a:S^d\to B(G)$.
Here is the construction behind this assertion. Choose a point of
$B(G)$ and send an antipodal pair to that point and its image under
the involution. Join the two images by a path, and use the involution
to define the image of the opposite semicircle. Continue by extending
over one hemisphere and defining the other by symmetry. The extension
at each stage is possible because $B(G)$ is $(d-1)$-connected.
This is the argument in the proof of
\cite[Proposition~5.3.2(iv), pp.~97--98]{Matousek}.

Let $q=d+2$, and sum the weights of equally colored vertices to obtain
$x_i=\sum_{\kappa(v)=i}\lambda_v$ and
$y_i=\sum_{\kappa(v)=i}\mu_v$ for $i\in[q]$.
Properness makes the supports of $x$ and $y$ disjoint. Thus
\[
 h(\lambda,\mu)=\frac{x-y}{\|x-y\|}
 \quad\text{lies on the unit sphere in }
 \bigl\{z\in\mathbb R^q:\textstyle\sum_i z_i=0\bigr\}.
\]
This is an equivariant map to $S^d$. The composite $h\circ a$
is an antipodal self-map of $S^d$, hence has odd degree and is
nonzero on $H_d(-;\mathbb F_2)$
\cite[Proposition~2.4.1(iii) and the following notes, pp.~43--46]{Matousek}.

To connect this conclusion to the coloring map, let
$p(\lambda,\mu)=\lambda\in|\N(G)|$. This is well-defined because
any vertex in the support of $\mu$ is a common neighbor of the
support of $\lambda$. Its colored image is $x$ on the simplex
boundary. Put $b=(1/q,\ldots,1/q)$ and let
$r(x)=(x-b)/\|x-b\|$ identify this boundary with the sphere.
The segment from $x-b$ to $x-y$ never passes through zero: any
coordinate with $y_i>0$ has $x_i=0$ and remains strictly negative.
Normalizing the segment gives $h\simeq r\circ\kappa_{\N}\circ p$.
If $(\kappa_{\N})_*$
were zero, $(h\circ a)_*$ would therefore be zero, a contradiction.
\end{proof}
\subsection{One notation for the two configurations}\label{ss:common-notation}
The pentagon and rook arguments have already been stated separately.
For their geometric proofs, use a radius $c$ and write $\C_c(N,k)$
for the complex whose nonempty faces are families of $k$-sets with a
common Johnson center at distance at most $c$. For $n>c$,
\[
                         \N(G_{n,c+1})=\C_c(2n,n).
\]
This is the same complementation identity used in the two lower bounds.

For $c=1,2$, put $q=4c+1$. On a $q$-point set $D$, use the pentagon
or rook graph, and write $A_i$ for the neighbors of $i$ and
$Z_i=D\sm(A_i\cup\{i\})$. In both cases
\[
        |A_i|=|Z_i|=2c,\qquad |A_i\cap Z_i|=0,\qquad
        |A_i\cap Z_j|=c\quad(i\ne j).
\]
Choose $I$ disjoint from $D$. If $|I|=k-2c$, use the labels
$I\cup A_i$; if $|I|=k-(2c+1)$, use $I\cup(D\sm A_i)$.
Their centers are obtained in the same way from $Z_j$.
Denote either boundary cycle by $z_{I,D}$, using only choices for
which the specified sizes are possible.

\begin{proposition}[Connectivity and generators]\label{prop:topology}
For $c\in\{1,2\}$ and $1\le k<N$, the complex $\C_c(N,k)$ is
$(4c-2)$-connected. Its homology group
$H_{4c-1}(\C_c(N,k);\F_2)$ is spanned by the classes $[z_{I,D}]$.
\end{proposition}

Taking $N=2n$ and $k=n$ gives Propositions~\ref{prop:pentagon-topology}
and~\ref{prop:rook-topology}. Indeed, complementation of an $n$-set
of the first type gives one of the second type, and conversely.
For $c=1$, every ingredient in the proof is in this section.
For $c=2$, only the final step uses the comparison theorem from
Section~\ref{sec:comparison}.

\subsection{The hypersimplex and its caps}\label{ss:hypersimplex}
We now work with arbitrary $1\le k<N$ and integer radius $c\ge1$.
For $S\subseteq[N]$ write
$x(S)=\sum_{i\in S}x_i$, and let $\one_S$ denote its indicator vector.
In the hypersimplex $\Delta(N,k)$ define the cap
\[
 P_{c,Z}=\{x\in\Delta(N,k):x(Z)\ge k-c\},\qquad
 U_c=\bigcup_{|Z|=k}P_{c,Z}.
\]
The integral vertices of $P_{c,Z}$ are precisely the labels at distance
at most $c$ from $Z$. Sending $A$ to $\one_A$ and extending affinely
over each simplex defines
\begin{equation}\label{eq:affine-map}
                  f_c:|\C_c(N,k)|\longrightarrow U_c.
\end{equation}
It is well-defined because all labels of a face have a common center
$Z$, and their indicator vectors lie in the convex cap $P_{c,Z}$.

A coordinate face $F$ is specified by disjoint sets $I,O\subseteq[N]$:
coordinates in $I$ are fixed to one, those in $O$ to zero, and the
remaining set $D$ is free. Put $m=|D|$ and $\ell=k-|I|$. Then
$F$ is a copy of $\Delta(m,\ell)$. It is a point if $\ell=0$ or $m$;
otherwise it has dimension $m-1$. Define
\[
        \rho(F)=\frac{\ell(m-\ell)}m,
        \qquad\rho(F)=0\text{ for a point}.
\]
Call $F$ \emph{retained} when $\rho(F)\le c$, and let $Q_c$ be the union
of retained faces. For a point $x$, let $F(x)$ be its smallest coordinate
face. Its free coordinates are exactly those with $0<x_i<1$.

\begin{lemma}[The face model]\label{lem:polyhedral}
\begin{enumerate}[(a)]
\item A face is contained in $U_c$ if and only if it is retained.
Every face of a retained face is retained.
\item There is a strong deformation retraction $U_c\to Q_c$ which
preserves each cap at every time. On an omitted face whose proper faces
are retained, the final retraction is radial projection from its
barycenter to its boundary.
\item $Q_c$ is $(4c-2)$-connected.
\end{enumerate}
\end{lemma}

\begin{proof}
On a coordinate face, the function $\max_{|Z|=k}x(Z)$ is convex and
invariant under permutations of the free coordinates. Averaging over
these permutations shows that its minimum occurs at the barycenter
$b_F$. Its value there is $k-\rho(F)$. This proves the first assertion
of (a). Fixing a further coordinate to zero or one decreases $\rho$,
so retained faces form a subcomplex of the face structure.

If $F$ is omitted, its barycenter lies outside every cap. Push points
of $F\cap U_c$ radially away from $b_F$ until they reach $\partial F$.
This motion preserves any cap that contains the starting point: the
linear score $x(Z)$ increases along that ray, since its value at $b_F$
is below $k-c$. Process omitted faces in decreasing dimension.
Each motion fixes the boundary of its face, so the motions fit together
and fix $Q_c$. This proves (b).

If $m\le4c$, then $\rho(F)\le m/4\le c$. Hence $Q_c$ contains every face
of dimension at most $4c-1$. The contractible polytope $\Delta(N,k)$
is obtained from $Q_c$ by attaching cells of dimension at least $4c$.
Cellular approximation and the relative homotopy sequence give
$\pi_j(Q_c)=0$ for $j\le4c-2$ \cite[Chapter~4]{Hatcher}.
\end{proof}

For $c=1,2$, the first omitted faces have $4c+1$ free coordinates.
Substitution into the inequality $\ell(m-\ell)>cm$ gives
\begin{center}
\begin{tabular}{@{}cccc@{}}
\toprule
Radius & Cycle dimension & First omitted faces & Configuration\\
\midrule
$1$ & $3$ & $\Delta(5,2),\ \Delta(5,3)$ & pentagon\\
$2$ & $7$ & $\Delta(9,4),\ \Delta(9,5)$ & rook graph\\
\bottomrule
\end{tabular}
\end{center}

\begin{lemma}[Boundary generators]\label{lem:boundary-generators}
Let $c=1$ or $2$, and put $d=4c-1$. Over $\Z$ or $\F_2$, the group
$H_d(Q_c)$ is generated by the boundary classes of the omitted faces
listed in the table.
\end{lemma}
\begin{proof}
The relative cellular chain complex of $(\Delta(N,k),Q_c)$ has no cells
in dimensions at most $d$. Thus its $(d+1)$st homology is a quotient
of the free group on omitted $(d+1)$-faces. The boundary homomorphism
of the pair identifies this group with $H_d(Q_c)$ and sends each face
to its boundary class. Higher-dimensional cells may impose relations
among these classes; they do not change the generating set.
\end{proof}

\subsection{The fractional complex and the affine map}\label{ss:fractional}
The nerve $Y_c$ of the caps has as faces the nonempty families $\sigma$
for which $P_{c,\sigma}:=\bigcap_{A\in\sigma}P_{c,A}$ is nonempty.
The points of $P_{c,\sigma}$ are precisely the common fractional
radius-$c$ centers of $\sigma$. An integral center supplies a fractional
one, giving an inclusion $i_c:\C_c(N,k)\hookrightarrow Y_c$.
Because cap intersections are convex, the nerve theorem gives
$|Y_c|\simeq U_c\simeq Q_c$.

The inclusion and the affine map have the same homological information,
up to the natural identifications from the two covers. We record this
fact to avoid confusing the two maps later.

\begin{lemma}[Nerve comparison]\label{lem:nervecomp}
For $\kk=\Z$ or $\F_2$, there are isomorphisms
\[
 \alpha:H_*(\C_c(N,k);\kk)\longrightarrow H_*(\C_c(N,k);\kk),
 \qquad \beta:H_*(Y_c;\kk)\longrightarrow H_*(U_c;\kk)
\]
such that $(f_c)_*=\beta\circ(i_c)_*\circ\alpha$.
In particular, $(f_c)_*$ is injective in a given degree if and only if
$(i_c)_*$ is.
\end{lemma}
\begin{proof}
Let $L_Z$ be the full simplex on the labels within distance $c$ of $Z$.
These simplices cover $\C_c(N,k)$. Their intersections are simplices;
by symmetry of Johnson distance, the intersection indexed by a family
$\varsigma$ is nonempty exactly when $\varsigma\in\C_c(N,k)$.
Thus the nerve of this cover is another copy of $\C_c(N,k)$.
The cap cover of $U_c$ has nerve $Y_c$, and $f_c(L_Z)\subseteq P_{c,Z}$.

Write $C=\C_c(N,k)$. The auxiliary spaces for the two covers are
\[
 X=\bigcup_{\varsigma\in C}|\Delta^\varsigma|\times
                \Bigl|\bigcap_{Z\in\varsigma}L_Z\Bigr|,
 \qquad
 X'=\bigcup_{\varsigma\in Y_c}|\Delta^\varsigma|\times P_{c,\varsigma}.
\]
All four coordinate projections are homotopy equivalences by the nerve
theorem for these finite good covers \cite{Bjorner}, using compatible
triangulations and \cite[Proposition~4G.1 and Section~4.G, Exercise~4,
pp.~458--460]{Hatcher}.
The map $(s,t)\mapsto(s,f_c(t))$ from $X$ to $X'$ commutes with the
projections. Using the first projections identifies its map on homology
with $(i_c)_*$; using the second identifies it with $(f_c)_*$.
This gives the stated formula.
\end{proof}

\subsection{The degree of a configuration}\label{ss:degree-one}
We return to the cycles $z_{I,D}$ of Section~\ref{ss:common-notation}.
Their integral vertices lie on a coordinate face of the appropriate
type in the table above. The following argument treats both small
graphs at once.

\begin{lemma}[Degree-one representatives]\label{lem:degree-one}
For $c=1,2$, the affine map $f_c$, followed by the retraction
$r_c:U_c\to Q_c$, maps the configuration boundary associated with
each omitted coordinate face of dimension $4c$ (the faces listed in
Section~\ref{ss:hypersimplex}) homeomorphically onto that face's boundary. In particular, the classes
$(f_c)_*[z_{I,D}]$ span $H_{4c-1}(U_c;\F_2)$.
\end{lemma}
\begin{proof}
Suppress the fixed coordinates and first use the labels $A_i$ on
$q=4c+1$ points. Let $M$ be the adjacency matrix of the pentagon or rook
graph. Their degree and common-neighbor counts give
\[
                           M^2=c\mathrm{Id}-M+cJ,
\]
where $J$ is the all-one matrix. If $Mv=0$, regularity gives
$\one^{\top}v=0$, and the displayed identity then gives $cv=0$.
Thus $M$ is invertible. Its rows $\one_{A_i}$ are affinely independent,
and their average is $(2c/q)\one$, the barycenter of $\Delta(q,2c)$.
Their convex hull is a full-dimensional simplex containing that
barycenter in its interior.

The affine map carries the abstract boundary cycle onto the boundary of
this simplex. Every proper face of the omitted coordinate face is
retained, so Lemma~\ref{lem:polyhedral}(b) identifies $r_c$ here with
radial projection. Such a projection maps the boundary of the inner
simplex homeomorphically onto the boundary of the coordinate face.
For the complementary labels $D\sm A_i$, replace each row vector by
$\one-\one_{A_i}$; the same argument applies. The spanning assertion
follows from Lemma~\ref{lem:boundary-generators}.
\end{proof}

\subsection{The short rounding argument for radius one}\label{ss:rounding-one}
The following lemma is all the rounding needed for six colors.
Its last case is simpler than the corresponding case of the general
five-constraint theorem. A reader following radius one may skip
Section~\ref{sec:rounding} after this proof and continue at
Section~\ref{ss:proof-topology}.

\begin{lemma}[Radius-one rounding]\label{lem:radius-one-rounding}
At most five $k$-sets with a common fractional radius-one center
have a common integral center. Thus $\C_1(N,k)$ and $Y_1$ have
the same four-skeleton.
\end{lemma}
\begin{proof}
Choose a vertex $x$ of the feasible polytope. At most six coordinates
are fractional: the sum equality and at most five tight subset
inequalities must have full rank on them. Fix the integral coordinates.
Let $I$ and $O$ be those fixed to one and zero, and let $D$ be the
remaining $m\le6$ coordinates, of total weight $\ell=k-|I|$.
If $m=0$, we are done. Complementing the coordinates and the original
sets if necessary, assume $1\le\ell\le m/2$.

For a label $A$, put $T=A\cap D$. Its residual constraint is
\[
             x(T)\ge k-1-|A\cap I|=|T|+|A\cap O|-1.
\]
An empty $T$ gives an automatic constraint. Otherwise $x(T)<|T|$
forces $A\cap O=\varnothing$, and consequently $|T|=k-|A\cap I|\ge\ell$.
If $T$ is proper in $D$, positivity outside $T$ gives $x(T)<\ell$,
so $|T|\le\ell$. If $T=D$, then $\ell\ge m-1$; together with
$\ell\le m/2$, this forces $(m,\ell)=(2,1)$, and the constraint follows
from $x(D)=\ell$. After discarding automatic constraints, all remaining
ones therefore have the form
\[
                         |T|=\ell,\qquad x(T)\ge\ell-1.
\]
We need an $\ell$-set meeting every such $T$ in at least $\ell-1$ points.
For $\ell=1$, any singleton works.

For $\ell=2$, there are at most five pairs $T_1,\ldots,T_s$.
Suppose that no pair meets all of them, and put
$H_j=\{i:j\notin T_i\}$ for $j\in D$.
These missing-index sets are nonempty and pairwise intersecting.
A singleton $H_j=\{i\}$ would force every $H_h$ to contain $i$,
so $T_i$ would be empty. If $H_j=\{i,r\}$, every $H_h$ meets
$\{i,r\}$; hence $T_i,T_r$ are disjoint and both miss the positive
coordinate $j$. This would give $x(T_i)+x(T_r)<2$, a contradiction.
Thus $|H_j|\ge3$ for all $j$, and
\[
 s\le\sum_i x(T_i)=\sum_j x_j(s-|H_j|)\le2(s-3)<s.
\]
This contradiction supplies the required pair.

The only remaining possibility is $(m,\ell)=(6,3)$.
Any two constraint triples meet in at least two points: otherwise
their scores sum to less than four, because their union has score at
most three and their intersection has score less than one.
Choose any one of the constraint triples. It meets all the others in
at least two points and is the desired integral solution.
If there are no constraints, choose any triple.
\end{proof}

\subsection{Rounding five subset constraints}\label{sec:rounding}

The next theorem rounds a fractional center while preserving any fixed
coordinates. We need this stronger form for the radius-two comparison.
Its proof only involves systems on at most six free coordinates.

\begin{theorem}\label{thm:rounding}
Let $m\ge1$ and $\ell$ be integers, let $T_1,\dots,T_s\subseteq[m]$ with
$s\le5$, and let $b_1,\dots,b_s$ be integers. If the polytope
\[
 \Pi=\Bigl\{x\in[0,1]^m:\ \sum_{j}x_j=\ell,\quad x(T_i)\ge b_i\ \
 (1\le i\le s)\Bigr\}
\]
is nonempty, then it contains a point of $\{0,1\}^m$.
\end{theorem}

The number five is sharp. On $[4]$, with $\ell=2$, the six constraints
$x(T)\ge1$ for $T\in\binom{[4]}2$ are satisfied by
$(\frac12,\frac12,\frac12,\frac12)$, whereas every $2$-subset of $[4]$
misses its complementary pair.

\begin{proof}
The idea is to pass to a vertex of $\Pi$, which has at most six
fractional coordinates, and then to treat the three possible shapes of
the residual problem by counting. The proof is divided into four steps.

\step{Step 1 (reduction to a strictly fractional vertex).}
Let $x$ be a vertex of $\Pi$, let $I=\{j:x_j=1\}$, $O=\{j:x_j=0\}$, and
$J=[m]\sm(I\cup O)$. The restriction $x|_J$ lies in the polytope
$\Pi_J\subseteq[0,1]^J$ defined by
\[
 \sum_{j\in J}y_j=\ell-|I|,\qquad y(T_i\cap J)\ge b_i-|T_i\cap I|
 \quad(1\le i\le s),
\]
a system of the same kind, and a $0/1$ point of $\Pi_J$, extended by the
values of $x$ on $I\cup O$, is a $0/1$ point of $\Pi$. Moreover $x|_J$ is
a vertex of $\Pi_J$, since $\Pi_J\times\{x|_{I\cup O}\}$ is a face of
$\Pi$. At a vertex the tight constraints have full rank; at $x|_J$ no
box constraint is tight, so the equation together with the tight subset
inequalities has rank $|J|$. There are at most $s+1\le6$ of these, so
$|J|\le6$. (If $J$ is empty, $x$ itself is a $0/1$ point.)

It therefore suffices to treat the following situation: $m\le6$, and
$\Pi$ has a vertex $x$ with $0<x_j<1$ for every $j$. Then, as just
observed,
\begin{equation}\label{eq:rank}
 \text{the equation and the subset inequalities tight at }x
 \text{ have rank }m .
\end{equation}
Replacing $x$ by $\one-x$ preserves this situation: the equation becomes
$\sum_jy_j=m-\ell$, and $x(T)\ge b$ becomes
$y([m]\sm T)\ge(m-\ell)-|T|+b$, again an integral lower bound; $0/1$
points correspond to $0/1$ points. Since $0<\ell<m$, we may therefore
assume $1\le\ell\le m/2$. The remaining cases are $\ell=1$; $\ell=2$ with
$4\le m\le6$; and $(m,\ell)=(6,3)$.

\step{Step 2 (the case $\ell=1$).}
Let $x(T)\ge b$ be one of the constraints. If $T=[m]$ then
$b\le x([m])=1=|\{j\}\cap T|$ for every $j$. If $T\ne[m]$ then
$x(T)=1-x([m]\sm T)<1$, so $b\le0$. Hence every singleton satisfies all
constraints. 

\step{Step 3 (the case $\ell=2$).}
A constraint with $b\le0$, or with $T=[m]$ (where $b\le2$), is satisfied
by every $2$-subset. For $T\ne[m]$ one has $x(T)<2$, hence $b\le1$. Let
$T_1,\dots,T_t$ ($t\le5$) be the sets carrying constraints
$x(T_i)\ge1$ with $T_i\ne[m]$. It remains to find a pair meeting every
$T_i$. Suppose that there is none, and put
\[
                 H_j=\{i:\ j\notin T_i\}\qquad(j\in[m]).
\]
A pair $\{j,j'\}$ fails exactly when $H_j\cap H_{j'}\ne\varnothing$. Hence
the sets $H_j$ are nonempty (a point $j$ with $H_j=\varnothing$ could be
paired with any other point) and pairwise intersecting. We claim that
$|H_j|\ge3$ for every $j$. If $H_j=\{i\}$, then $i$ belongs to every
$H_{j'}$, so $T_i=\varnothing$ and $x(T_i)=0<1$, which is impossible. If
$H_j=\{i,i'\}$, then every $H_{j'}$ meets $\{i,i'\}$, so $T_i$ and
$T_{i'}$ are disjoint, and neither contains $j$; hence
$x(T_i)+x(T_{i'})\le2-x_j<2$, which is again impossible. Now count with
weights:
\[
 6=3\sum_jx_j\le\sum_jx_j|H_j|=\sum_{i=1}^t\bigl(2-x(T_i)\bigr)\le t\le5 .
\]
This is a contradiction.

\step{Step 4 (the case $(m,\ell)=(6,3)$).}
Call a constraint $x(T)\ge b$ \emph{trivial} if every $3$-subset $S$ of
$[6]$ satisfies $|S\cap T|\ge b$, that is, if $b\le\max(0,|T|-3)$. For
$\varnothing\ne T\ne[6]$ the strict inequalities $0<x_j<1$ give
$x(T)<\min(|T|,3)$, so $b\le\min(|T|,3)-1$. Comparing, one finds that a
nontrivial constraint has one of the forms
\[
 x(e)\ge1\ (|e|=2),\qquad x(T)\ge1\ (|T|=3),\qquad x(T)\ge2\ (|T|=3),
 \qquad x(T)\ge2\ (|T|=4).
\]

\emph{First case: some constraint is $x(A)\ge2$ with $|A|=3$.} We claim
that the $3$-set $A$ satisfies every constraint. Suppose that
$|A\cap T|<b\le x(T)$ for some constraint, and put $a=|A\cap T|$. Since
$x(T\sm A)\le x([6]\sm A)=3-x(A)\le1$, one has $x(T)<a+1\le b$ if
$a\ge1$, using $x(T\cap A)<a$; this is impossible. So $a=0$, and then
$1\le b\le x(T)\le x([6]\sm A)\le1$. Positivity of the coordinates forces
$T=[6]\sm A$, and both constraints $x(A)\ge2$ and $x([6]\sm A)\ge1$ are
tight. Their normal vectors add up to $\one$, the normal vector of the
equation. Thus the equation and the at most five tight subset
inequalities have rank at most five, contradicting \eqref{eq:rank}. This
proves the claim.

\emph{Second case: no such constraint.} Count the $3$-subsets violating
each nontrivial constraint. For $x(e)\ge1$ they are the four $3$-sets
disjoint from $e$; for $x(T)\ge1$ with $|T|=3$ it is the single set
$[6]\sm T$; for $x(T)\ge2$ with $|T|=4$ they are the four $3$-sets
containing the pair $e=[6]\sm T$. There are twenty $3$-subsets, so five
constraints can exclude all of them only if all five are of the first or
the last form and their excluded sets are pairwise disjoint. Call $e$ a
\emph{positive} pair in the first case and a \emph{negative} pair in the
last. Two positive pairs with a common point are both avoided by the
complement of their union; two negative pairs with a common point are
both contained in their union; a positive pair $e$ disjoint from a
negative pair $e'$ is avoided by the union of $e'$ and a point outside
$e\cup e'$. Hence the positive pairs are pairwise disjoint, the negative
pairs are pairwise disjoint, and every positive pair meets every negative
pair. Three positive pairs would partition $[6]$, and no pair meets three
disjoint pairs; so if there are three positive pairs there is no negative
one, and likewise with the signs exchanged. Hence there are at most four
such constraints, and some $3$-subset satisfies all of them. 
\end{proof}

\begin{corollary}\label{cor:rounding}
Let $c\ge1$, and let $\sigma$ contain at most five $k$-subsets of $[N]$.
If $x$ is a fractional radius-$c$ center, then $\sigma$ has an integral
center $Z$ with $\one_Z\in F(x)$. In particular, $\C_c(N,k)$ and $Y_c$
have the same four-skeleton.
\end{corollary}
\begin{proof}
If $x$ is integral, use its support as $Z$. Otherwise let $I$ be the
fixed-one coordinates of $F(x)$ and $D$ its free coordinates. Apply
Theorem~\ref{thm:rounding} on $D$, with sum $k-|I|$ and constraints
\[
                 y(A\cap D)\ge k-c-|A\cap I|\qquad(A\in\sigma).
\]
They are satisfied by $x|_D$. An integral solution is the indicator of
a set $W\subseteq D$; then $Z=I\cup W$ is the required center.
\end{proof}

The rounding is constructive for a given rational feasible point.
First fix every coordinate already equal to zero or one, and compute
a vertex of the remaining polytope by linear programming, successively
minimizing each coordinate and fixing its optimal value. The rank
argument in Step~1 of Theorem~\ref{thm:rounding} leaves at most six
fractional coordinates. Fix all other coordinates of this vertex and
test the assignments on the remaining ones having the required sum.
There are at most $\binom63=20$ candidates, and
Theorem~\ref{thm:rounding} guarantees that one satisfies every constraint.
This gives an algorithm polynomial in the binary encoding length of
the system and the given point, preserving every coordinate that was
integral in that point. It solves this rounding problem; it does not
supply an algorithm for finding the fair allocations.

\subsection{Proof of the common topological proposition}\label{ss:proof-topology}
We can now prove Proposition~\ref{prop:topology} for both radii, using
the radius-two injectivity theorem proved in Section~\ref{sec:comparison}.
Put $C=\C_c(N,k)$ and $d=4c-1$.

For $c=1$, Lemma~\ref{lem:radius-one-rounding} identifies the four-skeleta of
$C$ and $Y_1$. Their chain groups and boundary maps agree through
degree four, so the inclusion is an isomorphism on homology through
degree three. For $c=2$, Theorem~\ref{thm:comparison}, together with
Propositions~\ref{prop:93} and~\ref{prop:84}, makes the inclusion
injective on homology through degree seven. Both assertions hold over
$\Z$ and $\F_2$.

In either case, $C$ and $Y_c$ have the same two-skeleton, so their
fundamental groups agree. Since $Y_c\simeq Q_c$ is $(d-1)$-connected,
$C$ is simply connected. Its reduced integral homology injects into
that of $Y_c$ and hence vanishes through degree $d-1$. Successive
applications of the Hurewicz theorem give that $C$ is $(d-1)$-connected.

Finally, Lemma~\ref{lem:nervecomp} makes $(f_c)_*$ injective on
$H_d(C;\F_2)$. By Lemma~\ref{lem:degree-one}, the images of the specified
cycles span $H_d(U_c;\F_2)$. Given $\xi\in H_d(C;\F_2)$, choose a sum of
these cycles with the same image as $\xi$. Their difference has zero
image and is therefore zero. This proves the generation assertion.

\section{The uniform box-complex index bound}\label{sec:general-index}

We prove Theorem~\ref{thm:index-intro}. The upper bound comes from an
explicit sphere map and does not use the local homology calculations.
Matching lower bounds for $r=2,3$ follow from
Proposition~\ref{prop:topology}. Write $c=r-1$, and assume $n\ge2c+1$.

\subsection{The index bound and its meaning}

Use the weighted-pair model $B(G)=\operatorname{Hom}(K_2,G)$
from Section~\ref{ss:nonvanishing-proof}. A map to a sphere is
\emph{equivariant} if exchanging the two sides sends its value
to its negative. This model and the simplicial box complex in
Matou\v{s}ek's book are $\Z_2$-homotopy equivalent
\cite[Theorem~9, p.~539]{CsorbaSimple}, so their indices agree.

\begin{lemma}[The sphere map]\label{thm:general-index}
For every $c\ge1$ and $n\ge2c+1$, there is an explicit equivariant map
\[
              B(G_{n,c+1})\longrightarrow S^{4c-1}.
\]
Consequently, $\operatorname{ind}_{\mathbb Z_2}B(G_{n,c+1})\le4c-1$,
and $\N(G_{n,c+1})$ is not $(4c-1)$-connected.
\end{lemma}

The restriction $n\ge2c+1$ matters: $G_{2,2}=K_6$ has box complex
equivariantly homotopy equivalent to $S^4$, hence index four
\cite[Proposition~4.3]{BabsonKozlovHom}.

Within the hierarchy of topological chromatic bounds~\cite{MatousekZiegler},
the usual index bound is
\[
 \chi(G)\ge\operatorname{ind}_{\mathbb Z_2}B(G)+2
 \qquad\text{\cite[Theorem~5.9.3, p.~130]{Matousek}}.
\]
Lemma~\ref{thm:general-index} shows that this bound can certify
at most $4c+1$ colors in our range of $n$. An upper bound on the
index does not itself supply a chromatic lower bound.
For $c=1,2$, however, the connectivity proved in Proposition~\ref{prop:topology} gives
the matching index lower bound and hence the exact values
\begin{equation}\label{eq:exact-small-indices}
 \operatorname{ind}_{\mathbb Z_2}B(G_{n,c+1})=4c-1,
 \qquad
 \operatorname{conn}\N(G_{n,c+1})=4c-2
 \qquad(c=1,2).
\end{equation}
Here $\operatorname{conn}X$ is the largest integer $t$ for which
$X$ is $t$-connected.

\subsection{Reducing to a fixed number of coordinates}

First we reduce $n$ to $2c+1$. A graph homomorphism is a map on
vertices that preserves adjacency. Such a map induces an equivariant
map on $B(G)$: transfer the weight at each vertex to its image,
adding weights when several vertices have the same image.
The two resulting supports still form a complete bipartite subgraph.
We use the deletion map behind Jafari and Alipour's monotonicity
inequality~\cite[Theorem~3.1, p.~73]{AJ}, including its proof here.

\begin{lemma}[Balanced deletion]\label{lem:balanced-deletion}
For $n\ge m\ge c+1$, there is a graph homomorphism
$G_{n,c+1}\to G_{m,c+1}$.
\end{lemma}

\begin{proof}
It suffices to reduce $k$ to $k-1$ when $k\ge c+2$.
Delete two coordinates $a,b$ from $[2k]$, and order the remaining
set $U$. For a $k$-set $A$, define a $(k-1)$-subset of $U$ by
\[
 p(A)=
 \begin{cases}
 A\cap U, & |A\cap\{a,b\}|=1,\\
 (A\cap U)\cup\{\min(U\setminus A)\}, & |A\cap\{a,b\}|=2,\\
 A\setminus\{\min A\}, & |A\cap\{a,b\}|=0.
 \end{cases}
\]
We claim that $|p(A)\cap p(C)|\le|A\cap C|$ for all $A,C$.
If neither set contains both deleted coordinates, the map only
removes elements. If both contain them, their intersection loses
two elements and gains at most two. If one contains both and the
other contains exactly one, it loses one and gains at most one.

The only remaining case is that $A$ contains both and $C$ contains
neither. Put $u=\min(U\setminus A)$ and $v=\min C$.
A new common element is possible only if $u\in C$, in which case
$v\le u$. If $v=u$, that element is deleted from $C$.
If $v<u$, the definition of $u$ gives $v\in A$, so deleting $v$
removes an old common element and offsets the possible gain.
This proves the claim. Adjacent inputs therefore have images
intersecting in at most $c$ elements. The images cannot coincide,
since each has $k-1>c$ elements. Iterating proves the lemma.
\end{proof}

It is now enough to construct the sphere map for $G_{2c+1,c+1}$
and compose it with the map induced by
Lemma~\ref{lem:balanced-deletion}. This reduces the construction
to $4c+2$ coordinates, independently of the original ground-set size.

\subsection{A sphere map from averaged indicator vectors}

Set $m=2c+1$ and $N=2m=4c+2$. At a point $(\lambda,\mu)$ of
$B(G_{m,c+1})$, average the indicator vectors on each side:
\[
 x=\sum_A\lambda_A\mathbf1_A,
 \qquad y=\sum_C\mu_C\mathbf1_C.
\]
The sums range over the $m$-subsets of $[N]$.
Every coordinate lies in $[0,1]$, and $\sum_jx_j=\sum_jy_j=m$.
Moreover, every pair carrying positive weights is an edge, so
\begin{equation}\label{eq:index-cross-product}
 x\cdot y=\sum_{A,C}\lambda_A\mu_C|A\cap C|\le c.
\end{equation}
Thus the averaged vectors cannot be too close. To quantify this,
put $v=x-y$ and $u=(x+y)/2$. Since $\sum_j u_j=m$,
Cauchy--Schwarz gives $\|u\|^2\ge m^2/N=m/2=c+1/2$.
Together with~\eqref{eq:index-cross-product}, this yields
\begin{equation}\label{eq:index-separation}
       \|v\|^2=4\bigl(\|u\|^2-x\cdot y\bigr)\ge2.
\end{equation}
All norms here are Euclidean.

The coordinates of $v$ sum to zero. Normalizing $v$ would therefore
give a map to $S^{N-2}$, one dimension above our target.
The quantitative separation in~\eqref{eq:index-separation} lets
us remove one more dimension: discard the first coordinate and
subtract the mean of the remaining coordinates. Explicitly, put
\[
 w_j=v_j+\frac{v_1}{N-1}\quad(2\le j\le N).
\]
Then $\sum_{j=2}^N w_j=0$, and expanding the squares gives
\[
 \|w\|^2=\|v\|^2-\frac{N}{N-1}v_1^2
       \ge2-\frac{N}{N-1}
       =\frac{N-2}{N-1}>0,
\]
where we used $|v_1|\le1$. We may therefore define
\[
 \Phi(\lambda,\mu)=\frac{w}{\|w\|}.
\]
This is a continuous map to the unit sphere of
$\{w\in\mathbb R^{N-1}:\sum_j w_j=0\}$, a vector space of
dimension $N-2$. Its unit sphere is $S^{N-3}=S^{4c-1}$.
Exchanging $\lambda$ and $\mu$ exchanges $x$ and $y$, hence sends
$v$, $w$, and $\Phi$ to their negatives. The map is equivariant,
proving the index assertion of Lemma~\ref{thm:general-index}.

For the connectivity assertion, recall that $B(G)$ and $\N(G)$
are homotopy equivalent as ordinary spaces. A $t$-connected free
$\mathbb Z_2$-space has index at least $t+1$
\cite[Proposition~5.3.2(iv), pp.~96--98]{Matousek};
Section~\ref{ss:nonvanishing-proof} described the successive extensions over hemispheres
that prove this fact. If $\N(G_{n,c+1})$ were $(4c-1)$-connected,
its box complex would have index at least $4c$, contradicting the
map just constructed. This completes the proof of
Lemma~\ref{thm:general-index} and, using Proposition~\ref{prop:topology}, of
\eqref{eq:exact-small-indices}. This proves Theorem~\ref{thm:index-intro}.

\paragraph{Where the extension to larger $c$ remains open.}
The polyhedral deformation of Lemma~\ref{lem:polyhedral} holds for every $c$:
its retained-face model $Q_c$ is $(4c-2)$-connected. Transferring
this conclusion to the neighborhood complex requires the separate
discrete-to-polyhedral comparison, which we have proved only for
$c=1,2$. Even a matching connectivity theorem for larger $c$
would give only $4c+1$ colors by the standard bound. To obtain
$4c+2$ by our method would additionally require suitable homology
generators and local coloring constraints that kill their images.
\subsection{Chromatic growth from nonlinear discrepancy}\label{ss:chromatic-growth}

We prove Corollary~\ref{cor:general-allowance} and compare its
chromatic consequence with the index bound. The lower bounds here
hold for every $n\ge r$, without the earlier restriction $n\ge2r-1$.

A set function is \emph{$1$-Lipschitz} if adding or removing one
element changes its value by at most one. The two-part case of
\cite[Theorem~8]{HMMS26} says that, for any $q$ such functions on
$2^{[m]}$, some $S\subseteq[m]$ satisfies
\[
 |f_i(S)-f_i(S^c)|<D(q)
 \qquad(i\in[q]).
\]

\begin{proof}[Proof of Corollary~\ref{cor:general-allowance}]
Suppose that $G_{n,c+1}$ has a proper $q$-coloring. Composing it
with the retraction gives a proper coloring of $\Gamma_{c,c}(2n)$.
For each nonempty color class $\mathcal I_i$, use its distance function
$f_i(S)=\min_{T\in\mathcal I_i}|S\mathbin{\triangle}T|$,
where $\triangle$ denotes symmetric difference.
These functions are $1$-Lipschitz; repeat them to obtain $q$ if needed.
Choose $S$ from the discrepancy theorem and let $i$ be its color.
Since $f_i(S)=0$, there is $T\in\mathcal I_i$ with
$|S^c\mathbin{\triangle}T|<D(q)\le c+1$. Integrality gives
\[
       |S\cap T|+|S^c\cap T^c|=|S^c\mathbin{\triangle}T|\le c.
\]
Thus $S,T$ are adjacent and have the same color. They are distinct,
since $S=T$ would make the displayed distance $2n>c$.
This contradiction proves the chromatic bound.
Theorem~\ref{thm:balanced-threshold} now gives balanced EF$c$ when
$n>c$. The small-bundle and dummy-good arguments in
Section~\ref{sec:fair} cover every remaining number of goods.
\end{proof}

For $r\ge4$, put $q=\lfloor r^2/(4\ln r+2)\rfloor\ge2$.
Since $2q\le r^2$, one has
$D(q)^2\le q(2+4\ln r)\le r^2$. Hence, for every $n\ge r$,
\begin{equation}\label{eq:chromatic-growth}
 \chi(G_{n,r})\ge
 1+\left\lfloor\frac{r^2}{4\ln r+2}\right\rfloor
 =\Omega(r^2/\log r).
\end{equation}
The growth rate is inherited from nonlinear discrepancy
\cite{HMMS26,DF25}; the retraction transfers it to the balanced graphs.
For comparison, the Hadamard construction of \cite{BCK} gives
$\chi(G_{n,r})\le2h$ whenever a Hadamard matrix of order $h$ exists
and $2n\ge h>4r^2$. Choosing a power of two with
$4r^2<h\le8r^2$ gives $\chi(G_{n,r})\le16r^2$ for $n\ge4r^2$.
This upper bound requires sufficiently large $n$: at $n=r$ the graph
is complete on $\binom{2r}{r}$ vertices.

More explicitly, the discrepancy consequence exceeds the index ceiling
for every $r\ge56$. Take $q=4r-3$ and put
$F(r)=r^2-2(4r-3)(1+\ln(8r-6))$. The estimate
$\ln442<1219/200$ gives $F(56)>0$ and $F'(56)>0$, while
\[
 F'(r)=2r-16-8\ln(8r-6),\qquad
 F''(r)=2-\frac{64}{8r-6}>0\quad(r\ge56).
\]
For completeness, the logarithmic estimate follows from
$e^6>\sum_{j=0}^{15}6^j/j!>403$ and
$403(1+19/200+(19/200)^2/2)>442$.
Thus $D(4r-3)\le r$ for $r\ge56$, and
$\chi(G_{n,r})\ge4r-2$ for every $n\ge r$.
For $n\ge2r-1$, this is above the ceiling $4r-3$ of the standard
index bound in Theorem~\ref{thm:index-intro}.

\section{Integral and fractional radius-two centers}\label{sec:comparison}

Fix $1\le k<N$. Throughout this section the radius is two, and we
abbreviate $\C=\C_2(N,k)$, $Y=Y_2$, $U=U_2$, $Q=Q_2$,
$P_Z=P_{2,Z}$, $P_\sigma=\bigcap_{A\in\sigma}P_A$, $i=i_2$, and $f=f_2$.
Coefficients $\kk$ are either $\Z$ or $\F_2$.
Throughout this section a face used as a family of labels is nonempty.
Reduced homology in degree $-1$ vanishes exactly when a complex has
at least one vertex; groups in smaller degrees are zero. A complex is
\emph{acyclic through degree $d$} if, with the stated coefficients,
its reduced homology vanishes in degrees $-1$ through $d$.

The goal is injectivity of the inclusion on homology through degree
seven. Its two finite inputs are proved in Appendices~\ref{sec:nine}
and~\ref{sec:eight}.

\begin{theorem}\label{thm:comparison}
Suppose that
\begin{equation}\label{eq:local-hyp}
 \Hr_t(\C_2(9,3);\Z)=\Hr_t(\C_2(8,4);\Z)=0\qquad(0\le t\le7).
\end{equation}
Then $i_*\colon H_j(\C;\kk)\to H_j(Y;\kk)$ is injective for
$0\le j\le7$.
\end{theorem}

\begin{remark}
The idea of the proof is to record, together with each simplex of labels,
a simplex of centers. Three things are needed. A cycle of $\C$ must be
liftable to such recorded data, with centers chosen compatibly on common
faces; this requires the spaces of admissible centers of a family to be
acyclic in a range of degrees that shrinks as the family grows
(Lemma~\ref{lem:lift}). A boundary in $Y$ must lift to a boundary of
recorded data; here all fractional centers are admissible and form a
contractible set (Lemma~\ref{lem:epsilon}). Finally, recorded data must be
converted back into chains of $\C$ without altering a lifted cycle
(Lemma~\ref{lem:replace}); this requires certain local complexes to be
acyclic through degree seven, and it is here that \eqref{eq:local-hyp}
enters. Because injectivity in degree seven concerns fillings of
dimension eight, the conversion must be defined through degree eight.
\end{remark}

\subsection{Hulls of integral centers}\label{sec:hulls}

For families of six or more labels a fractional center need not round to
an integral one. In the comparison argument such families will instead be
equipped with a space of centers built from their integral centers only.
The mediator below will fill every triangle of integral centers.

For a face $\sigma$ of $\C_2(N,k)$ let $\cZ_\sigma$ be the set of its
centers, and define
\[
 T_\sigma=\bigcup_{F\ \mathrm{retained}}
       \conv\{\one_Z:\ Z\in\cZ_\sigma,\ \one_Z\in F\},
\]
the union over the retained faces $F$ of the convex hulls of those
integral centers which lie in $F$ (empty hulls are omitted). Thus
$T_\sigma\subseteq Q$ is a finite union of polytopes. For a nonempty face
$\sigma$ of $Y$ we also put $D_\sigma=P_\sigma\cap Q$, the set of
fractional centers lying in $Q$.

\begin{lemma}[Segments of centers]\label{lem:hull-basic}
Let $\sigma$ be a face of $\C_2(N,k)$.
\begin{enumerate}[(a)]
\item Every $x\in T_\sigma$ lies in the convex hull of the integral
centers of $\sigma$ lying in $F(x)$. In particular $F(x)$ contains an
integral center of $\sigma$.
\item $T_\sigma\subseteq D_\sigma$, and $T_\sigma\subseteq T_\tau$ whenever
$\varnothing\ne\tau\subseteq\sigma$.
\item If $Z,Z'\in\cZ_\sigma$, then $d_J(Z,Z')\le4$, and the segment from
$\one_Z$ to $\one_{Z'}$ lies in $T_\sigma$.
\end{enumerate}
\end{lemma}

\begin{proof}
(a) Write $x=\sum\lambda_Z\one_Z$ with $\lambda_Z>0$, the $Z$ being
centers in a retained face. A coordinate in which $x$ equals $0$ or $1$
has the same value in every $\one_Z$ occurring, so all of them lie in
$F(x)$.
(b) Each $\one_Z$ with $Z\in\cZ_\sigma$ lies in the convex set
$P_\sigma$, and $\cZ_\sigma\subseteq\cZ_\tau$.
(c) For $A\in\sigma$ one has $d_J(Z,Z')\le d_J(Z,A)+d_J(A,Z')\le4$. The
smallest face containing $\one_Z$ and $\one_{Z'}$ has the $2d_J(Z,Z')\le8$
coordinates of $Z\triangle Z'$ as free coordinates, so it is retained and
contains the segment. 
\end{proof}

The essential point is that every triangle of integral centers can be
filled inside $T_\sigma$. This rests on the following lemma.

\begin{lemma}[Mediator]\label{lem:mediator}
Let $Z_1,Z_2,Z_3$ be $k$-sets which have a common center. There is a
$k$-set $W$ such that
\begin{enumerate}[(1)]
\item $d_J(W,A)\le2$ for every $k$-set $A$ which is a center of
$\{Z_1,Z_2,Z_3\}$;
\item for each pair $i,j$, the smallest face containing $\one_W$,
$\one_{Z_i}$ and $\one_{Z_j}$ is retained.
\end{enumerate}
\end{lemma}

\begin{proof}
If the smallest face containing $\one_{Z_1},\one_{Z_2},\one_{Z_3}$ is
retained, then $W=Z_1$ has both properties. Suppose it is omitted. We
construct $W$ in three steps.

\step{Step 1 (blocks).}
Divide the coordinates according to their membership in the three sets:
\begin{center}
\begin{tabular}{@{}lll@{}}
\toprule
block & points belonging to & size\\
\midrule
$C$ & all three sets & $|C|$\\
$X_i$ & $Z_i$ only & $x_i$\\
$Y_i$ & the two sets other than $Z_i$ & $y_i$\\
$E$ & none of the sets & $|E|$\\
\bottomrule
\end{tabular}
\end{center}
The free coordinates of the face spanned by the three sets are those in
the six blocks $X_i,Y_i$. Since $Z_i$ consists of $C$, $X_i$ and the two
blocks $Y_j$ with $j\ne i$, and all three sets have $k$ elements, there is
an integer $t$ with
\[
 x_i=y_i+t\ (i=1,2,3),\qquad \ell:=k-|C|=y_1+y_2+y_3+t,\qquad
 m:=\sum_i(x_i+y_i)=2\ell+t .
\]
The face spanned by the three sets is of type $\Delta(m,\ell)$.
Complementing all sets interchanges $C$ with $E$ and $X_i$ with $Y_i$,
replaces $t$ by $-t$, and preserves distances and retention; so we may
assume $t\ge0$. Relabel so that $y_3=\max(y_1,y_2,y_3)$. Since
$0\le t\le\ell$, one has
$\rho=\ell(\ell+t)/(2\ell+t)\le2\ell/3$, and the face is omitted only
if $\ell\ge4$.

\step{Step 2 (a median).}
Give each coordinate the weight $w_q$ equal to the number of the sets
$Z_i$ containing it. For a $k$-set $A$ put
$S(A)=\sum_i d_J(A,Z_i)=3k-\sum_{q\in A}w_q$. Let $W$ consist of $C$, the
three blocks $Y_i$, and $t$ points of $X_3$ (possible since
$x_3=y_3+t$). Then $|W|=|C|+(\ell-t)+t=k$, the set $W$ consists of $k$
coordinates of largest weight, and
\[
          S(W)=3\ell-2(\ell-t)-t=\ell+t
\]
is the least value of $S$. A center $A$ of the three sets has
$S(A)\le6$; hence $\ell+t\le6$, and $t\le2$ because $\ell\ge4$. Now let
$A$ be any center of the three sets, and $p=d_J(A,W)$. Among the $p$
points of $W\sm A$ at most $t$ have weight one and the others have weight
at least two, while the $p$ points of $A\sm W$ have weight at most one.
Hence
\[
            S(A)-S(W)\ge(2p-t)-p=p-t ,
\]
and $p\le S(A)-S(W)+t\le6-\ell\le2$. This proves (1).

\step{Step 3 (the three triangles).}
Let $\{h,i,j\}=\{1,2,3\}$. The pair $Z_i,Z_j$ has distance
$L_h=x_i+y_j=\ell-y_h$, which is at most four because both sets lie at
distance at most two from a common center. It spans a face of type
$\Delta(2L_h,L_h)$. Adding $\one_W$ creates a new free coordinate only
where $Z_i$ and $Z_j$ agree and $W$ differs from them. This happens only
for $h=3$, at the $t$ chosen points of $X_3$. Hence the three faces in
(2) have $2L_1$, $2L_2$ and $2L_3+t$ free coordinates. If $t=0$ all of
these are at most eight. If $t>0$ then $L_3$ is the smallest of the
three distances, and $3L_3\le L_1+L_2+L_3=2\ell+t\le12-t<12$; so
$L_3\le3$ and $2L_3+t\le8$. A face with at most eight free coordinates is
retained. 
\end{proof}

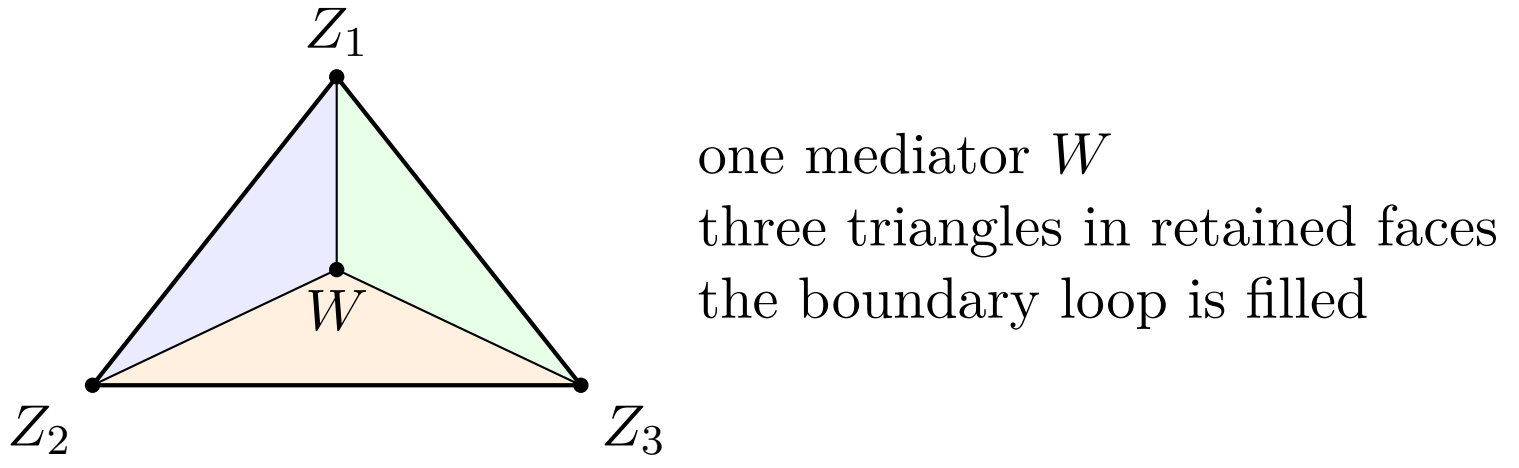
\begin{figure}[ht]
\centering
\begin{tikzpicture}[scale=.85,every node/.style={font=\normalsize}]
\coordinate (a) at (0,1.5);\coordinate (b) at (-1.9,-.9);\coordinate (c) at (1.9,-.9);
\coordinate (w) at (0,0);
\fill[blue!8] (a)--(b)--(w)--cycle;
\fill[orange!12] (b)--(c)--(w)--cycle;
\fill[green!10] (c)--(a)--(w)--cycle;
\draw[thick] (a)--(b)--(c)--cycle;
\draw (w)--(a) (w)--(b) (w)--(c);
\foreach \v/\lab/\pos in {a/Z_1/above,b/Z_2/below left,c/Z_3/below right,w/W/below}
 \fill (\v) circle (1.7pt) node[\pos] {$\lab$};
\node[align=left,anchor=west] at (2.65,.3) {one mediator $W$\\three triangles in retained faces\\the boundary loop is filled};
\end{tikzpicture}
\caption{The mediator supplies a simplicial filling of a triangle of
centers. This is an incidence diagram: the three filled triangles lie
in retained coordinate faces, even when the original three-center face
is omitted.}
\label{fig:mediator}
\end{figure}
\begin{proposition}[Integral center hulls]\label{prop:hull}
For every face $\sigma$ of $\C_2(N,k)$ the space $T_\sigma$ is nonempty
and simply connected. In particular
$\Hr_0(T_\sigma;\kk)=H_1(T_\sigma;\kk)=0$ for $\kk=\Z,\F_2$.
\end{proposition}

\begin{proof}
In this proof we identify each center $Z$ with its indicator vector
$\one_Z$. A center of $\sigma$ gives a point of $T_\sigma$. By
Lemma~\ref{lem:hull-basic}(a) every point $x$ of $T_\sigma$ is joined by a
segment inside $\conv(\cZ_\sigma\cap F(x))\subseteq T_\sigma$ to a center
$Z_x$ in $F(x)$, and by Lemma~\ref{lem:hull-basic}(c) any two centers are
joined by a segment in $T_\sigma$. Hence $T_\sigma$ is path connected.

Triangulate $T_\sigma$ so that each of the polytopes
$\conv(\cZ_\sigma\cap F)$ is a subcomplex. By simplicial approximation
every loop is homotopic to an edge loop \cite[Theorem~2C.1]{Hatcher}. For
each vertex $x$ of the triangulation choose a center $Z_x$ in $F(x)$. An
edge $[x,y]$ lies in some $\conv(\cZ_\sigma\cap F)$; then $F(x)$ and
$F(y)$ are faces of $F$, so $\one_{Z_x}$ and $\one_{Z_y}$ lie in the same
convex set, and straight-line interpolation moves the edge onto the
segment $[\one_{Z_x},\one_{Z_y}]$ inside $T_\sigma$. Thus every loop is
homotopic to a closed polygon whose vertices are centers
$Z^{(0)},Z^{(1)},\dots,Z^{(p)}=Z^{(0)}$ and whose sides are the segments
of Lemma~\ref{lem:hull-basic}(c).

Using the segments from $Z^{(0)}$ to the other vertices, such a polygon
is a product of triangular loops $Z^{(0)}\to Z\to Z'\to Z^{(0)}$, so it
suffices to show that each of these is null-homotopic. Every label of
$\sigma$ is a common center of $Z^{(0)},Z,Z'$. Let $W$ be the mediator of
Lemma~\ref{lem:mediator}. By property (1), every label of $\sigma$ lies at
distance at most two from $W$, so $W\in\cZ_\sigma$. By property (2), the
three triangles $\conv(\one_W,\one_{Z^{(0)}},\one_Z)$,
$\conv(\one_W,\one_Z,\one_{Z'})$ and $\conv(\one_W,\one_{Z'},\one_{Z^{(0)}})$
lie in retained faces, hence in $T_\sigma$. Mapping the cone on the
boundary of a triangle affinely onto these three triangles gives a
null-homotopy of the loop. Hence $T_\sigma$ is simply connected. The
statement about homology follows from the Hurewicz isomorphism in degree
one and the universal coefficient theorem. 
\end{proof}

\subsection{Local complexes of centers}\label{ss:carriers}
For a retained face $F$ let $\cK_F$ be the subcomplex of $\C$ consisting
of the families having a center among the integral vertices of $F$; the
labels themselves need not lie in $F$. Clearly $\cK_F\subseteq\cK_G$ when
$F\subseteq G$. Let $\C_2(F)\subseteq\cK_F$ be the subcomplex of families
of integral vertices of $F$ having a center among the integral vertices
of $F$. Deleting the fixed coordinates preserves Johnson distances, so
$\C_2(F)$ is isomorphic to $\C_2(m,\ell)$ when $F$ is of type
$\Delta(m,\ell)$.

\begin{lemma}[Coordinate-face retractions]\label{lem:face-retraction}
For every face $F$ of $\Delta(N,k)$ there is a map $p_F$ from
$\binom{[N]}k$ onto the set of integral vertices of $F$ which fixes the
integral vertices of $F$ and satisfies
$d_J(p_F(A),p_F(B))\le d_J(A,B)$ for all $A,B$.
\end{lemma}

\begin{proof}
We fix the coordinates of $I\cup O$ one at a time, and it suffices to
treat one step. Let $F'$ be a face with free set $D'$, let $j\in D'$, and
suppose the face $F''=F'\cap\{x_j=0\}$ is nonempty. Order $D'$, and for
an integral vertex $Z$ of $F'$ define $p_j(Z)=Z$ if $j\notin Z$, and
$p_j(Z)=(Z\sm\{j\})\cup\{a(Z)\}$ if $j\in Z$, where $a(Z)$ is the least
element of $D'\sm(Z\cup\{j\})$; it exists because $F''$ is nonempty. We
claim $|p_j(Z)\cap p_j(Z')|\ge|Z\cap Z'|$. This is clear if neither set
contains $j$, and if exactly one does, since then $j$ is not a common
element. If both contain $j$, the common element $j$ is lost; if
$a(Z)=a(Z')$ it is replaced by this element, and if, say, $a(Z)<a(Z')$
then $a(Z)\in Z'$ by minimality of $a(Z')$, so $a(Z)$ becomes a common
element. Thus $p_j$ is $1$-Lipschitz, maps onto the integral vertices of
$F''$, and fixes them. Fixing $j$ to the value $1$ is the same operation
applied to complements, complementation being an isometry from $k$-sets
to $(N-k)$-sets. Composing these maps, starting from $F'=\Delta(N,k)$,
gives $p_F$. 
\end{proof}

\begin{lemma}[Local acyclicity]\label{lem:carriers}
Assume \eqref{eq:local-hyp}. Then $\Hr_t(\cK_F;\kk)=0$ for every retained
face $F$ and $-1\le t\le7$.
\end{lemma}

\begin{proof}
\step{Step 1.} The inclusion $\C_2(F)\subseteq\cK_F$ is a homotopy
equivalence. Indeed, if $Z$ is an integral vertex of $F$ and a center of
$\sigma\in\cK_F$, then $p_F(Z)=Z$ is a center of $p_F(\sigma)$ and of
$\sigma\cup p_F(\sigma)$, by Lemma~\ref{lem:face-retraction}. So $p_F$ is a
simplicial retraction of $\cK_F$ onto $\C_2(F)$, and its composite with
the inclusion is contiguous to the identity: the two images of each
simplex together span a simplex. Contiguous simplicial maps are homotopic
\cite[\S3.5]{Spanier}.

\step{Step 2.} Let $F$ be of type $\Delta(m,\ell)$. If
$\min(\ell,m-\ell)\le2$, any two integral vertices of $F$ are at distance
at most two, and $\C_2(F)$ is a full simplex. Complementation within $F$
gives $\C_2(m,\ell)\cong\C_2(m,m-\ell)$, so let $3\le\ell\le m/2$. The
condition $\ell(m-\ell)\le2m$ then leaves $\ell=3$ with $6\le m\le9$, and
$(m,\ell)=(8,4)$; for $\ell\ge5$ one has
$\ell(m-\ell)-2m=(\ell-2)m-\ell^2\ge\ell^2-4\ell>0$.

\step{Step 3.} For $m\le8$ let $F_m$ be the face of $\Delta(9,3)$ with
fixed-zero set $\{m+1,\dots,9\}$. By Lemma~\ref{lem:face-retraction},
$p_{F_m}$ maps every face of $\C_2(9,3)$ to a face of
$\C_2(F_m)\cong\C_2(m,3)$, and it fixes $\C_2(F_m)\subseteq\C_2(9,3)$. So
$\C_2(m,3)$ is a retract of $\C_2(9,3)$, and its reduced homology is a
direct summand of that of $\C_2(9,3)$. Together with
\eqref{eq:local-hyp} this gives $\Hr_t(\C_2(F);\Z)=0$ for $t\le7$ in all
cases, and the universal coefficient theorem gives the same over $\F_2$.

\end{proof}

\subsection{Choosing admissible center spaces}\label{ss:centers}
For a nonempty face $\sigma$ of $Y$ recall that $D_\sigma=P_\sigma\cap Q$.
The deformation of Lemma~\ref{lem:polyhedral}(b) maps the nonempty convex
set $P_\sigma$ into itself and ends in $D_\sigma$, which it fixes; hence
$D_\sigma$ is contractible. For a face $\sigma$ of $\C$ define
\[
   E_\sigma=\begin{cases} D_\sigma, & |\sigma|\le5,\\
                          T_\sigma, & |\sigma|\ge6. \end{cases}
\]
The choice must provide both enough vanishing homology and an integral
center in the smallest coordinate face of each point. For at most five
labels, $D_\sigma$ is contractible and Corollary~\ref{cor:rounding}
supplies that integral center. For larger families this rounding
guarantee is unavailable, so we use $T_\sigma$, which has the
integral-center property by construction. Its simple connectivity
(Proposition~\ref{prop:hull}) suffices because the required homology
range shrinks to $t\le7-|\sigma|\le1$.

\begin{lemma}[Admissible centers]\label{lem:E}
Let $\sigma$ be a face of $\C$.
\begin{enumerate}[(a)]
\item $E_\sigma\subseteq D_\sigma$; if $\varnothing\ne\tau\subseteq\sigma$
then $D_\sigma\subseteq D_\tau$ and $E_\sigma\subseteq E_\tau$.
\item $\Hr_t(E_\sigma;\kk)=0$ for $-1\le t\le7-|\sigma|$.
\item For every $x\in E_\sigma$ the face $F(x)$ contains an integral center
of $\sigma$.
\end{enumerate}
\end{lemma}

\begin{proof}
(a) Use Lemma~\ref{lem:hull-basic}(b); when $|\tau|\le5<|\sigma|$ one has
$E_\sigma=T_\sigma\subseteq D_\sigma\subseteq D_\tau=E_\tau$.
(b) For $|\sigma|\le5$ the set $D_\sigma$ is contractible. For
$|\sigma|\ge6$ the range is $t\le1$, and Proposition~\ref{prop:hull}
applies.
(c) For $|\sigma|\ge6$ this is Lemma~\ref{lem:hull-basic}(a); for
$|\sigma|\le5$ it is Corollary~\ref{cor:rounding}. 
\end{proof}

The required range shrinks with the number of labels:
\begin{center}
\begin{tabular}{@{}lll@{}}
\toprule
$|\sigma|$ & Chosen space & Property used in a seven-cycle lift\\
\midrule
at most $5$ & $D_\sigma$ & contractible; five-label rounding applies\\
$6$ & $T_\sigma$ & connected, with $H_1=0$\\
$7$ & $T_\sigma$ & connected\\
$8$ & $T_\sigma$ & nonempty\\
\bottomrule
\end{tabular}
\end{center}
This is why simple connectivity of the integral center hulls suffices.

Part (c) is the bridge between the two kinds of centers: a point of
$E_\sigma$ certifies that $\sigma$ belongs to $\cK_{F(x)}$.

Choose a triangulation $\fT$ of $Q$ in which every retained face, every
$D_\sigma$ ($\sigma\in Y$) and every $T_\sigma$ ($\sigma\in\C$) is a
subcomplex; it exists because these are finitely many finite unions of
polytopes \cite[Chapter~2]{RS}. Every simplex $a$ of $\fT$ lies in a
retained face; let $F(a)$ be the smallest face of $\Delta(N,k)$
containing $a$. It is retained, and $F(a')\subseteq F(a)$ for every face
$a'$ of $a$; for a vertex $x$ of $\fT$ this agrees with the earlier
notation $F(x)$.

\subsection{Keeping centers with chains}\label{ss:recorded}
Let $V=\binom{[N]}k$. Let $C_*(\Delta^V)$ be the oriented simplicial chain
complex of the full simplex on $V$, and $C_*(\fT)$ that of $\fT$, both
with coefficients in $\kk$ and without augmentation. In the tensor
product, with differential
\[
 \partial(\sigma\otimes a)=\partial\sigma\otimes a
              +(-1)^{\dim\sigma}\sigma\otimes\partial a ,
\]
let $B$ be the span of the elements $\sigma\otimes a$ with $\sigma\in Y$
and $a$ a simplex of $D_\sigma$, and let $B^E\subseteq B$ be the span of
those with $\sigma\in\C$ and $a\subseteq E_\sigma$. Since
$D_\sigma\subseteq D_\tau$ for nonempty $\tau\subseteq\sigma\in Y$, and by
Lemma~\ref{lem:E}(a), both are subcomplexes. A generator $\sigma\otimes a$
records a family of labels together with a simplex of admissible centers
for it; its degree is $\dim\sigma+\dim a$, and we call $\dim\sigma$ its
\emph{label dimension}. (Geometrically, $B$ is the cellular chain complex
of $\bigcup_{\sigma\in Y}|\Delta^\sigma|\times D_\sigma$.)

Let $\aug\colon C_*(\fT)\to\kk$ be the augmentation, which sends each
vertex to $1$. The chain maps $\mathrm{id}\otimes\aug$ give
\[
 \varepsilon_Y\colon B\to C_*(Y),\qquad
 \varepsilon_\C\colon B^E\to C_*(\C),\qquad
 \sigma\otimes a\mapsto\begin{cases}\sigma,&\dim a=0,\\
                                     0,&\dim a>0,\end{cases}
\]
and $\varepsilon_Y=i\circ\varepsilon_\C$ on $B^E$.

\begin{lemma}[Filling recorded data]\label{lem:epsilon}
If $b\in B$ is a cycle and $\varepsilon_Y(b)$ is a boundary in $C_*(Y)$,
then $b$ is a boundary in $B$.
\end{lemma}

\begin{proof}
Suppose first that $\varepsilon_Y(b)=0$, and let $p$ be the largest label
dimension occurring in $b$. Write the part of $b$ of label dimension $p$
as $\sum_\sigma\sigma\otimes c_\sigma$, with $\dim\sigma=p$ and
$c_\sigma$ a chain of $D_\sigma$. The terms of $\partial b$ of label
dimension $p$ are $(-1)^p\sum_\sigma\sigma\otimes\partial c_\sigma$, so
each $c_\sigma$ is a cycle. If $c_\sigma$ has degree zero, the vanishing of
$\varepsilon_Y(b)$ says that its augmentation is zero. Since $D_\sigma$ is
contractible, $c_\sigma=\partial d_\sigma$ for a chain $d_\sigma$ of
$D_\sigma$. Now
\[
 b-\partial\Bigl(\sum_\sigma(-1)^p\sigma\otimes d_\sigma\Bigr)
 =b-\sum_\sigma\sigma\otimes c_\sigma
   -\sum_\sigma(-1)^p\,\partial\sigma\otimes d_\sigma
\]
is a cycle with no terms of label dimension $p$ or more, and its image
under $\varepsilon_Y$ is still zero, since every $d_\sigma$ has positive
degree. Repeating this, one reaches zero, so $b$ is a boundary.

In general, write $\varepsilon_Y(b)=\partial v$. Choosing for each
simplex $\sigma$ occurring in $v$ a vertex of $D_\sigma$ produces
$u\in B$ with $\varepsilon_Y(u)=v$. Then $b-\partial u$ is a cycle with
$\varepsilon_Y(b-\partial u)=0$, and the first case applies. 
\end{proof}

\begin{lemma}[Lifting an integral cycle]\label{lem:lift}
Let $0\le j\le7$. For every $j$-cycle $z$ of $\C$ there is a cycle
$\tilde z\in B^E$ of degree $j$ with $\varepsilon_\C(\tilde z)=z$.
\end{lemma}

\begin{proof}
Write $z=\sum_\sigma c_\sigma\sigma$ with $\dim\sigma=j$. Since
$|\sigma|=j+1\le8$, Lemma~\ref{lem:E}(b) shows that $E_\sigma$ is
nonempty; choose a vertex $x_\sigma$ of $\fT$ in $E_\sigma$, and put
$\tilde z_0=\sum_\sigma c_\sigma\,\sigma\otimes x_\sigma$. Then
$\varepsilon_\C(\tilde z_0)=z$, and $\partial\tilde z_0$ has label
dimension $j-1$.

Suppose that for some $s\ge0$ a chain $\tilde z_s\in B^E$ of degree $j$
has been found with $\varepsilon_\C(\tilde z_s)=z$ and with $\partial\tilde
z_s$ of label dimension at most $p=j-s-1$. If $p<0$ we are done. Otherwise
write the part of $\partial\tilde z_s$ of label dimension $p$ as
$\sum_\tau\tau\otimes w_\tau$, with $\dim\tau=p$ and $w_\tau$ a chain of
degree $s$ in $E_\tau$. The terms of label dimension $p$ in
$0=\partial\partial\tilde z_s$ are
$(-1)^p\sum_\tau\tau\otimes\partial w_\tau$, so each $w_\tau$ is a cycle.
For $s=0$ its augmentation is the coefficient of $\tau$ in
$\partial\varepsilon_\C(\tilde z_0)=\partial z=0$. Since
$s=j-1-p\le6-p=7-|\tau|$, Lemma~\ref{lem:E}(b) gives a chain $y_\tau$ of
$E_\tau$ with $\partial y_\tau=w_\tau$. Put
\[
   \tilde z_{s+1}=\tilde z_s-(-1)^p\sum_\tau\tau\otimes y_\tau .
\]
Since $\partial(\tau\otimes y_\tau)=\partial\tau\otimes y_\tau
+(-1)^p\tau\otimes w_\tau$, the terms of label dimension $p$ cancel in
$\partial\tilde z_{s+1}$, and the new terms have label dimension $p-1$
(they vanish if $p=0$, the chain complexes being unaugmented). The image
under $\varepsilon_\C$ is unchanged. After at most $j+1$ steps one obtains
the required cycle. 
\end{proof}

\begin{lemma}[Returning to integral chains]\label{lem:replace}
Assume \eqref{eq:local-hyp}. There are $\kk$-linear maps
$\phi\colon B_j\to C_j(\C)$ for $0\le j\le8$ such that
\begin{enumerate}[(i)]
\item $\partial\phi=\phi\partial$ on $B_j$ for $1\le j\le8$;
\item $\phi=\varepsilon_\C$ on $B^E_j$;
\item $\phi(\sigma\otimes a)$ is a chain of $\cK_{F(a)}$ for every
generator $\sigma\otimes a$.
\end{enumerate}
\end{lemma}

\begin{proof}
We define $\phi$ on generators by induction on the degree. On $B^E$ we set
$\phi=\varepsilon_\C$. This is compatible with (iii): a generator
$\sigma\otimes a$ of $B^E$ with $\dim a>0$ is sent to $0$, and a generator
$\sigma\otimes x$ with $x$ a vertex is sent to $\sigma$, which belongs to
$\cK_{F(x)}$ by Lemma~\ref{lem:E}(c). It is compatible with (i), because
$\varepsilon_\C$ is a chain map and $B^E$ is a subcomplex.

Every generator of degree zero has the form $\{A\}\otimes x$ with
$x\in D_{\{A\}}=E_{\{A\}}$, so it lies in $B^E$ and is sent to the vertex
$A$. Now let $1\le n\le8$, suppose $\phi$ has been defined in degrees
below $n$, and let $b=\sigma\otimes a$ be a generator of degree $n$ not
in $B^E$. The chain $\phi(\partial b)$ lies in $\cK_{F(a)}$: the terms
$\tau\otimes a$ of $\partial b$ are sent into $\cK_{F(a)}$, and the terms
$\sigma\otimes a'$, with $a'$ a facet of $a$, into
$\cK_{F(a')}\subseteq\cK_{F(a)}$. It is a cycle: for $n\ge2$ because
$\partial\phi(\partial b)=\phi(\partial\partial b)=0$ by (i) in degree
$n-1$, and for $n=1$ because $\phi$ sends every generator of degree zero
to a single vertex, so that $\phi(\partial b)$ has augmentation zero. By
Lemma~\ref{lem:carriers}, $\Hr_{n-1}(\cK_{F(a)};\kk)=0$, so we may choose
$\phi(b)\in C_n(\cK_{F(a)})$ with $\partial\phi(b)=\phi(\partial b)$. This
completes the induction and the proof.
\end{proof}

\subsection{Proof of Theorem~\ref{thm:comparison}}\label{ss:proof-comparison}
Let $0\le j\le7$, and let $z$ be a $j$-cycle of $\C$ whose image is a
boundary in $Y$, say $i(z)=\partial v$. By Lemma~\ref{lem:lift} there is
a cycle $\tilde z\in B^E_j$ with $\varepsilon_\C(\tilde z)=z$. Then
$\varepsilon_Y(\tilde z)=i(z)=\partial v$, and Lemma~\ref{lem:epsilon}
gives $w\in B_{j+1}$ with $\tilde z=\partial w$. Since $j+1\le8$,
Lemma~\ref{lem:replace} gives
\[
 z=\varepsilon_\C(\tilde z)=\phi(\tilde z)=\phi(\partial w)=\partial\phi(w).
\]
Thus $z$ is a boundary in $\C$. 

\clearpage\appendix
\section{Filling from fibers}\label{sec:filling}

The two local computations rest on one chain-level device, which we now
describe, together with two standard facts about nerves. In this section
and the next two, homology has integer coefficients, and chain complexes
of simplicial complexes are \emph{augmented}: the empty simplex
$[\varnothing]$ spans degree $-1$, and $\partial[v]=[\varnothing]$ for a
vertex $v$. Thus $\Hr_{-1}(L)=0$ means that $L$ has at least one vertex.
For complexes $K,L$ on disjoint vertex sets, their \emph{join} is
\[
 K*L=\{\sigma\cup\tau:\ \sigma\in K,\ \tau\in L\}.
\]
Either component face may be empty, so the join contains both factors.
Joining with a one-vertex complex gives a cone.

Let $K$ be a finite simplicial complex whose vertex set is divided into
\emph{base} vertices and \emph{fiber} vertices. For a face $R$ of $K$
consisting of base vertices ($R=\varnothing$ allowed), the \textbf{fiber}
of $R$ is the simplicial complex
\[
 K_R=\{S:\ S\text{ is a set of fiber vertices and }R\cup S\in K\},
\]
which always contains the empty set.

\begin{lemma}[Filling from fibers]\label{lem:filling}
Let $d\ge0$. Suppose that $\Hr_t(K_R)=0$ whenever $R$ is a face of base
vertices and $-1\le t\le d-|R|$. Then $\Hr_j(K)=0$ for $0\le j\le d$.
\end{lemma}

\begin{proof}
Order the vertices of $K$ so that the base vertices come first. For an
ordered face $[R]$ of base vertices and a chain $c$ of $K_R$, let
$[R]*c$ be the chain obtained by placing the vertices of $R$ in front of
those of each simplex of $c$, with $[R]*[\varnothing]=[R]$. For
$|R|\ge1$ one has
\[
     \partial([R]*c)=(\partial[R])*c+(-1)^{|R|}[R]*\partial c ,
\]
where $(\partial[R])*c$ is formed in the same way and $[\varnothing]*c=c$.

Let $z$ be a $j$-cycle of $K$ with $0\le j\le d$, and let $p$ be the
largest number of base vertices of a simplex occurring in $z$. Collect
the simplices with exactly $p$ base vertices according to their base part:
this part of $z$ is $\sum_{|R|=p}[R]*z_R$, where $z_R$ is a chain of $K_R$
of degree $j-p\ge-1$. In $\partial z$ the simplices with $p$ base
vertices are exactly those of $(-1)^p\sum_R[R]*\partial z_R$; hence each
$z_R$ is a cycle. Since $-1\le j-p\le d-p$, the hypothesis gives chains
$b_R$ of $K_R$ with $\partial b_R=z_R$, and
\[
 z-\partial\sum_{|R|=p}(-1)^p[R]*b_R
 =z-\sum_R[R]*z_R-\sum_R(-1)^p(\partial[R])*b_R
\]
is a cycle of $K$ (every simplex of $[R]*b_R$ belongs to $K$ by the
definition of $K_R$) all of whose simplices have fewer than $p$ base
vertices. Repeating, one reaches a cycle of $K_\varnothing$ of degree
$j\le d$, which bounds by hypothesis. Hence $z$ is a boundary. 
\end{proof}

The fibers that occur will be unions of full simplices. For a finite set
$S$ let $\Delta(S)$ be the full simplex on $S$ (so
$\Delta(\varnothing)=\{\varnothing\}$).

\begin{lemma}[{\cite[Theorem~10.6]{Bjorner}, \cite[Corollary~4G.3]{Hatcher}}]\label{lem:nerve}
Let $L=\bigcup_{w\in W}\Delta(S_w)$ be a union of full simplices with
nonempty vertex sets $S_w$, indexed by a finite set $W$. Then $L$ is
homotopy equivalent to its \textbf{witness nerve}, the complex on $W$
whose faces are the nonempty sets $W'\subseteq W$ with
$\bigcap_{w\in W'}S_w\ne\varnothing$.
\end{lemma}

\begin{lemma}\label{lem:skeleton}
Let $M$ be a simplicial complex on a finite vertex set $W$, and let
$t\ge0$. Suppose that every subset of $W$ with at most $t+1$ elements is
a face. Then $\Hr_s(M)=0$ for $s<t$. Call a $(t+2)$-element subset of $W$
\emph{missing} if it is not a face. If for every missing set $S$ there is
$w\in W\sm S$ such that $S'\cup\{w\}$ is a face for every $(t+1)$-element
subset $S'\subseteq S$, then also $\Hr_t(M)=0$.
\end{lemma}

\begin{proof}
The complex $M$ and the full simplex on $W$ have the same chain
groups and boundary maps through degree $t$, which gives the first
assertion. A $t$-cycle $z$ of $M$ bounds a $(t+1)$-chain
$\sum_Sc_S[S]$ in the full simplex. For a missing $S$ replace $[S]$ by
the cone $w*\partial[S]$, a chain of $M$ by hypothesis, with
$\partial(w*\partial[S])=\partial[S]$. The resulting chain of $M$ has
boundary $z$. 
\end{proof}

\section{The nine-point complex}\label{sec:nine}

In this section we prove that $\C_2(9,3)$ is acyclic through degree
seven. For $3$-sets $A,Z$ one has $d_J(A,Z)\le2$ if and only if
$A\cap Z\neq\varnothing$. Hence a family of triples is a face of
$\C_2(9,3)$ exactly when it has a \emph{transversal} of three points, that
is, a $3$-set meeting each of its members. We abbreviate sets by strings
of digits, writing $123$ for $\{1,2,3\}$.

For a family $R$ of subsets of $[8]$ let $\tau(R)$ be the least size of a
transversal of $R$ contained in $[8]$, and let $W(R)$ be the set of
$3$-subsets of $[8]$ which are transversals of $R$. For a set $W$ of
triples in $[8]$ let $\N_2(W)$ be the complex on $W$ whose faces are the
subfamilies having a transversal of at most two points of $[8]$.

\subsection{Triples on eight points}
\begin{figure}[ht]
\centering
\begin{tikzpicture}[x=1cm,y=.8cm,
 triple/.style={draw,rounded corners,minimum width=12mm,minimum height=6.5mm}]
 \node[triple] (t1) at (-2,1.2) {$123$};
 \node[triple] (t2) at (2,1.2) {$124$};
 \node[triple] (t3) at (0,-1.2) {$345$};
 \node[triple] (t4) at (5.1,0) {$678$};
 \draw (t1) to[bend left=22] node[above,font=\normalsize] {$1$} (t2);
 \draw (t1) to[bend right=13] node[below,font=\normalsize] {$2$} (t2);
 \draw (t1) -- node[left,font=\normalsize] {$3$} (t3);
 \draw (t2) -- node[right,font=\normalsize] {$4$} (t3);
 \node[font=\normalsize] at (0,-1.95) {private point $5$};
 \node[font=\normalsize,align=center] at (5.1,-.95) {private points $6,7,8$};
\end{tikzpicture}
\caption{The family $\cX$. An edge records a point lying in the two
triples it joins.}
\label{fig:X}
\end{figure}
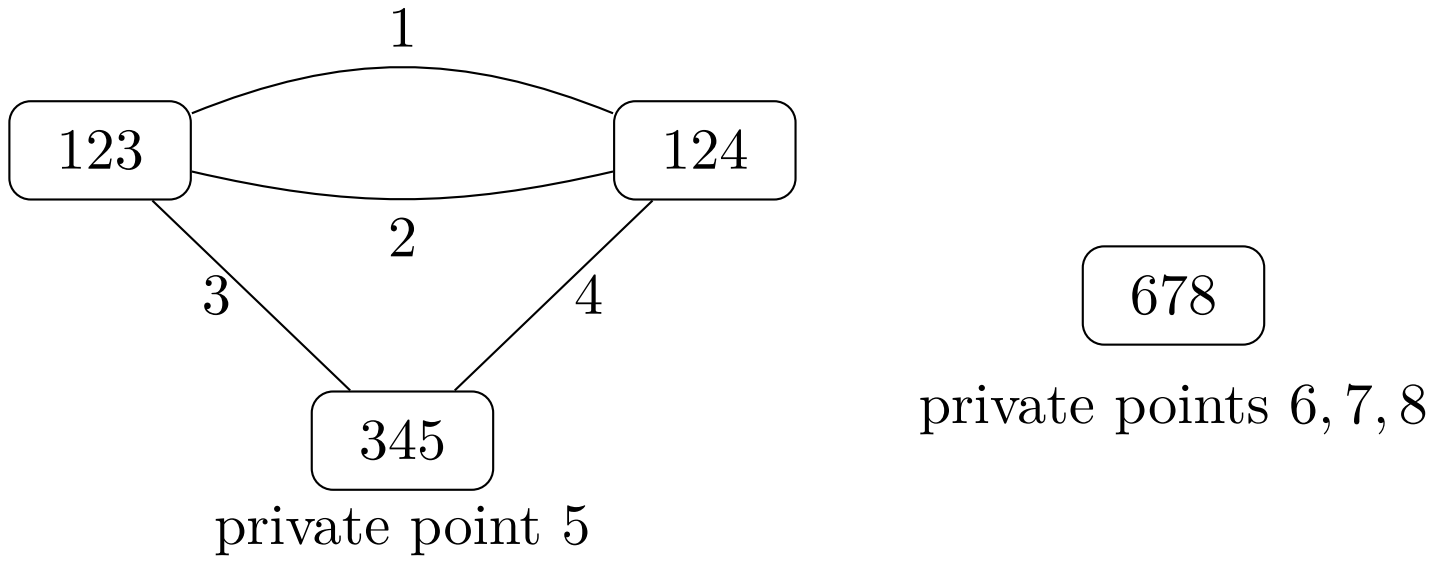

Let
\[
              \cX=\{123,\ 124,\ 345,\ 678\}.
\]
Its incidences are shown in Figure~\ref{fig:X}: the first three triples
pairwise intersect, in $12$, $3$ and $4$, and the fourth is disjoint from
them. A transversal of $\cX$ must meet $678$, and no point lies in all of
$123,124,345$; so $\tau(\cX)=3$.

\begin{lemma}[Triples on eight points]\label{lem:triples}
Let $R$ be a family of distinct triples in $[8]$.
\begin{enumerate}[(a)]
\item If $|R|\le3$ then $\tau(R)\le2$.
\item If $|R|=4$ and $\tau(R)=3$, then $R$ is the image of $\cX$ under a
permutation of $[8]$.
\item The triples $B\notin\cX$ for which some three members of $\cX$
together with $B$ have no transversal of two points are exactly
$125,135,145,235,245$.
\item If $|R|\le5$ then $|W(R)|\ge10$.
\end{enumerate}
\end{lemma}

\begin{proof}
(a) Three triples have nine incidences on eight points, so two of them
share a point $x$; then $x$ and a point of the third triple form a
transversal.

(b) No point lies in three members of $R$, since such a point together
with a point of the fourth member would be a transversal. Form a
multigraph on the four members, with one edge for each point lying in two
of them. If $e$ points lie in two members and $s$ points in one, then
$2e+s=12$ and $e+s\le8$, so $e\ge4$. Two edges without a common end would
give a two-point transversal, so the edges pairwise share an end. Hence
they all pass through one member, or all join members of a single
triangle. A member has three points, so it carries at most three edges,
and the first alternative is impossible. In a triangle the degree sum is
$2e\ge8$, while each degree is at most $3$; so $e=4$, the multiplicities
on the three sides are $2,1,1$ (a multiplicity $3$ would make two members
equal, and multiplicities $2,2,0$ would give a member of degree four), and
the fourth member carries no edge. Thus three members have the
shape of $123,124,345$, the fifth point of their union lies in only one of
them, and the fourth member consists of the three remaining points.

(c) Let $\cX'$ consist of three members of $\cX$ such that $\cX'\cup\{B\}$
has no two-point transversal. By (b) it has the shape of $\cX$: three
pairwise intersecting members with intersections of sizes $2,1,1$, no
point in three of them, and a fourth member disjoint from these. If
$\cX'=\{123,124,345\}$, then $B$ is the disjoint member, so $B=678$.
Otherwise $678\in\cX'$ is the disjoint member, and $B\subseteq12345$
completes the other two members of $\cX'$ to the intersecting part. If
these are $123,124$, then $B$ meets each of them in one point and avoids
$1,2$, so $B=345$. If they are $123,345$, then $B$ avoids $3$ and either
contains $12$ and one of $4,5$, or contains $45$ and one of $1,2$; this
gives $124,125,145,245$. If they are $124,345$, one obtains in the same way
$123,125,135,235$. Removing the members of $\cX$ leaves the five triples
stated.

(d) Adding triples can only remove transversals, so we may assume that
$R=\{R_1,\dots,R_5\}$. For a triple $Z\subseteq[8]$ let $m_Z$ be the
number of $i$ with $R_i\cap Z=\varnothing$. Each $R_i$ is disjoint from
$\binom53=10$ triples, so $\sum_Zm_Z=50$. Put
$e=\sum_{m_Z>0}(m_Z-1)$; then exactly $50-e$ triples have $m_Z>0$, and
$|W(R)|=56-50+e=6+e$. We show $e\ge4$. This is clear if some $m_Z=5$.
Otherwise $\binom{m_Z}2\le2(m_Z-1)$ whenever $m_Z>0$, and
\[
 2e\ge\sum_Z\binom{m_Z}2=\sum_{i<j}\binom{2+|R_i\cap R_j|}3
 \ge\sum_{i<j}|R_i\cap R_j| ,
\]
since $[8]\sm(R_i\cup R_j)$ has $2+|R_i\cap R_j|$ points. The last sum
equals $\sum_x\binom{\delta_x}2$, where $\delta_x$ is the number of
members containing $x$; as $\sum_x\delta_x=15$ over eight points and
$\binom\delta2\ge\delta-1$, it is at least $7$. Hence $e\ge4$. 
\end{proof}

\begin{lemma}[Two-point transversals]\label{lem:N2}
Let $W$ be a nonempty set of triples in $[8]$. Then
$\Hr_0(\N_2(W))=H_1(\N_2(W))=0$. If $|W|\ge10$, then also
$H_2(\N_2(W))=0$.
\end{lemma}

\begin{proof}
By Lemma~\ref{lem:triples}(a), every subfamily with at most three members
is a face, and Lemma~\ref{lem:skeleton} with $t=2$ gives the first
assertion. A missing four-element subfamily is a copy of $\cX$, by
Lemma~\ref{lem:triples}(b). By Lemma~\ref{lem:triples}(c), at most nine
elements of $W$ (the four members and the five triples listed there) fail
to be cone vertices for it, so for $|W|\ge10$ a cone vertex exists, and
Lemma~\ref{lem:skeleton} gives $H_2(\N_2(W))=0$. 
\end{proof}

\subsection{The fibers}
Apply Lemma~\ref{lem:filling} to $K=\C_2(9,3)$, with the triples in $[8]$
as base vertices and the triples containing $9$ as fiber vertices. We
identify the fiber vertex $\{9\}\cup X$ with the pair $X\subseteq[8]$.

\begin{lemma}[Fibers over triple families]\label{lem:93-fibers}
Let $R$ be a face of $\C_2(9,3)$ consisting of triples in $[8]$.
\begin{enumerate}[(a)]
\item If $\tau(R)\le2$, then $K_R$ is the full simplex on $\binom{[8]}2$.
\item If $\tau(R)=3$, then $W(R)\ne\varnothing$ and
\[
 K_R=\bigcup_{Z\in W(R)}\Delta\bigl(\{X:\ X\cap Z\ne\varnothing\}\bigr)
 \ \simeq\ \N_2(W(R)).
\]
\end{enumerate}
\end{lemma}

\begin{proof}
(a) If $T\subseteq[8]$ is a transversal of $R$ with $|T|\le2$, then
$T\cup\{9\}$ meets $R$ and all fiber vertices.
(b) A transversal of $R$ together with some fiber vertices, having at
most three points, cannot contain $9$, since $9$ lies in no member of $R$
and removing it would leave a transversal of $R$ with at most two points.
So it is a member of $W(R)$ meeting each of the pairs $X$. This proves the
displayed equality; $W(R)\ne\varnothing$ because $R$ is a face. A set
$W'\subseteq W(R)$ has a common vertex $X$ in the simplices exactly
when a set of at most two points meets every member of
$W'$, so the witness nerve is $\N_2(W(R))$, and
Lemma~\ref{lem:nerve} applies. 
\end{proof}

\subsection{The exceptional fiber}
The degree budget of Lemma~\ref{lem:filling} is most demanding for the
family $\cX$, where degrees up to three are required.

\begin{lemma}[The exceptional fiber]\label{lem:exceptional}
$\Hr_t(K_\cX)=0$ for $-1\le t\le3$.
\end{lemma}

\begin{proof}
Put $A=12345$, $J=678$, and let $\Gamma$ be the set of pairs in $A$
meeting each of $123,124,345$:
\[
            \Gamma=\{13,\ 14,\ 15,\ 23,\ 24,\ 25,\ 34\}.
\]
We regard $\Gamma$ as a graph on $A$; every point of $A$ lies on an edge
of $\Gamma$.

\step{Step 1 (the centers).}
A member of $W(\cX)$ meets $678$. It cannot contain two points of $J$,
since no point lies in all of $123,124,345$. Hence
\[
     W(\cX)=\{P\cup\{e\}:\ P\in\Gamma,\ e\in J\},
\]
twenty-one triples; we call $e$ the \emph{type} of the center $P\cup\{e\}$.

\step{Step 2 (three cones).}
For $e\in J$ let $K_e\subseteq K_\cX$ consist of the families of pairs met
by a single center of type $e$. Then $K_\cX=K_6\cup K_7\cup K_8$. Any pair
containing $e$ meets every center of type $e$; hence each $K_e$, and each
$K_e\cap K_{e'}$ ($e\ne e'$), is a cone with apex $\{e,e'\}$. Put
$\Theta=K_6\cap K_7\cap K_8$. The Mayer--Vietoris sequence for augmented
chains shows, in turn, that $K_6\cup K_7$ is acyclic; that
$\Hr_t((K_6\cup K_7)\cap K_8)\cong\Hr_{t-1}(\Theta)$, because
$(K_6\cup K_7)\cap K_8=(K_6\cap K_8)\cup(K_7\cap K_8)$ is a union of two
cones meeting in $\Theta$; and finally that
\[
              \Hr_t(K_\cX)\cong\Hr_{t-2}(\Theta)\qquad\text{for all }t .
\]
It therefore suffices to show that $\Theta$ is nonempty and connected and
that $H_1(\Theta)=0$.

\step{Step 3 (the vertices of $\Theta$).}
A family $\mathcal S$ of pairs lies in $\Theta$ exactly when for each
$e\in J$ some $P_e\in\Gamma$ satisfies: $P_e\cup\{e\}$ meets every member
of $\mathcal S$. A pair inside $J$ misses every center of the third type.
Hence the vertices of $\Theta$ are the ten pairs inside $A$, which we call
\emph{inner}, and the fifteen pairs $\{x,e\}$ with $x\in A$, $e\in J$,
which we call \emph{cross} vertices.

\step{Step 4 (the inner part).}
Let $\Theta_A$ be the subcomplex induced on the inner vertices. A family
of inner pairs lies in $\Theta$ exactly when a single edge $P\in\Gamma$
meets all of them (use $P$ for all three types; conversely the witness
for type $6$ works). So $\Theta_A=\bigcup_{P\in\Gamma}\Delta(\{\text{inner
pairs meeting }P\})$, and by Lemma~\ref{lem:nerve} it is homotopy
equivalent to the complex on $\Gamma$ whose faces are the sets of edges
having a transversal of two points of $A$. Three edges always have one:
three disjoint pairs would need six points, so two of them share a point,
which together with a point of the third edge is a transversal. By
Lemma~\ref{lem:skeleton}, $\Theta_A$ is connected and $H_1(\Theta_A)=0$.

\step{Step 5 (the cross vertices).}
Let $v=\{x,e\}$ be a cross vertex. Three observations hold.
\begin{enumerate}[(i)]
\item The inner pairs containing $x$, together with $v$, form a face of
$\Theta$: use an edge of $\Gamma$ through $x$ for all three types.
\item If an inner vertex $Y$ is adjacent to $v$, there is an edge
$P\in\Gamma$ through $x$ with $\{v,Y,P\}\in\Theta$. Indeed, choose
$e'\in J\sm\{e\}$; the witness $P_{e'}\cup\{e'\}$ for $\{v,Y\}$ must meet
$v$ in $x$, so $P=P_{e'}$ contains $x$ and meets $Y$, and using $P$ for all
three types shows $\{v,Y,P\}\in\Theta$.
\item If $v'=\{x',e'\}$ is a cross vertex adjacent to $v$, there is an
inner vertex $c$ with $\{v,v',c\}\in\Theta$. If $x=x'$ take for $c$ any
edge of $\Gamma$ through $x$. If $x\ne x'$, choose $e''\in J\sm\{e,e'\}$;
the witness of type $e''$ for $\{v,v'\}$ must contain $x$ and $x'$, so
$\{x,x'\}\in\Gamma$, and $c=\{x,x'\}$ works with this edge used for all
three types.
\end{enumerate}

\step{Step 6 (conclusion).}
By (i) every cross vertex is adjacent to an inner vertex, and $\Theta_A$ is
connected; so $\Theta$ is nonempty and connected. Let $\gamma$ be a closed
edge path in $\Theta$. Replace each edge $vv'$ between cross vertices by
the path $vcv'$ of (iii); the difference is the boundary of the triangle
$\{v,v',c\}$. Each remaining visit to a cross vertex $v$ has the form
$YvY'$ with $Y,Y'$ inner. Consider the graph on the inner neighbors of
$v$ in which $Y_1,Y_2$ are joined when $\{v,Y_1,Y_2\}\in\Theta$. By (i) the
inner pairs through $x$ form a complete subgraph, and by (ii) every
vertex is joined to one of them; so this graph is connected, and contains
a path $Y=Y_0,Y_1,\dots,Y_q=Y'$. The triangles $\{v,Y_l,Y_{l+1}\}$ show
that $YvY'$ is homologous, relative to its ends, to the path
$Y_0Y_1\cdots Y_q$ in $\Theta_A$. After these replacements $\gamma$ is
homologous to a cycle of $\Theta_A$, which bounds there. Since every
integral $1$-cycle is a sum of closed edge paths, $H_1(\Theta)=0$. 
\end{proof}

\begin{proposition}[The nine-point vanishing]\label{prop:93}
$\Hr_j(\C_2(9,3);\Z)=0$ for $0\le j\le7$.
\end{proposition}

\begin{proof}
By Lemma~\ref{lem:filling} with $d=7$ it suffices to show
$\Hr_t(K_R)=0$ for $-1\le t\le7-|R|$, for every face $R$ of triples in
$[8]$. If $\tau(R)\le2$, which includes all $R$ with $|R|\le3$ by
Lemma~\ref{lem:triples}(a), the fiber is a full simplex. Let $\tau(R)=3$;
then $K_R\simeq\N_2(W(R))$ with $W(R)\ne\varnothing$, and $|R|\ge4$. If
$|R|\ge6$, only $t\le1$ is required, and Lemma~\ref{lem:N2} applies. If
$|R|=5$, then $t\le2$ is required, and Lemma~\ref{lem:triples}(d) and
Lemma~\ref{lem:N2} apply. If $|R|=4$, then $R$ is a copy of $\cX$ by
Lemma~\ref{lem:triples}(b), and Lemma~\ref{lem:exceptional} supplies all
degrees $t\le3$. 
\end{proof}

\section{The eight-point complex}\label{sec:eight}

In this section we prove that $\C_2(8,4)$ is acyclic through degree
seven. We identify $[8]$ with $\Omega=\F_2^3$; the labels are the seventy
$4$-subsets of $\Omega$. The fourteen affine planes play a special role:
they will be the fiber vertices in Lemma~\ref{lem:filling}, and a Fourier
transform places them on the fourteen signed coordinate axes of $\R^7$.

The uniform fiber bound handles families with at least five nonplane
labels. For smaller families, a degree shift leaves three borderline
configurations, treated by the majority, one-missing-plane, and
ten-center lemmas. The table in the proof of
Proposition~\ref{prop:84} records how these inputs fit together.

\subsection{Signed vectors}
For $u,x\in\Omega$ let $u\cdot x\in\F_2$ be the dot product and
$\chi_u(x)=(-1)^{u\cdot x}$. For a $4$-set $A$ define $v_A\in\R^7$, with
coordinates indexed by $u\in\Omega\sm\{0\}$, by
\[
          v_A(u)=\tfrac12\sum_{x\in A}\chi_u(x).
\]
Let $\cP$ be the set of affine planes (cosets of $2$-dimensional
subspaces); the other $56$ labels are \emph{nonplanes}. Put
$s(A)=\sum_{x\in A}x\in\Omega$, and for $a\ne0$ let
$S_a=\{u:\ u\cdot a=1\}$, a set of four coordinates. We write $e_u$ for
the standard basis vectors of $\R^7$ and $A^c=\Omega\sm A$.

\begin{lemma}[The Fourier coordinates]\label{lem:signed}
\begin{enumerate}[(a)]
\item $v_A\cdot v_Z=2|A\cap Z|-4$. Hence $\|v_A\|^2=4$,
$v_{A^c}=-v_A$, the map $A\mapsto v_A$ is injective, and $Z$ is a center
of $A$ if and only if $v_A\cdot v_Z\ge0$. If
$v_{Z_1}+\dots+v_{Z_t}=0$ then $t$ is even.
\item $A$ is a plane if and only if $s(A)=0$. The plane
$\{x:\ u\cdot x=b\}$ has the vector $2(-1)^be_u$, so the planes
correspond to the fourteen vectors $\pm2e_u$.
\item If $A$ is a nonplane and $a=s(A)$, then $v_A$ is supported on $S_a$,
with entries $\pm1$ whose product is $-1$. Each of the $7\cdot8=56$ sign
patterns of this kind is $v_A$ for exactly one nonplane $A$.
\item Distinct supports $S_a\ne S_b$ meet in two coordinates. Two sign
patterns on the same support, each with product $-1$, agree in an even
number of coordinates.
\end{enumerate}
\end{lemma}

\begin{proof}
(a) For $g\colon\Omega\to\R$ put $\hat g(u)=\sum_xg(x)\chi_u(x)$.
Orthogonality of characters gives
$\sum_{u\in\Omega}\hat g(u)\hat h(u)=8\sum_xg(x)h(x)$ and the inversion
formula $g=\frac18\sum_u\hat g(u)\chi_u$. For $g_A=\one_A-\frac12$ one has
$\hat g_A(0)=0$ and $\hat g_A(u)=2v_A(u)$ for $u\ne0$; hence
$4\,v_A\cdot v_Z=8\sum_xg_Ag_Z=8(|A\cap Z|-2)$. Also $g_{A^c}=-g_A$. By
inversion $v_A$ determines $g_A$, and if $\sum_iv_{Z_i}=0$ then
$\sum_ig_{Z_i}=0$, so that $\sum_i\one_{Z_i}$ is the constant $t/2$, which
must be an integer.

(b), (c) The product of the four signs $\chi_u(x)$, $x\in A$, equals
$\chi_u(s(A))$. If $u\cdot s(A)=1$, an odd number of them is $-1$ and
$v_A(u)=\pm1$; if $u\cdot s(A)=0$, an even number is $-1$ and
$v_A(u)\in\{0,\pm2\}$. If $a=s(A)\ne0$, the four coordinates of $S_a$
already contribute $4=\|v_A\|^2$, so all other entries vanish. If
$s(A)=0$, all entries lie in $\{0,\pm2\}$, so exactly one is nonzero, say
$v_A=2(-1)^be_u$; then $\chi_u(x)=(-1)^b$ for all $x\in A$, and $A$ is the
plane $\{x:u\cdot x=b\}$. Conversely the four points $p,p+v,p+w,p+v+w$ of
a plane sum to $0$. This proves (b) and the first part of (c).

Translating $A$ by $y$ multiplies $v_A(u)$ by $\chi_u(y)$ and does not
change $s(A)$; the product of these factors over $S_a$ is
$\chi_{\sum_{u\in S_a}u}(y)=1$, because $S_a$ is an affine plane in the
space of characters and its four elements sum to $0$. An invertible linear
map $g$ of $\Omega$ satisfies $v_{gA}(u)=v_A(g^{\top}u)$, and so
permutes coordinates. Nonplanes are exactly the affinely independent
$4$-sets (an affine dependence among four distinct points of $\Omega$
says that their sum is $0$), and the affine group acts transitively on
them; so the sign product is the same for all nonplanes. For $A=\{000,100,010,001\}$ one finds
$a=(1,1,1)$, $S_a=\{100,010,001,111\}$, and entries $1,1,1,-1$; the
product is $-1$. Finally there are $\binom84-14=56$ nonplanes, $7\cdot8$
such sign patterns, and $A\mapsto v_A$ is injective.

(d) The conditions $u\cdot a=u\cdot b=1$ are independent, so they have two
solutions. On a common support, the coordinatewise product of two
patterns of product $-1$ has product $+1$; so they differ, and hence
agree, in an even number of coordinates. 
\end{proof}

\subsection{Exclusions}\label{ss:exclusions}
For a label $Z$, its \textbf{exclusion} is the set of planes for which $Z$
is not a center:
\[
          \Ex(Z)=\{p\in\cP:\ v_p\cdot v_Z<0\}.
\]
By Lemma~\ref{lem:signed}, the plane with vector $2\eta e_u$ ($\eta=\pm1$)
lies in $\Ex(Z)$ exactly when $\eta\,v_Z(u)<0$. Thus $\Ex(p)=\{p^c\}$ for
a plane $p$, while a nonplane excludes four planes, one on each coordinate
of its support, of sign opposite to the entry there; the exclusion of a
nonplane determines its vector. For a set $\mathcal S$ of labels put
$\Ex(\mathcal S)=\bigcup_{Z\in\mathcal S}\Ex(Z)$, and say that
$\mathcal S$ \emph{covers} $\cP$ if $\Ex(\mathcal S)=\cP$. The following
reformulation is used throughout:
\begin{equation}\label{eq:cover}
\begin{minipage}{0.86\textwidth}
$\mathcal S$ covers $\cP$ if and only if, at every coordinate $u$, the
entries $v_Z(u)$, $Z\in\mathcal S$, include a positive and a negative
number.
\end{minipage}
\end{equation}

\begin{lemma}[Three and four exclusions]\label{lem:exclusions}
\begin{enumerate}[(a)]
\item For any three labels,
$|\Ex(Z_1)\cup\Ex(Z_2)\cup\Ex(Z_3)|\le11$. In particular three labels
never cover $\cP$.
\item If four labels $Z_1,\dots,Z_4$ cover $\cP$, they are nonplanes in
\textbf{star position}: there are seven distinct coordinates $u_0$ and
$u_{ij}=u_{ji}$ ($1\le i<j\le4$) such that the support of $v_{Z_i}$ is
$\{u_0\}\cup\{u_{ij}:j\ne i\}$, the two nonzero entries at each $u_{ij}$
are opposite, and the four entries at $u_0$ are two $+1$'s and two
$-1$'s. In particular $\sum_iv_{Z_i}=0$.
\end{enumerate}
\end{lemma}

\begin{proof}
(a) By \eqref{eq:cover}, covering requires two nonzero entries at each of
the seven coordinates, whereas three labels have at most twelve nonzero
entries. For the sharper bound, a union of size twelve requires three
nonplanes with pairwise disjoint exclusions. Two exclusions are disjoint
exactly when the two vectors have opposite signs at every common
coordinate. If two of the three vectors have the same support, they are
therefore opposite; the third vector meets this support in at least two
coordinates, and there it agrees in sign with one of them. If the three
supports are distinct, then $v_{Z_i}\cdot v_{Z_j}=-2$ for $i\ne j$ (two
common coordinates, opposite signs), and
$\|v_{Z_1}+v_{Z_2}+v_{Z_3}\|^2=12-12=0$, contrary to
Lemma~\ref{lem:signed}(a).

(b) If one label is a plane, the supports have total size at most
$1+12<14$, which is impossible by \eqref{eq:cover}. Let $a_i=s(Z_i)$ and
$\Phi(u)=(u\cdot a_1,\dots,u\cdot a_4)\in\F_2^4$. The map $\Phi$ is
linear, and $u$ lies in the support of $v_{Z_i}$ exactly when
$\Phi(u)_i=1$. By \eqref{eq:cover}, $\Phi(u)$ has weight at least two for
every $u\ne0$. So $\Phi$ is injective, and its image is a
$3$-dimensional subspace, the kernel of $t\mapsto\sum_{i\in I}t_i$ for a
nonempty $I\subseteq[4]$. This kernel contains the unit vectors $f_i$,
$i\notin I$, of weight one; hence $I=[4]$, and the image consists of the
vectors of even weight: the six vectors $f_i+f_j$ and $(1,1,1,1)$. Put
$u_{ij}=\Phi^{-1}(f_i+f_j)$ and $u_0=\Phi^{-1}(1,1,1,1)$. By
\eqref{eq:cover} the two entries at $u_{ij}$ are opposite. The sixteen
nonzero entries have product $(-1)^4=1$ by Lemma~\ref{lem:signed}(c), and
the twelve entries at the $u_{ij}$ have product $(-1)^6=1$; so the four
entries at $u_0$ have product $1$, and since both signs occur there, two
are positive and two negative. 
\end{proof}

\begin{figure}[ht]
\centering
\begin{tikzpicture}[scale=1.6]
\node[circle,draw,inner sep=1.5pt] (1) at (0,1.2) {$1$};
\node[circle,draw,inner sep=1.5pt] (3) at (1.8,1.2) {$3$};
\node[circle,draw,inner sep=1.5pt] (2) at (1.8,0) {$2$};
\node[circle,draw,inner sep=1.5pt] (4) at (0,0) {$4$};
\draw[very thick] (1)--node[above,font=\normalsize]{$u_{13}$}(3)
   --node[right,font=\normalsize]{$u_{23}$}(2)
   --node[below,font=\normalsize]{$u_{24}$}(4)
   --node[left,font=\normalsize]{$u_{14}$}(1);
\draw[dashed] (1)--node[pos=.22,sloped,above,font=\normalsize,inner sep=1pt]{$u_{12}$}(2);
\draw[dashed] (3)--node[pos=.22,sloped,above,font=\normalsize,inner sep=1pt]{$u_{34}$}(4);
\node[font=\normalsize,align=left,anchor=west] at (2.3,.6)
 {$u_0$: common to all four supports\\[2pt]
  support of $Z_i$: $u_0$ and the three edges at $i$\\[2pt]
  bold $4$-cycle: support of $b^\pm$ (Lemma~\ref{lem:ten})};
\end{tikzpicture}
\caption{Star position. The coordinates $u_{ij}$ are the edges of the
complete graph on $\{1,2,3,4\}$; $u_0$ is the remaining coordinate.}
\label{fig:star}
\end{figure}
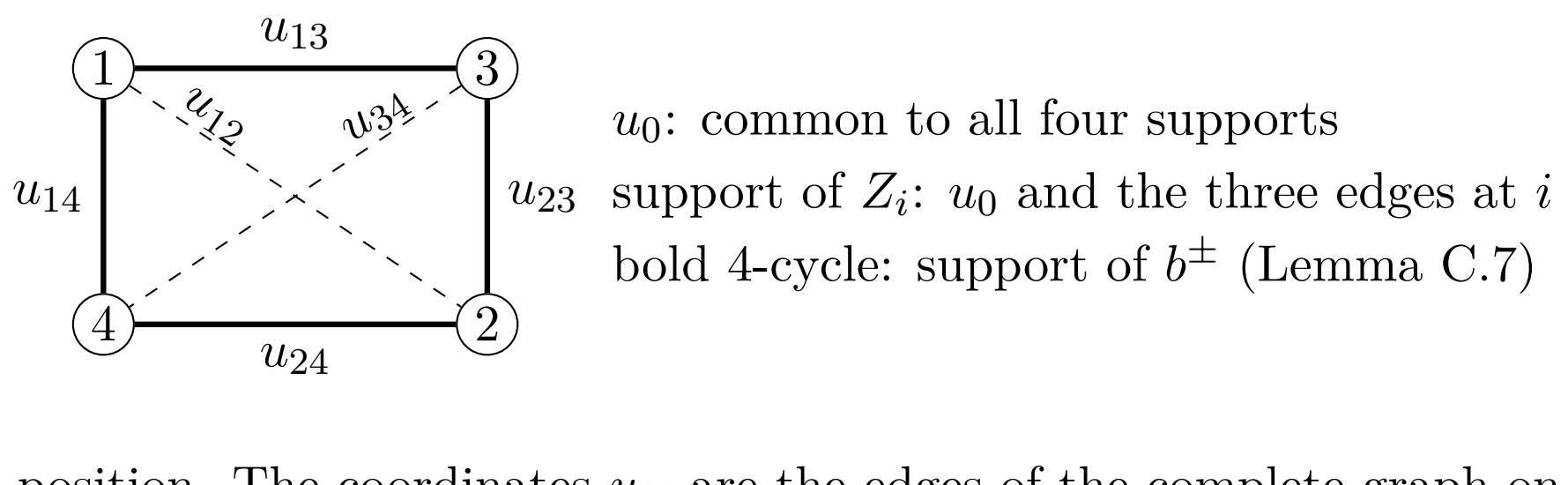

\subsection{The fibers and a uniform bound}
Apply Lemma~\ref{lem:filling} to $K=\C_2(8,4)$, with the nonplanes as
base vertices and the planes as fiber vertices. For a face $R$ of
nonplanes let $W(R)$ be its set of centers, which is nonempty. A set of
planes is a face of $K_R$ exactly when some center of $R$ excludes none of
them, so
\[
 K_R=\bigcup_{Z\in W(R)}\Delta\bigl(\cP\sm\Ex(Z)\bigr).
\]
By Lemma~\ref{lem:nerve}, $K_R$ is homotopy equivalent to the witness
nerve $\N_R$, the complex on $W(R)$ whose faces are the sets of centers
which do not cover $\cP$.

\begin{lemma}[A uniform fiber bound]\label{lem:uniform}
For every face $R$ of nonplanes ($R=\varnothing$ allowed),
$\Hr_t(K_R)=0$ for $-1\le t\le2$.
\end{lemma}

\begin{proof}
By Lemma~\ref{lem:exclusions}(a), every set of at most three centers is a
face of $\N_R$, so $\Hr_t(\N_R)=0$ for $t\le1$ by
Lemma~\ref{lem:skeleton}. Let $\{Z_1,\dots,Z_4\}\subseteq W(R)$ be missing.
By Lemma~\ref{lem:exclusions}(b) the $Z_i$ are in star position and
$\sum_iv_{Z_i}=0$. For $A\in R$ the four numbers $v_A\cdot v_{Z_i}$ are
nonnegative with sum $0$, so they vanish; hence
$v_A\cdot v_{Z_1^c}=-v_A\cdot v_{Z_1}=0$, and $Z_1^c\in W(R)$. It differs
from $Z_1,\dots,Z_4$, since its support is that of $Z_1$ while its vector
is $-v_{Z_1}$, and it is a cone vertex. Indeed, $\{Z_1^c,Z_2,Z_3,Z_4\}$ does not cover $\cP$, since at
$u_{12}$ the only nonzero entries are those of $Z_2$ and $Z_1^c$, and
these have the same sign. For $j\ne1$ choose $k\notin\{1,j\}$; then
$\{Z_1^c\}\cup(\{Z_1,\dots,Z_4\}\sm\{Z_j\})$ does not cover $\cP$, since at
$u_{jk}$ only $Z_k$ has a nonzero entry. Lemma~\ref{lem:skeleton} gives
$H_2(\N_R)=0$. 
\end{proof}

\subsection{The degree shift}
For the faces with at most four labels a finer analysis is needed. The
point is that the plane centers of $R$ contribute a contractible part,
which shifts degrees.

Let $R$ be a nonempty face of nonplanes, and put $m=|\Ex(R)|$, so that
$4\le m\le14$. A plane is a center of $R$ exactly when it lies outside
$\Ex(R)$. Let
\[
 \cP_0=\{p^c:\ p\in\cP\sm\Ex(R)\},\qquad \cP_1=\cP\sm\cP_0 ,
\]
so that $\cP_0$ consists of the exclusions of the plane centers and
$|\cP_1|=m$. Call a nonplane center $Z$ of $R$ \emph{essential} if
$\Ex(Z)\subseteq\cP_1$, and put
\[
   J_R=\bigcup_{Z\ \mathrm{essential}}\Delta\bigl(\cP_1\sm\Ex(Z)\bigr),
\]
a complex on $\cP_1$, possibly equal to $\{\varnothing\}$.

\begin{lemma}[The degree shift]\label{lem:shift}
Let $R$ be a nonempty face of nonplanes, with $m=|\Ex(R)|$.
\begin{enumerate}[(a)]
\item If $m<14$ and $R$ has no essential center, $K_R$ is contractible.
\item Otherwise $\Hr_t(K_R)\cong\Hr_{t+m-14}(J_R)$ for every $t$.
\item Suppose that $R$ has an essential center. Then $J_R$ has a vertex if
$m\ge5$. Moreover, for $m\ge5$, $J_R$ is homotopy equivalent to the
complex $\N'_R$ on the essential centers whose faces are the sets
$\mathcal S$ with $\Ex(\mathcal S)\ne\cP_1$; and $J_R$ is connected if
$m\ge9$, and satisfies $H_1(J_R)=0$ if $m\ge12$.
\end{enumerate}
\end{lemma}

\begin{proof}
A plane center $p$ contributes the simplex $\Delta(\cP\sm\{p^c\})$, and
the union of these is
\[
 D_R=\{S\subseteq\cP:\ S\not\supseteq\cP_0\},
\]
the join of the boundary of $\Delta(\cP_0)$ with $\Delta(\cP_1)$. If a
nonplane center $Z$ excludes some $p^c\in\cP_0$, its simplex lies in
$\Delta(\cP\sm\{p^c\})\subseteq D_R$. The simplex of an essential center
consists of the sets $S_0\cup S_1$ with $S_0\subseteq\cP_0$ and
$S_1\in\Delta(\cP_1\sm\Ex(Z))$. Hence
\[
        K_R=D_R\cup\{S_0\cup S_1:\ S_0\subseteq\cP_0,\ S_1\in J_R\}.
\]
(a) If $m<14$ there is a plane center, so $\cP_0\ne\varnothing$, and
$\cP_1\ne\varnothing$ because $m\ge4$. Without essential centers
$K_R=D_R$, which is a cone with any vertex of $\cP_1$ as apex.

(b) If $m=14$ there are no plane centers, every nonplane center is
essential, and $K_R=J_R$. If $m<14$, then $D_R$ is contractible, so
$\Hr_t(K_R)\cong H_t(K_R,D_R)$. The faces of $K_R$ not in $D_R$ are the
sets $\cP_0\cup S$ with $S\in J_R$ (including $S=\varnothing$). Ordering
$\cP_0$ first, the boundary of $[\cP_0]*[S]$ modulo $D_R$ is
$(-1)^{|\cP_0|}[\cP_0]*\partial[S]$, because deleting a vertex of $\cP_0$
gives a face of $D_R$. So the relative chain complex is the augmented
chain complex of $J_R$ shifted up by $|\cP_0|=14-m$.

(c) An exclusion of a nonplane has four elements, so for $m\ge5$ every
essential center contributes a nonempty simplex, and
Lemma~\ref{lem:nerve} identifies $J_R$ with $\N'_R$ up to homotopy. Two
exclusions have at most eight elements, and three have at most eleven by
Lemma~\ref{lem:exclusions}(a). So for $m\ge9$ every pair of essential
centers is a face of $\N'_R$, and for $m\ge12$ every triple is; now apply
Lemma~\ref{lem:skeleton}. 
\end{proof}

\subsection{Three borderline configurations}
Three configurations are not settled by part (c) of
Lemma~\ref{lem:shift}; the following lemmas treat them.

\begin{lemma}[Majority center]\label{lem:majority}
Let $Z_1,Z_2,Z_3$ be labels with
$|\Ex(Z_1)\cup\Ex(Z_2)\cup\Ex(Z_3)|=11$, and let $M$ be the set of points
lying in at least two of them. Then $M$ is a $4$-set, and
\begin{enumerate}[(1)]
\item $M$ is a center of every label of which $Z_1,Z_2,Z_3$ are centers;
\item $\Ex(M)\subseteq\Ex(Z_1)\cup\Ex(Z_2)\cup\Ex(Z_3)$;
\item for each pair $i,j$, the set $\Ex(M)\cup\Ex(Z_i)\cup\Ex(Z_j)$ is
strictly smaller than $\Ex(Z_1)\cup\Ex(Z_2)\cup\Ex(Z_3)$.
\end{enumerate}
\end{lemma}

\begin{proof}
Write $E_i=\Ex(Z_i)$, $v_i=v_{Z_i}$, and $E=E_1\cup E_2\cup E_3$. If some
$Z_i$ were a plane, $|E|\le9$; so all three are nonplanes. As
$|E_1|+|E_2|+|E_3|=12$, exactly one plane is excluded twice; relabel so
that $|E_1\cap E_2|=1$ and $E_3$ is disjoint from $E_1\cup E_2$. Two
exclusions share the plane on a coordinate $u$ exactly when the two
vectors have nonzero entries of the same sign at $u$. Thus $v_1$ and $v_2$
agree in sign at exactly one common coordinate; by
Lemma~\ref{lem:signed}(d) their supports are distinct, they have two
common coordinates, and $v_1\cdot v_2=0$. The vector $v_3$ is opposite to
$v_1$ and to $v_2$ at all common coordinates. If $v_3$ had the support of
$v_1$, then $v_3=-v_1$, and at the two coordinates common to $v_1$ and
$v_2$ the vector $v_3$ would agree in sign with $v_2$ once; the same
argument applies to the support of $v_2$. So the three supports are
distinct, and $v_1\cdot v_3=v_2\cdot v_3=-2$. By
Lemma~\ref{lem:signed}(a), $|Z_1\cap Z_2|=2$ and
$|Z_1\cap Z_3|=|Z_2\cap Z_3|=1$, and inclusion--exclusion gives
$|Z_1\cup Z_2\cup Z_3|=8+|Z_1\cap Z_2\cap Z_3|\le8$. Hence no point lies
in all three sets, their union is $\Omega$, $|M|=4$, and
$\one_M=\one_{Z_1}+\one_{Z_2}+\one_{Z_3}-\one$. Since
$\one_M-\frac12=\sum_i(\one_{Z_i}-\frac12)$, this gives
\[
                    v_M=v_1+v_2+v_3 .
\]
(1) If $v_A\cdot v_i\ge0$ for $i=1,2,3$, then $v_A\cdot v_M\ge0$.
(2) If $v_p\cdot v_M<0$, then $v_p\cdot v_i<0$ for some $i$.
(3) One has $v_M\cdot v_1=v_M\cdot v_2=2$ and $v_M\cdot v_3=0$. We claim
that $|\Ex(M)\cap E_i|\le2$ for $i=1,2$ and $E_3\not\subseteq\Ex(M)$. If
$M$ is a plane, $\Ex(M)$ has one element. If $M$ is a nonplane, it cannot
have the support of $v_i$ ($i=1,2$), since an inner product $2$ on a
common support of four coordinates would mean three agreements, contrary
to Lemma~\ref{lem:signed}(d); so the supports meet in two coordinates and
$|\Ex(M)\cap E_i|\le2$. If $E_3\subseteq\Ex(M)$, then the two exclusions
are equal, so $v_M=v_3$ and $v_1+v_2=0$, contrary to $v_1\cdot v_2=0$. Now
$\Ex(M)\cup E_1\cup E_2$ misses a plane of $E_3$, and
$\Ex(M)\cup E_1\cup E_3$ misses one of the three planes of $E_2\sm E_1$;
the case $\Ex(M)\cup E_2\cup E_3$ is symmetric. 
\end{proof}

\begin{lemma}[One missing plane]\label{lem:holes}
\begin{enumerate}[(a)]
\item Let $\mathcal A=\{A_1,\dots,A_4\}$ be nonplanes with
$\Ex(\mathcal A)=\cP\sm\{q\}$, where $q$ has the vector $2e_h$, so that
$q=\{x:h\cdot x=0\}$. Put $d=\sum_iv_{A_i}$ and
$w=\sum_i\one_{A_i}-2\cdot\one$, a function on $\Omega$. Then one of the
following holds:
\begin{enumerate}[(i)]
\item $d=4e_h$;
\item $d=2e_h+2\delta e_u$ for a coordinate $u\ne h$ and $\delta=\pm1$;
\item the vectors $v_{A_i}$ are linearly independent, and
$w=\delta_x-\delta_y$ with $x\in q$, $y\notin q$, where $\delta_x$ is the
indicator function of the point $x$.
\end{enumerate}
\item \emph{(Opposite holes.)} Let $\mathcal A$ and $\mathcal A'$ be sets
of four nonplanes with $\Ex(\mathcal A)=\cP\sm\{q\}$ and
$\Ex(\mathcal A')=\cP\sm\{q^c\}$ for some plane $q$. Then
$v_A\cdot v_{A'}<0$ for some $A\in\mathcal A$ and $A'\in\mathcal A'$.
\end{enumerate}
\end{lemma}

\begin{proof}
(a) Since $q$ is not excluded, all entries at $h$ are nonnegative, and
since $q^c$ is excluded, one of them is positive. Every other coordinate
carries entries of both signs. Define $\Phi$ as in the proof of
Lemma~\ref{lem:exclusions}(b). Now $\Phi(u)$ has weight at least two for
$u\notin\{0,h\}$ and at least one for $u=h$; so $\Phi$ is injective, and
its image is the kernel of $t\mapsto\sum_{i\in I}t_i$ for some nonempty
$I$. The image contains $4-|I|$ unit vectors, and at most one vector of
weight one is allowed; so $|I|\ge3$.

If $|I|=4$, the image consists of the vectors of even weight, and the
supports have the star shape of Lemma~\ref{lem:exclusions}(b), with
coordinates $u_0$ and $u_{ij}$; at each $u_{ij}\ne h$ the two entries are
opposite. If $h=u_0$, all four entries at $h$ are $+1$, and $d=4e_h$. If
$h=u_{ij}$, both entries at $h$ are $+1$, so the twelve entries at the
pair coordinates have product $(-1)^5=-1$. The sixteen entries have
product $1$, so the four entries at $u_0$ have product $-1$ and sum
$\pm2$. This gives (ii) with $u=u_0$.

If $|I|=3$, say $I=\{1,2,3\}$, the nonzero vectors of the image are $f_4$,
which must be $\Phi(h)$, the three vectors $f_i+f_j$ ($i<j\le3$), and the
three vectors $f_i+f_j+f_4$. At $h$ only $v_{A_4}$ is nonzero, with entry
$1$; at the coordinates of weight two the entries are opposite; at those
of weight three both signs occur among three entries $\pm1$. Hence $d$ has
four entries $\pm1$, with $d(h)=1$, and $\|d\|^2=4$. The function $w$ has
sum $0$ and transform $\hat w(u)=2d(u)$ for $u\ne0$; so
$8\sum_xw(x)^2=\sum_u\hat w(u)^2=16$. An integer-valued function with
sum $0$ and $\sum w^2=2$ has the form $\delta_x-\delta_y$, and
$2=\hat w(h)=\chi_h(x)-\chi_h(y)$ gives $x\in q$, $y\notin q$. Finally,
in a relation $\sum_ic_iv_{A_i}=0$ the coordinate $h$ gives $c_4=0$, the
coordinates $f_i+f_j$ give $c_1=c_2=c_3$, and then $c_1=0$ because three
vectors $v_A$ never sum to zero (Lemma~\ref{lem:signed}(a)).

(b) Complementing all sets negates all vectors, preserves inner products,
and replaces each excluded plane by its complement; so we may assume that
$q$ has the vector $2e_h$. Suppose that all sixteen inner products are
nonnegative. Every $v'=v_{A'}$, $A'\in\mathcal A'$, has $v'(h)\le0$, and
one of them is negative. If $\mathcal A$ is of type (i), then
$4v'(h)=d\cdot v'\ge0$, so $v'(h)=0$ for all $A'$, a contradiction. If
$\mathcal A$ is of type (ii), then $d\cdot v'\ge0$ gives
$\delta v'(u)\ge-v'(h)\ge0$ for all $A'$, so the plane with vector
$2\delta e_u$ is not excluded by $\mathcal A'$; as $u\ne h$ it differs
from $q^c$, a contradiction. The pair formed by the complements of the
members of $\mathcal A'$ and those of $\mathcal A$ satisfies the same
hypotheses, with the same $q$; so the same argument shows that the family
of complements of the members of $\mathcal A'$ is also of type (iii).
Complementation replaces $w'=\sum\one_{A'}-2\cdot\one$ by $-w'$. Hence
$w=\delta_x-\delta_y$ with $x\in q$, $y\notin q$, and
$w'=\delta_{x'}-\delta_{y'}$ with $x'\notin q$, $y'\in q$. Then $x\ne x'$ and $y\ne y'$, and
\[
 \sum_{A,A'}v_A\cdot v_{A'}=\tfrac14\sum_{u\ne0}\hat w(u)\hat w'(u)
 =2\sum_xw(x)w'(x)=-2\bigl([x=y']+[y=x']\bigr)\le0 .
\]
Here $[P]$ equals one when $P$ holds and zero otherwise.
So all sixteen inner products vanish, and the two families span
orthogonal $4$-dimensional subspaces of $\R^7$, which is impossible. 
\end{proof}

\begin{lemma}[Ten centers]\label{lem:ten}
Let $R=\{A_1,\dots,A_4\}$ consist of nonplanes which cover $\cP$. Then $R$
has exactly ten centers, and there are two disjoint four-element sets
$F_+,F_-$ of centers such that a set of at most five centers covers $\cP$
if and only if it contains $F_+$ or $F_-$.
\end{lemma}

\begin{proof}
For the current four labels, put
$\Phi(u)=(u\cdot s(A_1),\ldots,u\cdot s(A_4))\in\F_2^4$.
By Lemma~\ref{lem:exclusions}(b) the vectors $v_i=v_{A_i}$ are in star
position. Put $s_i=v_i(u_0)$ and $\varepsilon_{ij}=v_i(u_{ij})$, so that
$\varepsilon_{ji}=-\varepsilon_{ij}$, and relabel so that
$(s_1,s_2,s_3,s_4)=(1,1,-1,-1)$. Lemma~\ref{lem:signed}(c) gives
\begin{equation}\label{eq:star-signs}
          s_i\prod_{j\ne i}\varepsilon_{ij}=-1\qquad(i=1,\dots,4).
\end{equation}

\step{Step 1 (the centers).}
For a center $Z$ the four numbers $v_Z\cdot v_i$ are nonnegative with sum
$v_Z\cdot0=0$, so $v_Z$ is orthogonal to every $v_i$. A plane center would
need a coordinate outside all four supports, so the centers are
nonplanes. Under $\Phi$ the coordinates correspond to the nonzero vectors
of even weight in $\F_2^4$, and the seven supports to the seven nonzero
linear forms on that space: the forms $t_i$ give the four star supports
$S_i=\{u_0\}\cup\{u_{ij}:j\ne i\}$, and the forms
$t_i+t_j=t_k+t_l$ give the three $4$-cycles
$\{u_{ik},u_{il},u_{jk},u_{jl}\}$ of the complete graph, one for each
splitting $\{i,j\}\cup\{k,l\}$ of $[4]$ (Figure~\ref{fig:star}).

\emph{Star supports.} Let $Z$ have support $S_i$, with entry $t$ at $u_0$
and $y_j$ at $u_{ij}$. Orthogonality to $v_j$ ($j\ne i$), whose support
meets $S_i$ in $\{u_0,u_{ij}\}$, reads $ts_j+\varepsilon_{ji}y_j=0$, that
is, $y_j=ts_j\varepsilon_{ij}$. Orthogonality to $v_i$ is then
$t\sum_js_j=0$. The sign product is
$t\prod_{j\ne i}ts_j\varepsilon_{ij}=s_i\cdot(-s_i)=-1$ by
\eqref{eq:star-signs}, since $\prod_{j\ne i}s_j=s_i$. So for each $i$ and
each $t=\pm1$ there is exactly one center with support $S_i$ and entry $t$
at $u_0$; call it $w_i^t$.

\emph{Cycle supports.} Let $Z$ have the support
$\{u_{ik},u_{il},u_{jk},u_{jl}\}$, with entries $b(u_{xy})$, and put
$g(x,y)=\varepsilon_{xy}b(u_{xy})$, so that $g(y,x)=-g(x,y)$.
Orthogonality to $v_i$ reads $g(i,k)+g(i,l)=0$, and similarly at $j,k,l$:
at every vertex of the $4$-cycle $i\,k\,j\,l$ the two outgoing values
cancel. Hence $g$ is a unit circulation around the cycle, in one of its
two directions; then
$b(u_{ik})b(u_{il})b(u_{jk})b(u_{jl})=
\varepsilon_{ik}\varepsilon_{il}\varepsilon_{jk}\varepsilon_{jl}$.
Multiplying \eqref{eq:star-signs} for $i$ and $j$ and using
$\varepsilon_{ij}\varepsilon_{ji}=-1$ shows that this product equals
$-s_is_j$. A nonplane requires the value $-1$, that is, $s_i=s_j$. This
happens only for the splitting $\{1,2\}\cup\{3,4\}$. So there are exactly
two centers, $b^+$ and $b^-$, with support
$\{u_{13},u_{14},u_{23},u_{24}\}$, and none with the other two cycle
supports. In all there are ten centers.

\step{Step 2 (covers by at most five centers).}
Give $w_i^t$ the \emph{color} $ts_i$, and call
$\{w_i^+,w_i^-\}$ \emph{group} $i$. At $u_{ij}$, after multiplication by
$\varepsilon_{ij}$, the entry of $w_i^t$ is $ts_j=(ts_i)s_is_j$, and that
of $w_j^{t'}$ is $-t's_i=-(t's_j)s_is_j$. Hence, at $u_{ij}$, a center of
group $i$ and one of group $j$ supply both signs exactly when they have
the same color, while the two centers of one group always do. The
centers $b^\pm$ vanish at $u_0$, $u_{12}$ and $u_{34}$. By
\eqref{eq:cover}, a set covers $\cP$ when both signs occur at every
coordinate.

Let $\mathcal S$ be a set of at most five centers covering $\cP$.
Coverage at $u_{12}$ needs two centers from groups $1,2$, and at $u_{34}$
two from groups $3,4$; so $\mathcal S$ contains at least four $w$'s and at
most one $b$. Every group occurs. Indeed, suppose group $i$ is absent;
let $i'$ be its partner in $\{1,2\}$ or $\{3,4\}$, and $\{j,k\}$ the other
pair. Coverage at $u_{ii'}$ forces both centers of group $i'$. If group
$j$ were also absent, only a $b$ could contribute at $u_{ij}$, giving at
most one sign. So groups $j$ and $k$ occur. Without a $b$, coverage at
$u_{ij}$ and $u_{ik}$ forces both centers of groups $j$ and $k$, six
centers in all. With one $b$ there is exactly one center in each of the
groups $j,k$; coverage at $u_{jk}$ forces them to have the same color,
and then (as $s_j=s_k$) their weighted entries at $u_{ij}$ and $u_{ik}$
are equal. The center $b$ would have to supply the opposite sign at both
coordinates, so that $g(i,j)=g(i,k)\ne0$; but orthogonality of $b$ to
$v_i$ gives $g(i,j)+g(i,k)=0$. This contradiction shows that every group
occurs.

If $\mathcal S$ contains one center from each group, coverage at $u_{12}$
and $u_{34}$ forces equal colors within $\{1,2\}$ and within $\{3,4\}$.
If the two colors were different, all four entries at $u_0$
would be equal, since the entry of $w_i^t$ there is
$t=(ts_i)s_i$. The centers $b^\pm$ vanish at $u_0$, so
$\mathcal S$ would not cover $\cP$. Hence all four colors are equal. If
$\mathcal S$ consists of five $w$'s, one group is doubled; coverage at the
coordinates joining the three other groups forces their colors to be
equal, and the doubled group contains a center of that color. In both
cases $\mathcal S$ contains
\[
 F_+=\{w_1^+,w_2^+,w_3^-,w_4^-\}\quad\text{or}\quad
 F_-=\{w_1^-,w_2^-,w_3^+,w_4^+\},
\]
the four centers of color $+1$ or of color $-1$. Conversely $F_\pm$
covers $\cP$: colors agree on every $u_{ij}$, and the entries
$\pm s_i$ at $u_0$ are two of each sign. 
\end{proof}

\subsection{Assembly}

\begin{proposition}[The eight-point vanishing]\label{prop:84}
$\Hr_j(\C_2(8,4);\Z)=0$ for $0\le j\le7$.
\end{proposition}

\begin{proof}
By Lemma~\ref{lem:filling} with $d=7$ we must show $\Hr_t(K_R)=0$ for
$-1\le t\le7-r$, for every face $R$ of nonplanes, where $r=|R|$. For
$R=\varnothing$, a set of at most thirteen planes omits some plane $p^c$
and so lies in the simplex of the center $p$, while every label excludes
at least one plane; so $K_\varnothing$ is the boundary of the simplex on
$\cP$, a $12$-sphere, and is acyclic in the required range. For $r\ge5$
only $t\le2$ is required, and Lemma~\ref{lem:uniform} applies. Let
$1\le r\le4$, put $m=|\Ex(R)|$, and assume that $R$ has an essential
center or $m=14$, since otherwise $K_R$ is contractible
(Lemma~\ref{lem:shift}(a)). By Lemma~\ref{lem:shift}(b) the required
groups are $\Hr_s(J_R)$ with $s=t+m-14\le m-7-r$, and groups of degree
below $-1$ vanish. Also $m\le4r$, and $m\le11$ when $r=3$
(Lemma~\ref{lem:exclusions}(a)). The cases are as follows.
\begin{center}
\begin{tabular}{@{}cccl@{}}
\toprule
$r$ & $m$ & required $s$ & reason\\
\midrule
$1$ & $4$ & $s\le-4$ & nothing to prove\\
$2$ & $\le8$ & $s\le m-9\le-1$ & $J_R$ has a vertex ($m=8$)\\
$3$ & $\le10$ & $s\le m-10\le0$ & vertex ($m=9$); connected ($m=10$)\\
$3$ & $11$ & $s\le1$ & Lemma~\ref{lem:majority}\\
$4$ & $\le12$ & $s\le m-11\le1$ & vertex, connected, $H_1=0$
   (Lemma~\ref{lem:shift}(c))\\
$4$ & $13$ & $s\le2$ & Lemma~\ref{lem:holes}(b)\\
$4$ & $14$ & $t\le3$ (no shift) & Lemma~\ref{lem:ten}\\
\bottomrule
\end{tabular}
\end{center}
The entries not referring to a borderline lemma follow from
Lemma~\ref{lem:shift}(c). It remains to treat the three borderline cases.

\emph{Case $r=3$, $m=11$.} By Lemma~\ref{lem:shift}(c), $J_R$ is
connected, and we must show $H_1(\N'_R)=0$. Every pair of essential
centers is a face of $\N'_R$. A missing triple consists of essential
centers $Z_1,Z_2,Z_3$ whose exclusions cover $\cP_1$, so that
$|\Ex(Z_1)\cup\Ex(Z_2)\cup\Ex(Z_3)|=11$. Let $M$ be their majority set.
By Lemma~\ref{lem:majority}(1), $M$ is a center of every label of $R$. By
(2), $\Ex(M)\subseteq\cP_1$; a plane center has its exclusion in $\cP_0$,
so $M$ is an essential center. By (3), $M$ is a cone vertex for the
missing triple. Lemma~\ref{lem:skeleton} gives $H_1(\N'_R)=0$.

\emph{Case $r=4$, $m=13$.} Here $\Ex(R)=\cP\sm\{q\}$ for a plane $q$,
which is the only plane center, $\cP_0=\{q^c\}$ and
$\cP_1=\cP\sm\{q^c\}$. By Lemma~\ref{lem:shift}(c) we must still show
$H_2(\N'_R)=0$, and every triple of essential centers is a face. A missing
quadruple would consist of four essential centers, nonplanes whose
exclusions cover exactly $\cP\sm\{q^c\}$, and all of whose inner products
with the members of $R$ are nonnegative. This contradicts
Lemma~\ref{lem:holes}(b). So $\N'_R$ contains every set of at most four
essential centers, and $H_2(\N'_R)=0$ by Lemma~\ref{lem:skeleton}.

\emph{Case $r=4$, $m=14$.} Here $K_R=J_R\simeq\N_R$, and degrees $t\le2$
are covered by Lemma~\ref{lem:uniform}. By Lemma~\ref{lem:ten}, the faces
of $\N_R$ with at most five elements are exactly the sets of at most five
centers containing neither $F_+$ nor $F_-$. These are also the faces with
at most five elements of the join
\[
 \partial\Delta(F_+)*\partial\Delta(F_-)*\Delta(\{b^+,b^-\}),
\]
whose faces are the sets of centers containing neither $F_+$ nor $F_-$.
The two complexes have the same four-skeleton and hence the same third
homology; the join is a cone with apex $b^+$. So $H_3(\N_R)=0$. 
\end{proof}

\section*{Verification materials}
Every finite lemma has a written proof; no theorem depends on
computation. The source package contains four supplementary programs
and their outputs. Reproduction commands, software versions, expected
outputs, and the scope of each check are given in
\texttt{Checker\_README.txt} in the supplementary source archive.
The standard-library program
\texttt{verify\_local\_data.py} checks the pentagon and rook incidence
identities, the local coloring data in Sections~\ref{sec:pentagon}
and~\ref{sec:rook-local}, and both upper-bound constructions. It also checks
the Fourier identities and finite classifications in
Lemmas~\ref{lem:signed}, \ref{lem:exclusions}, \ref{lem:holes},
and~\ref{lem:ten}. The standard-library program
\texttt{check\_appendixC\_finite.py} adds an exhaustive check of the
majority-center Lemma~\ref{lem:majority} on all $2{,}016$ extremal
triples. The standard-library program \texttt{verify\_balancing.py}
checks the retraction and distance inequality in
Theorem~\ref{thm:exact-balancing} through ten goods.
The optional \texttt{check\_local\_coloring\_lemmas.py} checks the two
local coloring lemmas using \texttt{python-sat} and its Glucose3 solver;
it does not produce an independently checked unsatisfiability certificate.
None of these programs verifies the homology or chain comparison arguments.

\section*{AI methodology statement}
We used ChatGPT, including Codex, to assist with literature searches,
exposition and proof organization throughout the paper, the figures,
and the implementation of \texttt{verify\_local\_data.py} and
\texttt{verify\_balancing.py}. We used Claude to assist with a
pre-submission review of the exposition and with the implementation of
\texttt{check\_appendixC\_finite.py} and
\texttt{check\_local\_coloring\_lemmas.py}. The authors verified the
correctness and originality of all content, including references.

\bigskip
\noindent
Sharareh Alipour: \href{mailto:sh.alipour@newuu.uz}{\texttt{sh.alipour@newuu.uz}}\\
Amir Jafari: \href{mailto:a.jafari@newuu.uz}{\texttt{a.jafari@newuu.uz}}

\end{document}